\documentclass[11pt,a4paper]{article}
\usepackage[utf8]{inputenc}
\usepackage{jheppub}
\usepackage{amsmath}
\usepackage{amsthm}
\usepackage{graphicx}
\usepackage{framed}
\usepackage{float}
\usepackage{pdfpages}
\usepackage{amsfonts}
\usepackage{dsfont}
\usepackage{braket}
\usepackage{verbatim}
\usepackage{slashed}
\usepackage{epstopdf}
\usepackage{subfig}

\graphicspath{{./figures/}}

\newcommand{\be}{\begin{equation}}
\newcommand{\ee}{\end{equation}}

\newcommand{\bea}{\begin{eqnarray}}
\newcommand{\eea}{\end{eqnarray}}
\newcommand{\bfig}{\begin{figure}}
	\newcommand{\efig}{\end{figure}}
\newcommand{\bc}{\begin{center}}
	\newcommand{\ec}{\end{center}}

\newcommand{\f}[2]{\frac{#1}{#2}}

\newcommand{\eps}{{\varepsilon}}

\newcommand{\order}[1]{\mathcal{O}(#1)}
\newcommand{\del}{\partial}
\newcommand{\psibar}{\bar{\psi}}
\newcommand{\half}{\frac{1}{2}}

\newcommand{\mn}{{\mu\nu}}
\newcommand{\pslash}{\slashed{p}}

\newcommand{\chibar}{\bar{\chi}}
\newcommand{\xbar}{\bar{x}}
\newcommand{\ybar}{\bar{y}}

\newcommand{\gmu}{\gamma_\mu}
\newcommand{\gnu}{\gamma_\nu}

\newcommand{\mcF}{\mathcal{F}}

\newcommand{\cbar}{\bar{c}}
\newcommand{\dbar}{\bar{d}}
\newcommand{\lambdabar}{\bar{\lambda}}
\newcommand{\fbar}{\bar{f}}
\newcommand{\lapx}{\Delta_x}

\newcommand{\Psibar}{\bar{\Psi}}
\newcommand{\MSbar}{\overline{\text{MS}}}
\newcommand{\LQCD}{\Lambda_\text{QCD}}

\title{Nonminimal gradient flows in QCD-like theories}

\author{Marco Boers}

\affiliation{Van Swinderen Institute for Particle Physics and Gravity, University of Groningen, 9747 AG, The Netherlands}

\emailAdd{m.r.boers@rug.nl}

\abstract{The Yang-Mills gradient flow for QCD-like theories is generalized by including a fermionic matter term in the gauge field flow equation. We combine this with two different flow equations for the fermionic degrees of freedom. The solutions for the different gradient flow setups are used in the perturbative computations of the vacuum expectation value of the Yang-Mills Lagrangian density and the field renormalization factor of the evolved fermions up to next-to-leading order in the coupling. We find a one-parameter family of flow systems for which there exists a renormalization scheme in which the evolved fermion anomalous dimension vanishes to all orders in perturbation theory. The fermion number dependence of different flows is studied and applications to lattice studies are anticipated.
}

\begin{document}

\maketitle

\numberwithin{equation}{section}

\section{Introduction}

The past decade has seen a lot of interest in the Yang-Mills gradient flow. Its classical version was formulated in a pure mathematical context in \cite{Atiyah:1982fa,Donaldson:1985zz}, and the applications to quantum field theory followed in \cite{Narayanan:2006rf,Luscher:2009eq,Luscher:2010iy}. See also \cite{Lohmayer:2011si} for a nice review on `smearing', as the effect of the gradient flow on the fields involved is often referred to.

The gradient flow evolved fields have very particular renormalization properties, first demonstrated in \cite{Luscher:2011bx}, see also \cite{Hieda:2016xpq}. Correlators consisting of evolved gauge fields $B_\mu$ --- and possibly composite operators thereof --- do not require any renormalization beyond that of the usual renormalization of the fields and parameters of the nonevolved theory over which the path integral average is being taken, and correlators containing evolved fermion fields require a universal\footnote{With universal we mean that any correlator consisting of $n$ evolved fermion fields $\chi$, $n$ fields $\chibar$ and any number of Wilsonian normalized evolved gauge fields $B_\mu$ is rendered finite by multiplicative renormalization with $Z_\chi^{n}$, also if evolved composite operators are present. We will elaborate on this in section~\ref{sec:renormEvolvedFieldsYM}.} field renormalization factor \cite{Luscher:2013cpa}.

A much studied quantity is the observable
\begin{equation}\label{eq:E}
E(t,x)=g_0^2\,\mathcal{L}_{YM}\Big|_{A_\mu(x)=B_\mu(t,x)}
\end{equation}
with $\mathcal{L}_{YM}$ the Euclidean pure Yang-Mills Lagrangian density, $g_0$ the bare gauge coupling, and $t$ the gradient flow `time' which is of dimension $\text{length}^2$, see appendix \ref{app:notationconventions} for conventions used throughout this paper. Since $E(t,x)$ solely consists of evolved gauge fields\footnote{This is only true for Wilsonian normalized gauge fields. We choose to work with canonical instead of Wilsonian normalization from the outset. This will make for a neater comparison with stochastic quantization, and we are mainly concerned with perturbative analyses. All statements about Wilsonian normalized $B_\mu$ holds equally true for canonically normalized $B_\mu$ multiplied by $g_0$.}, it acquires no multiplicative renormalization factor. Its expectation value $\braket{E(t)}$ is used in lattice studies for e.g.~scale setting \cite{Sommer:2014mea} and defining finite volume running coupling schemes\footnote{\label{footnote:disagreement}There is some degree of disagreement in the lattice community on the validity of certain schemes defined in this way, especially in relatively large $N_f$ studies, see e.g.~\cite{Fodor:2017gtj,Hasenfratz:2017mdh}.} \cite{Ramos:2015dla}.

It is customary to apply the Yang-Mills gradient flow, defined by
\begin{equation}\label{eq:YMGFintro}
\del_t B_\mu(t,x)=-\frac{\delta S_{YM}}{\delta A_\mu(x)}\bigg|_{A_\mu(x)=B_\mu(t,x)},\hspace{1cm}B_\mu(0,x)=A_\mu(x)
\end{equation}
to QCD-like theories, where also (nonevolved) fermions are present. We consider this equation more in depth in section \ref{sec:GF basics}, but for now it is sufficient to note that the gradient flow \eqref{eq:YMGFintro} drives the gauge field $B_\mu(t,x)$ towards the stationary points of the Yang-Mills action for increasing flow time $t$. Since some properties of the expectation value of $E(t,x)$ in pure Yang-Mills theory are attributed to this fact \cite{Luscher:2010iy}, one might wonder what happens if one adds the fermionic term to the action in \eqref{eq:YMGFintro}, i.e.~$S_{YM}\rightarrow S_{YM}+S_F$ (see appendix \ref{app:euclideanQCDaction} for the explicit expressions). The consequences of this change are the subject of this paper.
\\
\\
There exists a close relation between the gradient flow and the Langevin equation used in stochastic quantization \cite{Parisi:1980ys}. Since it is customary to include the fermionic matter term in the Langevin equation when stochastically quantizing QCD-like theories, we will make this relation more explicit.

\subsection{Motivation from stochastic quantization}\label{sec:stochastic}
In stochastic quantization the basic idea is to introduce an extra scale $t$ --- we will shortly see how it is related to the flow time $t$ of \eqref{eq:YMGFintro} --- and a Gaussian noise field $\eta(t,x)$, and describe the $t$ evolution of some field $\phi(t,x)$ by the generalized Langevin equation
\begin{equation}\label{eq:generalizedscalarLangevin}
\del_t\phi(t,x)=-\int_y M(x,y) \frac{\delta S}{\delta \phi(t,y)} +\eta(t,x)
\end{equation}
where $M(x,y)$ is some kernel\footnote{The standard Langevin equation is the special case $M(x,y)=\delta^{(d)}(x-y)$.}, $S$ is the Euclidean action for the theory under study, and $\frac{\delta S}{\delta{\phi(t,y)}}$ is short for $\frac{\delta S}{\delta\phi(y)}\big|_{\phi(y)=\phi(t,y)}$. The Gaussian noise satisfies
\begin{equation}
\braket{\eta(t,x)\eta(s,y)}_\eta=2M(x,y)\delta(t-s), \hspace{1cm}\braket{\eta(t,x)}_\eta=0
\end{equation}
where the $\eta$ subscript on the angle brackets indicates that the average over $\eta$ is being taken, see e.g.~\cite{Damgaard:1987rr} for details. Note that the presence of the noise in \eqref{eq:generalizedscalarLangevin} makes the value of $\phi$ at large $t$ completely independent of the boundary condition at $t=0$, and a common choice is $\phi(0,x)=0$. We now imagine that the fields are coupled to some heat reservoir. It can be shown that in the large $t$ limit we reach equilibrium \cite{Damgaard:1987rr}, and all stochastic correlators converge to their Euclidean quantum field theory counterparts:
\begin{equation}
\braket{\phi(t,x)\phi(t,y)...}_\eta\xrightarrow{t\rightarrow\infty}\braket{\phi(x)\phi(y)...}
\end{equation}
Now, the gradient flow evolution can be understood as follows\footnote{This reasoning is based on \cite{Lohmayer:2011si}.}: at some large time $t=t_0$ we abruptly switch off the heat reservoir and the noise field. At some later time $t_1>t_0$, the fields will have evolved via the flow equation
\begin{equation}
\del_t\phi(t,x)=-\int_yM(x,y)\frac{\delta S}{\delta\phi(t,y)},\hspace{1cm}\phi(t_0,x)=\phi(x)
\end{equation}
which is exactly the gradient flow equation.
\\ \\
When stochastically quantizing pure Yang-Mills theory, the following Langevin equation is used \cite{Damgaard:1987rr} --- here we neglect the possible presence of a `gauge fixing term', usually called a \textit{Zwanziger term} in this context:
\begin{equation}\label{eq:YMLangevin}
\del_t A_\mu(t,x)=D_\nu F_{\nu\mu}(t,x)+\eta(t,x)
\end{equation}
with $D_\mu=\del_\mu+g_0[A_\mu,\cdot\,]$ the covariant derivative, and $F_\mn=\del_\mu A_\nu-\del_\nu A_\mu +g_0[A_\mu,A_\nu]$ the field strength. However, when considering QCD-like theories, a different Langevin equation is being used for the gauge field in the presence of fermionic matter:
\begin{equation}\label{eq:QCDgaugeLangevin}
\del_t A_\mu(t,x)=D_\nu F_{\nu\mu}(t,x)-g_0\psibar(t,x)\gamma_\mu T^a\psi(t,x)T^a+\eta(t,x)
\end{equation}
with $T^a$ the $SU(N)$ generators, where for the Langevin evolution of the fermions one can take \cite{Tzani:1986ps}
\begin{equation}\label{eq:QCDfermionLangevin}
\begin{split}
\del_t\psi(t,x)&=\slashed{D}^2\psi(t,x)-\slashed{D}\theta(t,x)\\ \del_t\psibar(t,x)&=\psibar(t,x)\overleftarrow{\slashed{D}}^2+\bar{\theta}(t,x)
\end{split}
\end{equation}
where $\theta,\bar{\theta}$ are Gaussian noise fields.

\subsection{Nonminimal gradient flows}

In gradient flow studies it is customary instead to use the Yang-Mills gradient flow of \eqref{eq:YMGFintro}, i.e.~\eqref{eq:YMLangevin} with\footnote{And with $A_\mu(t,x)\rightarrow B_\mu(t,x)$ with boundary condition $B_\mu(0,x)=A_\mu(x)$.} $\eta=0$, also when fermions are present. We set out to investigate what will change if we instead use \eqref{eq:QCDgaugeLangevin} with $\eta=0$ as the basis for the gradient flow, i.e.~the case where we include the fermion bilinear color non-singlet vector current in the right hand side of the flow equation. We will refer to this as the \textit{nonminimal gradient flow} equation.

For the evolution of the fermions we are considering two different flow equations. Firstly we use the one that is standard in gradient flow studies \cite{Luscher:2013cpa} using the operator $D^2$, and secondly we use the operator $\slashed{D}^2$ instead, i.e.~\eqref{eq:QCDfermionLangevin} with $\theta=\bar{\theta}=0$.

All flows must respect the symmetries of the nonevolved theory, and the Yang-Mills gradient flow is indeed the simplest possibility. However, the nonminimal cases we will investigate have the conceptual advantage that they are actually driving the gauge field towards the stationary points of the full QCD-like action, not just of the Yang-Mills action. 

That said, we also carry out the exercise of writing the most general flow equations for the gauge field and the fermions, respecting the symmetries of the nonevolved theory, for completeness.

We use the nonminimal flows to calculate the expectation value of the operator $E(t,x)$ from \eqref{eq:E} to next-to-leading order in the coupling, generalizing the result first presented in \cite{Luscher:2010iy}. We also generalize the demonstration of the renormalization properties of the evolved fields of \cite{Luscher:2011bx,Luscher:2013cpa} (see also \cite{Hieda:2016xpq}), and calculate the evolved fermions' field renormalization factor $Z_\chi$ for the different nonminimal flows. We find that there exists a one-parameter family of flow equation systems for which there is a renormalization scheme in which the evolved fermion anomalous dimension vanishes to all orders in perturbation theory. 

We solely consider massless fermions in this work. The quantity $\braket{E(t)}$ will have a mass dependence at next-to-leading order in the gauge coupling, see e.g.~\cite{Harlander:2016vzb} for the Yang-Mills gradient flow case. However, the $\MSbar$ expressions for $Z_\chi$ and the other renormalization properties discussed in this paper are independent of the presence of fermion masses. 

\subsection{Organization of the paper}

In section \ref{sec:GF basics} we review the basics of the Yang-Mills gradient flow (YMGF), the next-to-leading order in the gauge coupling result for $\braket{E}$ with $E$ in \eqref{eq:E}, and the renormalization properties of the evolved fields. In section \ref{sec:NMGF} we introduce the nonminimal gradient flow (NMGF), calculate $\braket{E}$, present the generalization of the arguments for the renormalization properties for the nonminimally evolved fields, and calculate the field renormalization factor of the evolved fermions. Subsequently, in section \ref{sec:Dslash^2} we include the effects of replacing the operator $D^2$ in the fermion flow equation with $\slashed{D}^2$, coining this case the `slashed' nonminimal gradient flow (sNMGF). In section \ref{sec:Nfdependence} we investigate the $N_f$ dependence of $\braket{E}$ with $N=3$, calculated with the three different flows. In section \ref{sec:generalized flows} we generalize the flow equations by putting them in the most general form compliant with all the symmetries of the nonevolved theory. We present our conclusions and outlook in section \ref{sec:conclusion}. Appendices \ref{app:notationconventions} to \ref{app:generalNresults} contain relevant conventions and calculations.

\section{Gradient flow basics}\label{sec:GF basics}
\subsection{Yang-Mills gradient flow}\label{sec:YMGF}
The Yang-Mills gradient flow equation is given by
\begin{equation}\label{eq:YMGF}
\del_t B_\mu(t,x)=-\frac{\delta S_{YM}}{\delta A_\mu(x)}\bigg|_{A_\mu(x)=B_\mu(t,x)}=D_\nu G_{\nu\mu}(t,x),\hspace{1cm} B_\mu(0,x)=A_\mu(x)
\end{equation}
where $D_\mu=\del_\mu+g_0[B_\mu,\cdot\,]$ and $G_\mn=\del_\mu B_\nu-\del_\nu B_\mu+g_0[B_\mu,B_\nu]$. For further conventions and the explicit expression for the action we refer the reader to appendices \ref{app:notationconventions} and \ref{app:euclideanQCDaction}, respectively.

When performing perturbative calculations, this flow equation is usually conveniently modified by adding a term which in the context of stochastic quantization is known as a Zwanziger term \cite{Damgaard:1987rr}:
\begin{equation}\label{eq:YMGFmodified}
\del_t B_\mu(t,x)=D_\nu G_{\nu\mu}(t,x)+\alpha_0D_\mu\del_\nu B_\nu(t,x),\hspace{1cm} B_\mu(0,x)=A_\mu(x)
\end{equation}
and subsequently setting $\alpha_0=1$ \cite{Luscher:2010iy}. The solution to this modified flow equation is related to the solution of \eqref{eq:YMGF} via a $t$-dependent gauge transformation (see appendix \ref{app:gaugetrans}), and thus observables will be independent of $\alpha_0$.
The solution to \eqref{eq:YMGFmodified} with $\alpha_0=1$ is given by
\begin{equation}\label{eq:YMGFmodifiedSolution}
B_\mu(t,x)=\int_y K_{t,\mu\nu}(x-y) A_\nu(y)+\int_y\int_0^tds\,K_{t-s,\mu\nu}(x-y) R_\nu(s,y)
\end{equation}
with $K_{t,\mn}(x)=\delta_\mn K_t(x)$, where the scalar kernel reads
\begin{equation}
K_t(x) =\int_p e^{ipx}e^{-tp^2} = \f{1}{(4\pi t)^{d/2}} e^{-\f{x^2}{4t}      }
\end{equation}
and 
\begin{equation}\label{eq:Rmu}
R_\mu=2g_0[B_\nu,\del_\nu B_\mu]-g_0[B_\nu,\del_\mu B_\nu]+g_0^2[B_\nu,[B_\nu,B_\mu]]
\end{equation}
We will often use the `exponential-of-Laplacian' notation, with the Laplacian $\Delta_x=\del_\mu\del_\mu$, in which
\begin{equation}
K_t(x-y)=e^{t\lapx}\delta^{(d)}(x-y)
\end{equation}
and
\begin{equation}\label{eq:YMGFB_muExpofLap}
B_\mu(t,x)=e^{t\lapx}A_\mu(x)+\int_0^tds\,e^{(t-s)\lapx}R_\mu(s,x)
\end{equation}

\subsection{Fermion flow}\label{sec:fermionflow}
The conventional flow equations for the fermions introduced in \cite{Luscher:2013cpa} read
\begin{equation}\label{eq:gfeqnfermions}
\begin{split}
\del_t\chi(t,x)&=\Delta\chi(t,x),\hspace{1cm}\chi(0,x)=\psi(x)\\
\del_t\chibar(t,x)&=\chibar(t,x)\overleftarrow{\Delta},\hspace{1cm}\chibar(0,x)=\psibar(x)
\end{split}
\end{equation}
where
\begin{equation}
\begin{split}
\Delta&=D^2,\hspace{1cm}D_\mu=\del_\mu+g_0B_\mu\\
\overleftarrow{\Delta}&=\overleftarrow{D}^2,\hspace{1cm}\overleftarrow{D}_\mu=\overleftarrow{\del}_\mu-g_0B_\mu
\end{split}
\end{equation}
In section \ref{sec:Dslash^2} we investigate the consequences of using $\slashed{D}^2$ instead of $D^2$, establishing a more direct connection to the stochastic quantization interpretation outlined in section \ref{sec:stochastic}.

Similarly to the addition of the Zwanziger term in \eqref{eq:YMGFmodified}, these flow equations are conveniently modified by adding an $\alpha_0$-dependent term:
\begin{equation}\label{eq:gfeqnfermionsmodified}
\begin{split}
\del_t\chi(t,x)&=\{\Delta-\alpha_0 g_0\del_\mu B_\mu(t,x)\}\,\chi(t,x),\hspace{1cm}\chi(0,x)=\psi(x)\\
\del_t\chibar(t,x)&=\chibar(t,x)\{\overleftarrow{\Delta}+\alpha_0g_0\del_\mu B_\mu(t,x)\},\hspace{1cm}\chibar(0,x)=\psibar(x)
\end{split}
\end{equation}
and the solutions to \eqref{eq:gfeqnfermionsmodified} are related to those of \eqref{eq:gfeqnfermions} by the same $t$-dependent gauge transformation as in the pure gauge case from section \ref{sec:YMGF} (see appendix \ref{app:gaugetrans}).

The solutions to \eqref{eq:gfeqnfermionsmodified} with $\alpha_0=1$ are given by
\begin{equation}\label{eq:gffermionsmodifiedsolution}
\begin{split}
\chi(t,x)&=e^{t\lapx}\psi(x)+\int_0^tds\,e^{(t-s)\lapx}\{\Delta' \chi(s,x)\}\\
\chibar(t,x)&=e^{t\lapx}\psibar(x)+\int_0^tds\,e^{(t-s)\lapx}\{\chibar(s,x)\overleftarrow{\Delta}'\}
\end{split}
\end{equation}
with
\begin{equation}\label{eq:Deltaprime}
\Delta'=2g_0B_\mu\del_\mu+g_0^2B_\mu B_\mu,\hspace{1cm}\overleftarrow{\Delta}'=-2g_0\overleftarrow{\del_\mu}B_\mu+g_0^2B_\mu B_\mu
\end{equation}

\subsection{Perturbative calculation of $\braket{E}$}\label{sec:perturbCalcEnonevolved}
For later comparison, we briefly review the perturbative calculation of $\braket{E}$ up to $\order{g_0^4}$ of \cite{Luscher:2010iy}. At this order in the coupling it only involves the evolved gauge fields given in \eqref{eq:YMGFB_muExpofLap}; the fermionic contribution will solely be coming from the nonevolved path integral averaging.

\begin{figure}[t]
	\centering
	\begin{minipage}{.33\textwidth}
		\centering
		\includegraphics[width=.8\linewidth]{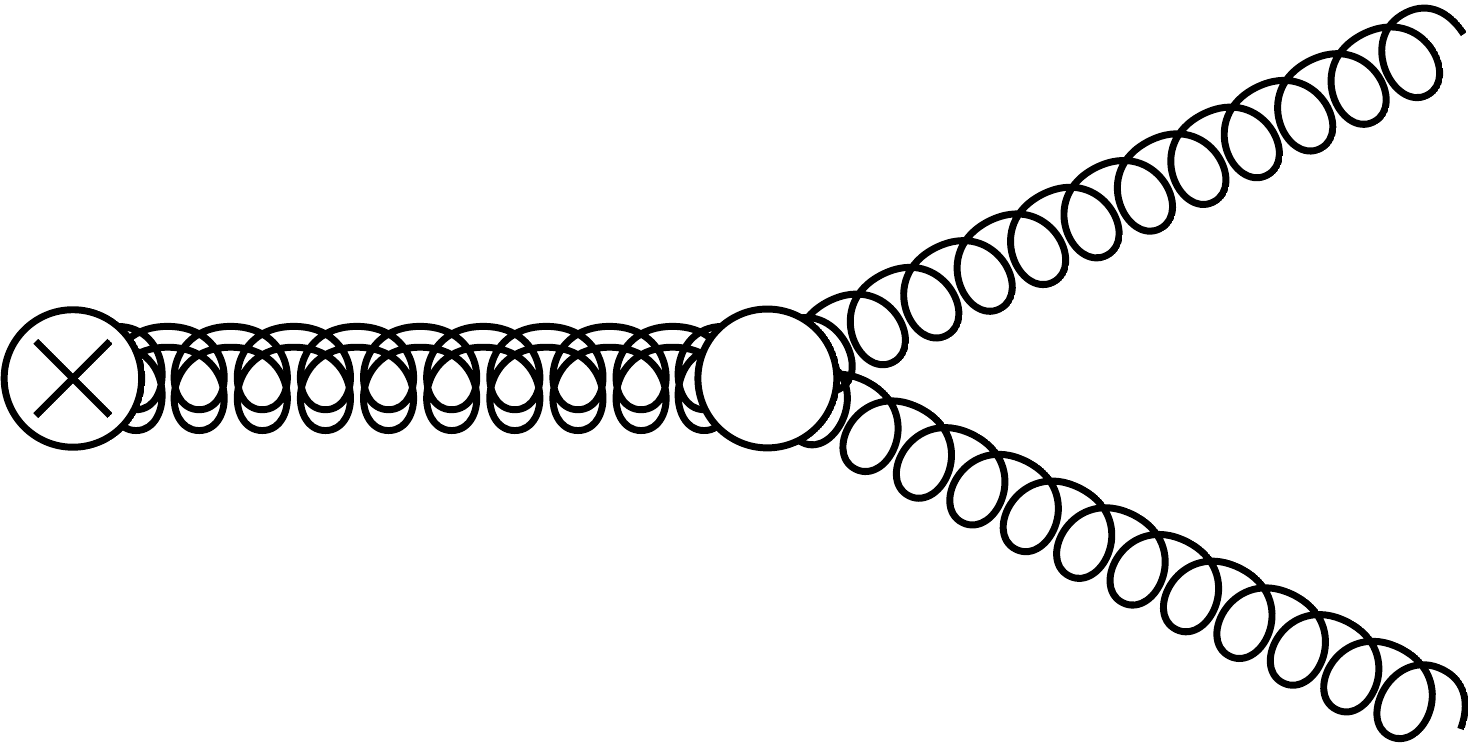}
		\caption*{(a)}
	\end{minipage}%
	\begin{minipage}{.33\textwidth}
		\centering
		\includegraphics[width=.8\linewidth]{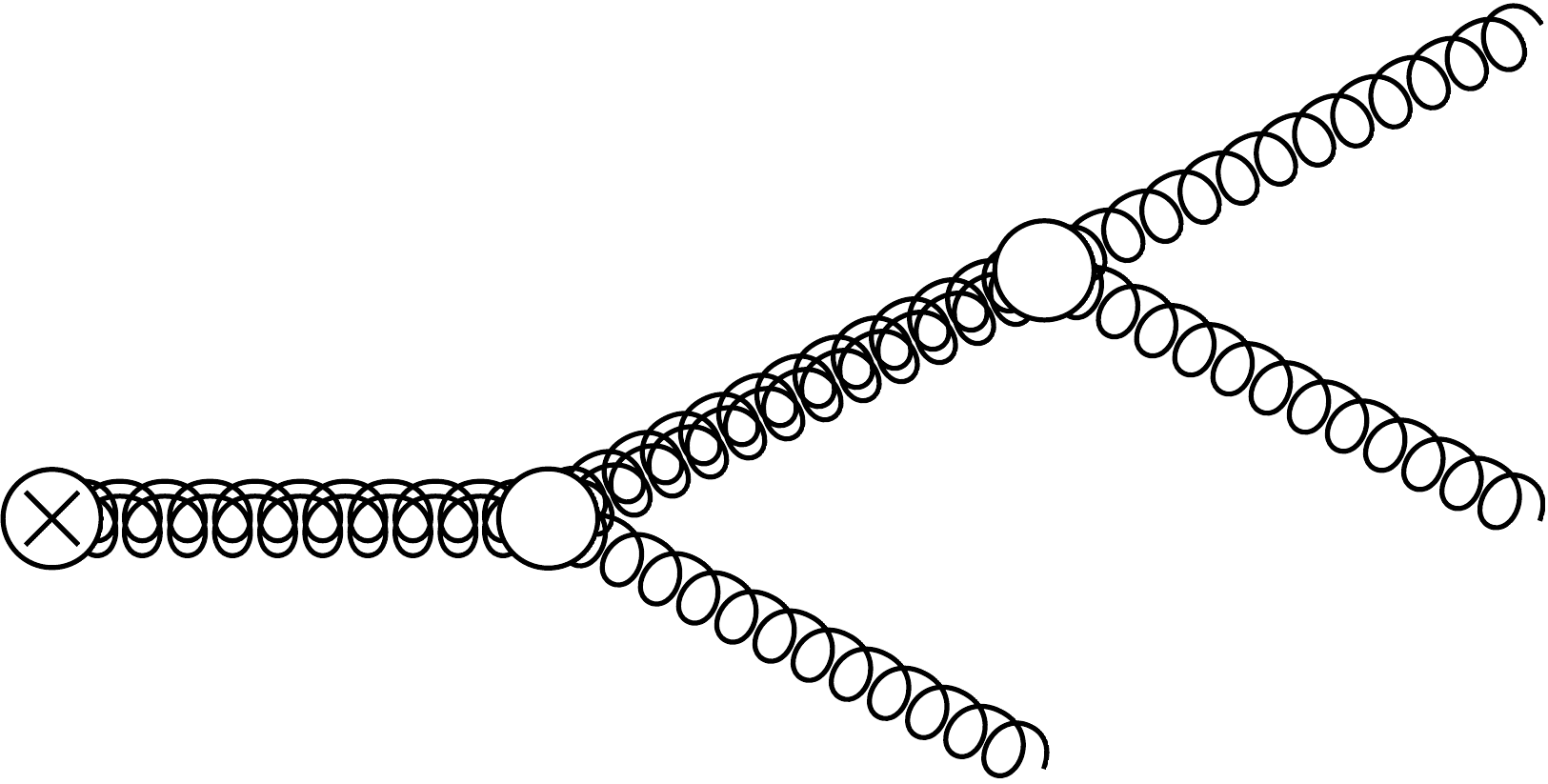}
		\caption*{(b)}
	\end{minipage}%
	\begin{minipage}{.33\textwidth}
		\centering
		\includegraphics[width=.8\linewidth]{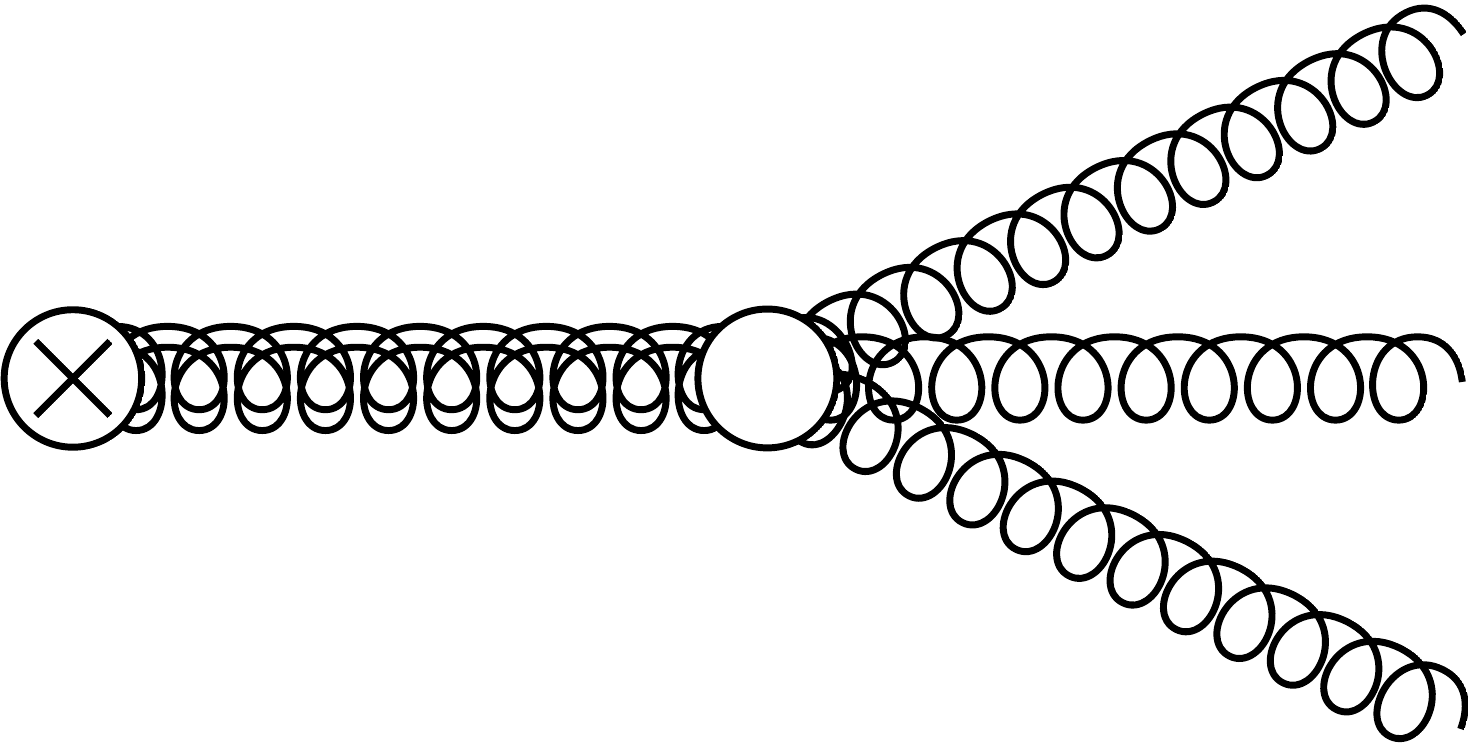}
		\caption*{(c)}
	\end{minipage}
	\caption{Diagrammatic representation of the first two nontrivial orders of the evolved gauge field's expansion in the coupling $g_0$ from \eqref{eq:YMgaugefieldexpansion}. (a) represents $B_{\mu,1}$ from \eqref{eq:YMgaugefieldexpansionfirstorders}, (b) collectively represents the first two terms in $B_{\mu,2}$ from \eqref{eq:YMgaugefieldexpansionfirstorders}, and (c) represents the last term in $B_{\mu,2}$ from \eqref{eq:YMgaugefieldexpansionfirstorders}. In this notation, double lines stand for kernels (`exponentials-of-Laplacian') together with a flow time integration, white blobs represent GF vertices, black blobs are the usual QCD vertices, and the crossed blobs are the external points.}
	\label{fig:YMgaugefieldexpansion}
\end{figure}

It is useful to write the evolved gauge field as a series in powers of the bare coupling:
\begin{equation}\label{eq:YMgaugefieldexpansion}
B_\mu(t,x)=\sum_{n=0}^{\infty}g_0^{n}B_{\mu,n}(t,x)
\end{equation}
with the first few orders iteratively given by
\begin{equation}\label{eq:YMgaugefieldexpansionfirstorders}
\begin{split}
B_{\mu,0}(t,x)&=e^{t\lapx}A_\mu(x)\\
B_{\mu,1}(t,x)&=\int_0^tds\,e^{(t-s)\lapx}\{[B_{\nu,0},2\del_\nu B_{\mu,0}-\del_\mu B_{\nu,0}]\}\\
B_{\mu,2}(t,x)&=\int_0^tds\,e^{(t-s)\lapx}\{[B_{\nu,1},2\del_\nu B_{\mu,0}-\del_\mu B_{\nu,0}]+[B_{\nu,0},2\del_\nu B_{\mu,1}-\del_\mu B_{\nu,1}]\\
&\hspace{1cm}+[B_{\nu,0},[B_{\nu,0},B_{\mu,0}]]\}
\end{split}
\end{equation}
These different orders can be conveniently expressed in diagrammatic form, see figure \ref{fig:YMgaugefieldexpansion}.
The observable $E$ from \eqref{eq:E} written in terms of the $B_\mu$ field reads:
\begin{equation}\label{eq:EexpandedinB}
\begin{split}
E&=\frac{g_0^2}{4}G^a_\mn G^a_\mn\\
&=\half g_0^2\del_\mu B_\nu^a(\del_\mu B_\nu^a-\del_\nu B_\mu^a)+g_0^3f^{abc}\del_\mu B_\nu^a B_\mu^b B_\nu^c+\frac{1}{4}g_0^4f^{abe}f^{cde}B_\mu^a B_\nu^b B_\mu^c B_\nu^d
\end{split}
\end{equation}
and the expectation value $\braket{E}$ can be computed order-by-order in the coupling \cite{Luscher:2010iy,Harlander:2016vzb}, using \eqref{eq:YMgaugefieldexpansionfirstorders} combined with the contributions stemming from the nonevolved path integral averaging. The result up to next-to-leading order in the coupling is given by (we use dimensional regularization with $d=4-2\eps$)
\begin{equation}
\braket{E}=\half g_0^2\frac{(N^2-1)}{(8\pi t)^{d/2}}(d-1)\left\{1+c_1g_0^2+\order{g_0^4}\right\}
\end{equation}
\begin{equation}
c_1=\frac{1}{(4\pi)^2}(4\pi)^{\eps}(8t)^{\eps}\left\{N\left(\frac{11}{3\eps}+\frac{52}{9}-3\log 3\right)-N_f\left(\frac{2}{3\eps}+\frac{4}{9}-\frac{4}{3}\log 2\right)+\order{\eps}\right\}
\end{equation}
This observable is rendered finite by the coupling renormalization of the nonevolved theory. In the $\MSbar$ scheme one has:
\begin{equation}\label{eq:couplingRenormalizationMSbar}
g_0^2=\mu^{2\eps}\left(4\pi e^{-\gamma_E}\right)^{-\eps}g^2(\mu) Z_g^2(g(\mu),\eps)
\end{equation}
with $g(\mu)$ the renormalized coupling, and
\begin{equation}\label{eq:b0etc}
Z_g^2(g(\mu),\eps)=1-\frac{\beta_0}{\eps}\frac{g^2(\mu)}{(4\pi)^2}+\order{g^4(\mu)},\hspace{1cm}\beta_0=\frac{11}{3}N-\frac{4}{3}T(R)N_f
\end{equation}
After renormalization\footnote{We would like to emphasize that the fact that $\braket{E}$ is rendered finite by the coupling renormalization at this order does not constitute an example of the `nonrenormalization' of the evolved gauge field $B_\mu$, to be discussed in section \ref{sec:renormEvolvedFieldsYM}. In order to achieve this, one has to use the 2-loop universality of $Z_g$ from \eqref{eq:b0etc} and calculate $\braket{E}$ to next-to-next-to-leading order. This has been done in \cite{Harlander:2016vzb}, where the `nonrenormalization' is indeed confirmed.} we express the renormalized coupling $\alpha(\mu)=\frac{g^2(\mu)}{4\pi}$ in terms of the renormalization group invariant coupling $\alpha(q)$ by inverting \eqref{eq:alpha(q)intermsofalpha(mu)}, and subsequently setting $q=(8t)^{-1/2}$ and $N=3$ we arrive at
\begin{equation}\label{eq:EN=3Luscher}
\braket{E}=\frac{3}{4\pi t^2}\alpha(q)\left\{1+k_1\alpha(q)+\order{\alpha^2}\right\},\hspace{1cm}k_1=1.0978+0.0075\times N_f
\end{equation}
For the renormalized result for general $N$ we refer the reader to appendix \ref{app:generalNresults}.

\subsection{Renormalization of evolved fields}\label{sec:renormEvolvedFieldsYM}

It has been shown in \cite{Luscher:2011bx} (see also \cite{Hieda:2016xpq}) that in Wilsonian normalization all correlators consisting of only $B_\mu$ fields are finite, i.e. they do not require renormalization beyond the usual renormalization of the parameters of the four dimensional theory over which the path integral average is being taken. When switching to canonical normalization, i.e.~taking $B_\mu\rightarrow g_0 B_\mu$, any correlator consisting of $m$ number of $B_\mu$ fields now automatically contains a factor of $g_0^m$, which will renormalize as given in \eqref{eq:couplingRenormalizationMSbar} in the case of $\MSbar$.

The fermion fields $\chi$ and $\chibar$ do require a renormalization factor \cite{Luscher:2013cpa}, namely
\begin{equation}
\chi_R(t,x)=Z_\chi^{1/2}\chi(t,x),\hspace{1cm}\chibar_R(t,x)=Z_\chi^{1/2}\chibar(t,x)
\end{equation}
with
\begin{equation}\label{eq:ZchiYM}
Z_\chi(g(\mu),\eps)=1+C_2(R)\frac{3}{\eps}\frac{g^2(\mu)}{(4\pi)^2}+\order{g^4}
\end{equation}
and the contributing self-energy diagrams together with their values are presented in appendix \ref{app:ZchiYMGF}. Any bare evolved correlator consisting of an arbitrary number of $g_0 B_\mu$'s, and $n$ number of $\chi$'s and $\chibar$'s is now rendered finite by multiplicative renormalization with $Z_\chi^{n}$, supplemented with the usual renormalization of the parameters\footnote{With `parameters' we mean the gauge coupling and the gauge fixing parameter. See appendix \ref{app:euclideanQCDaction} for the Euclidean action of the nonevolved theory.} of the nonevolved theory. An important feature is that this also holds true when some of the evolved fields have coinciding spacetime positions at strictly positive flow time; i.e.~composite operators of evolved fields do not acquire any additional renormalization factor, in contrast to the nonevolved case.
\\

The all order demonstration of these statements is found in \cite{Luscher:2011bx} and \cite{Hieda:2016xpq}. In the remainder of this section we present the general reasoning for this demonstration, without any intention or pretension of being complete. We nevertheless choose to include it here, since these will be the arguments that we generalize later on in sections \ref{sec:renormNMGF} and \ref{sec:renormsNMGF}.

The starting point of the demonstration from \cite{Luscher:2011bx} is to describe the theory in $d+1$ dimensions, the extra dimension being the flow time direction $t\in[0,\infty)$. The evolved fields are being considered as independent from the nonevolved elementary fields, apart from their implicit dependence through the boundary conditions. The $d+1$ dimensional `bulk' action is given by:
\begin{equation}
S_{bulk}=S_{G,fl}+S_{d\dbar}+S_{F,fl}
\end{equation}
with
\begin{align}
S_{G,fl}&=-2\int_0^\infty dt\int_x\text{tr}\left\{L_\mu(\del_t B_\mu-D_\nu G_{\nu\mu}-\alpha_0D_\mu\del_\nu B_\nu)\right\}\label{eq:gaugeBulkActionYM}\\
S_{F,fl}&=\int_0^\infty dt \int_x \left\{\lambdabar(\del_t-\Delta+\alpha_0g_0\del_\nu B_\nu)\chi+\chibar(\overleftarrow{\del}_t-\overleftarrow{\Delta}-\alpha_0g_0\del_\nu B_\nu)\lambda\right\}\label{eq:fermBulkAction}\\
S_{d\dbar}&=-2\int_0^\infty dt\int_x\text{tr}\left\{\dbar(\del_t d -\alpha_0D_\mu\del_\mu d)\right\}
\end{align}
where we have introduced the lie algebra valued lagrange multiplier field $L_\mu(t,x)=L_\mu^a(t,x)T^a$ with purely imaginary components, the two fermionic lagrange multipliers $\lambda(t,x)$ and $\lambdabar(t,x)$ which carry the same indices as the quark fields, and the bulk ghost fields $d(t,x)$ and $\dbar(t,x)$. The ghost $d$ has boundary condition
\begin{equation}
d(0,x)=c(x)
\end{equation}
while the other new fields do not obey any. Note that the lagrange multipliers $L_\mu$, $\lambda$ and $\lambdabar$ are there to enforce the flow equations upon variation. The $\dbar$ field acts as a lagrange multiplier enforcing the correct diffusion equation on the $d+1$ dimension ghost, such that the flow equations are invariant under an infinitesimal gauge transformation $\Lambda(t,x)=e^{\eps \omega(t,x)}$, with $\omega(t,x)=-g_0 d(t,x)$ (see appendix \ref{app:gaugetrans}). This generalizes the BRST symmetry to the $d+1$ dimensional bulk action:
\begin{equation}
\delta S_{bulk}=0
\end{equation}
with $\delta=\delta_{BRST}$, which can be checked using the BRST identities in appendix \ref{app:BRST}. Together with the BRST invariance of the $d$ dimensional QCD action, and the fact that the path integral measure is invariant, we have
\begin{equation}\label{eq:deltaBRSTO=0}
\braket{\delta O}=0
\end{equation}
with $O$ any combination of evolved and nonevolved elementary fields. Equation \eqref{eq:deltaBRSTO=0} has some important consequences. For our purposes the most important relation that follows, for reasons becoming obvious below, is
\begin{equation}\label{eq:BRSTimportantresult}
\lambda_0\braket{B_\mu^a(t,x)\del_\nu A_\nu^b(y)\del_\rho A_\rho^c(z)}=-\braket{(D_\mu d)^a(t,x)\cbar^b(y)\del_\rho A_\rho^c(z)}
\end{equation}

The next step in the demonstration is to exclude the possibility of counterterms. These are either localized in the $d+1$ dimensional $\mathds{R}^d\times(0,\infty)$ bulk or at the $d$ dimensional $\mathds{R}^d$ boundary \cite{Symanzik:1981wd}. We first focus on the former.

The impossibility of bulk divergences is shown by first eliminating the elementary nonevolved fields from the quadratic part of the bulk action, by plugging in
\begin{subequations}\label{eq:homogenousbcsubstitutions}
\begin{align}
B_\mu(t,x)&=e^{t\lapx}A_\mu(x)+b_\mu(t,x)\\
\chi(t,x)&=e^{t\lapx}\psi(x)+\Psi(t,x)\\
\chibar(t,x)&=e^{t\lapx}\psibar(x)+\Psibar(t,x)
\end{align}
\end{subequations}
where $b_\mu, \Psi, \Psibar$ obey homogenous boundary conditions. The terms involving the nonevolved elementary fields all drop out, since they solve the linearized flow equations. Therefore, the only nonzero 2-point bulk correlators are $\braket{bL}$, $\braket{\Psi\lambdabar}$, $\braket{\lambda\Psibar}$ and $\braket{d\dbar}$. By subsequently analyzing the interaction part of the bulk action it becomes clear that the only nonzero bulk correlators of these fields are trees, i.e.~they can never form loops, and thus will be finite. Therefore there will be no bulk counterterms.

The remaining possibility is the presence of boundary counterterms. Dimensional analysis together with symmetry requirements and ghost number conservation tells us that the only allowed new counterterms are of the form
\begin{equation}\label{eq:boundaryCTs}
\begin{split}
\text{Boundary CTs}&\sim\int_x\text{tr}\{z_1L_\mu(0,x)(A_R)_\mu(x)+z_2\dbar(0,x)c_R(x)\}\\
&\hspace{1cm}+z_3\int_x\{\lambdabar(0,x)\psi_R(x)+\psibar_R(x)\lambda(0,x)\}
\end{split}
\end{equation}
At this point, the importance of \eqref{eq:BRSTimportantresult} becomes apparent; the tree level computation of its renormalized version\footnote{Only the fields and parameters of the four dimensional theory are renormalized here, i.e. $\lambda_0\rightarrow\lambda$, $A_\mu\rightarrow(A_R)_\mu$, $c\rightarrow c_R$. See \cite{Luscher:2011bx} for the explicit computation.} forces $z_1=z_2=0$, i.e. there is no field renormalization for $L_\mu$ and $\dbar$, and thus no field renormalization for $B_\mu$ and $d$ (since there can be no bulk counterterms). However, there is no restriction on $z_3$, and consequently $\lambdabar$ and $\lambda$ do get a field renormalization factor
\begin{equation}
\lambdabar_R(t,x)=Z_\chi^{-1/2}\lambdabar(t,x),\hspace{1cm}\lambda_R(t,x)=Z_\chi^{-1/2}\lambda(t,x)
\end{equation}
which implies the reciprocal field renormalization factors for $\chi$ and $\chibar$ in order not to have a bulk counterterm.

This completes the reasoning for the case of correlators consisting of the evolved fields at distinct points in spacetime. In order to include composite operators it is sufficient to notice the following: the small --- or indeed vanishing --- distance limit of any number of evolved fields inside a correlator will always remain finite for $t>0$ due to the exponential suppression of the high frequency modes.

Missing details of the demonstration can be found in \cite{Luscher:2011bx} and \cite{Hieda:2016xpq}.

\section{Nonminimal gradient flow}\label{sec:NMGF}
\subsection{Defining the nonminimal gradient flow}
We now include fermionic matter in the flow equation for the gauge field, giving rise to the nonminimal gradient flow (NMGF) equation:
\begin{equation}\label{eq:NMGF2}
\del_t B_\mu=-\frac{\delta (S_{YM}+S_F)}{\delta B_\mu}\bigg|_{A_\mu=B_\mu,\psi=\chi,\psibar=\chibar}=D_\nu G_{\nu\mu}-g_0 j_\mu^aT^a,\hspace{1cm}B_\mu\big|_{t=0}=A_\mu
\end{equation}
where
\begin{equation}
j_\mu^a(t,x)=\chibar(t,x) \gmu T^a \chi(t,x)
\end{equation}
is the gauge covariant matter current. The quark fields $\chi$ and $\chibar$ are the same as in section \ref{sec:fermionflow}, apart from the fact that the $B_\mu$ field in their flow equations is now the solution of the NMGF equation \eqref{eq:NMGF2}. Importantly, the NMGF equation is gauge invariant, as shown in appendix \ref{app:gaugetrans}.

\begin{figure}[t]
	\centering
	\begin{minipage}{.33\textwidth}
		\centering
		\includegraphics[width=.8\linewidth]{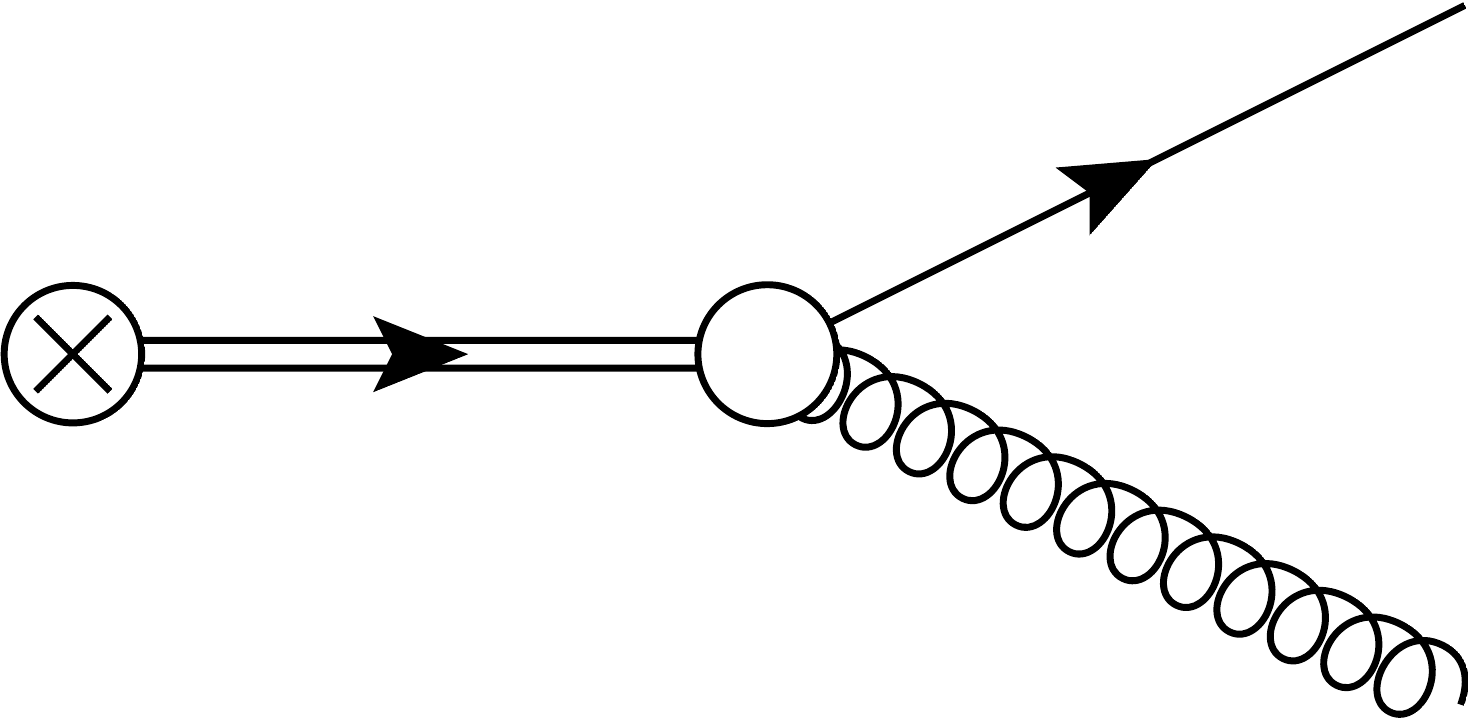}
		\caption*{(a)}
	\end{minipage}%
	\begin{minipage}{.33\textwidth}
		\centering
		\includegraphics[width=.8\linewidth]{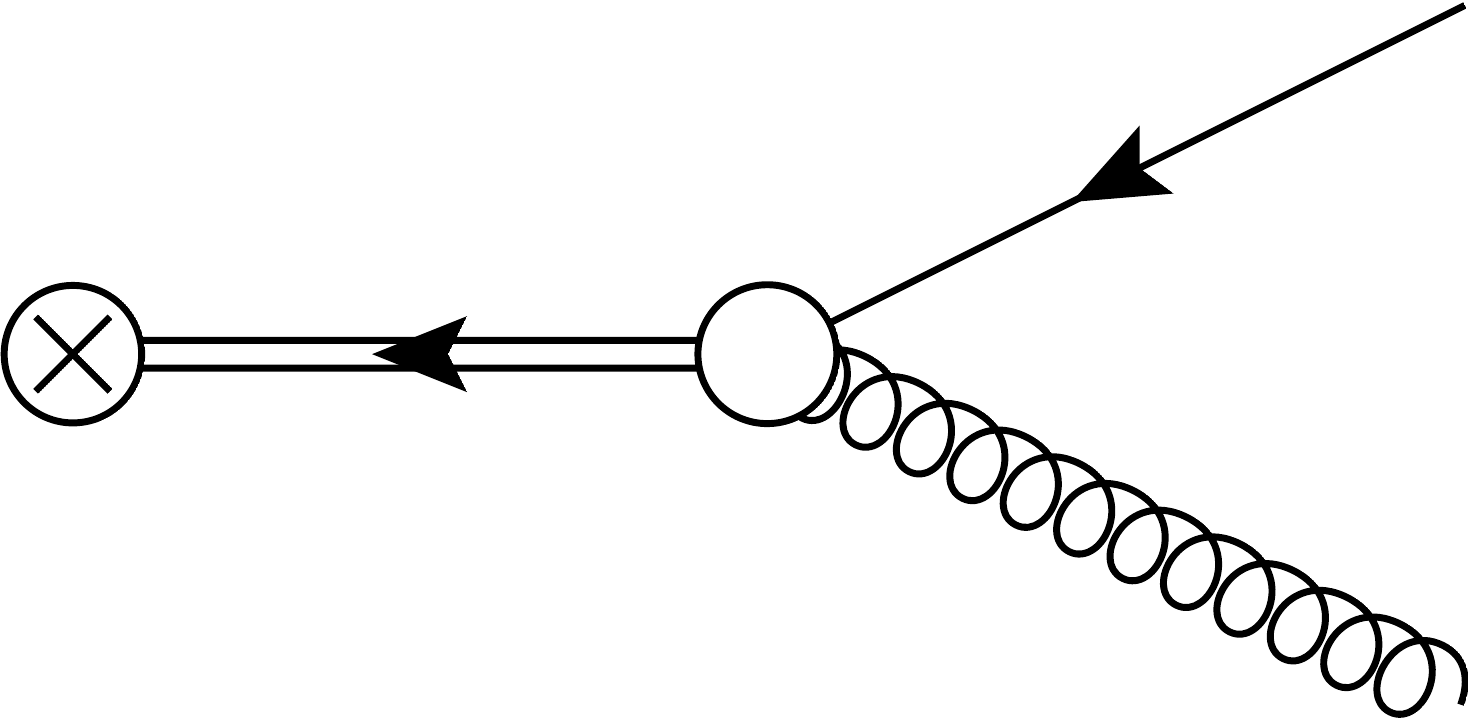}
		\caption*{(b)}
	\end{minipage}%
	\begin{minipage}{.33\textwidth}
		\centering
		\includegraphics[width=.8\linewidth]{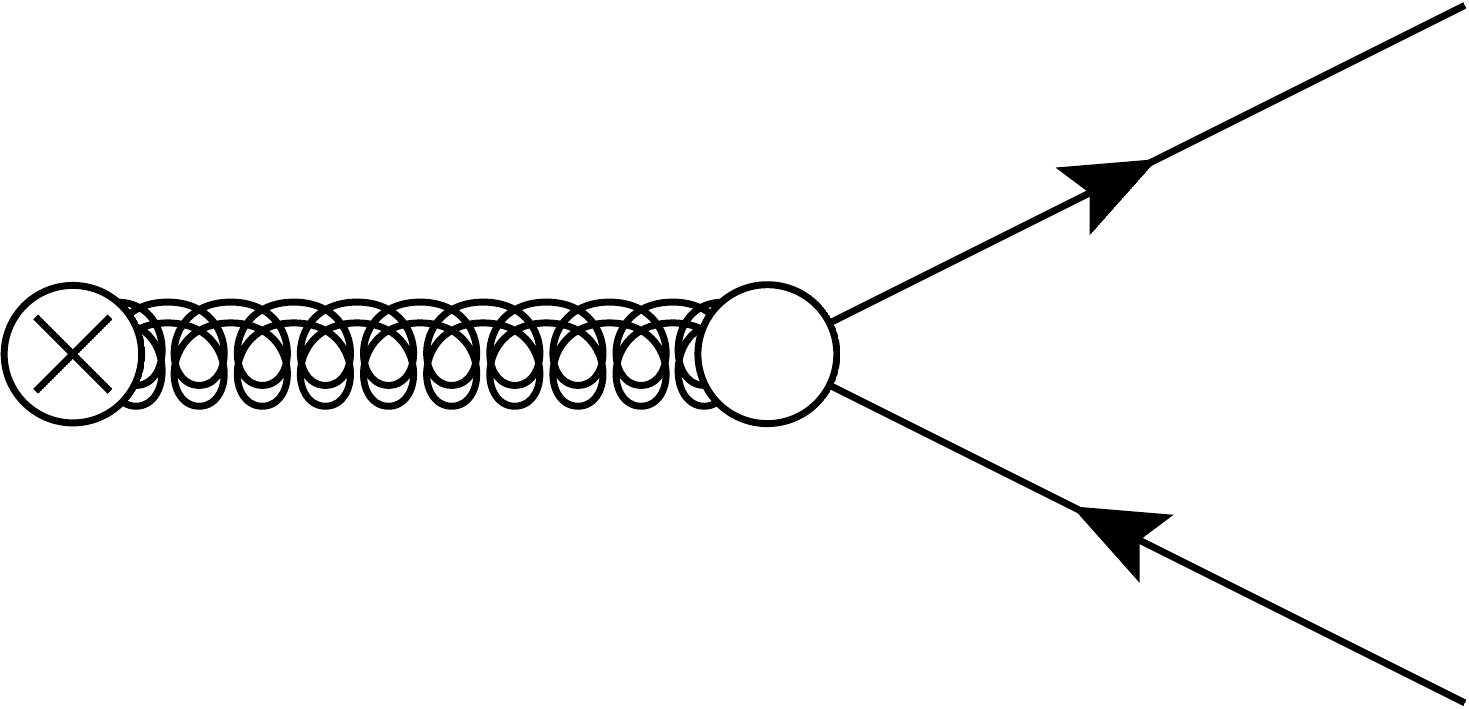}
		\caption*{(c)}
	\end{minipage}
	\begin{minipage}{.33\textwidth}
		\centering
		\includegraphics[width=.8\linewidth]{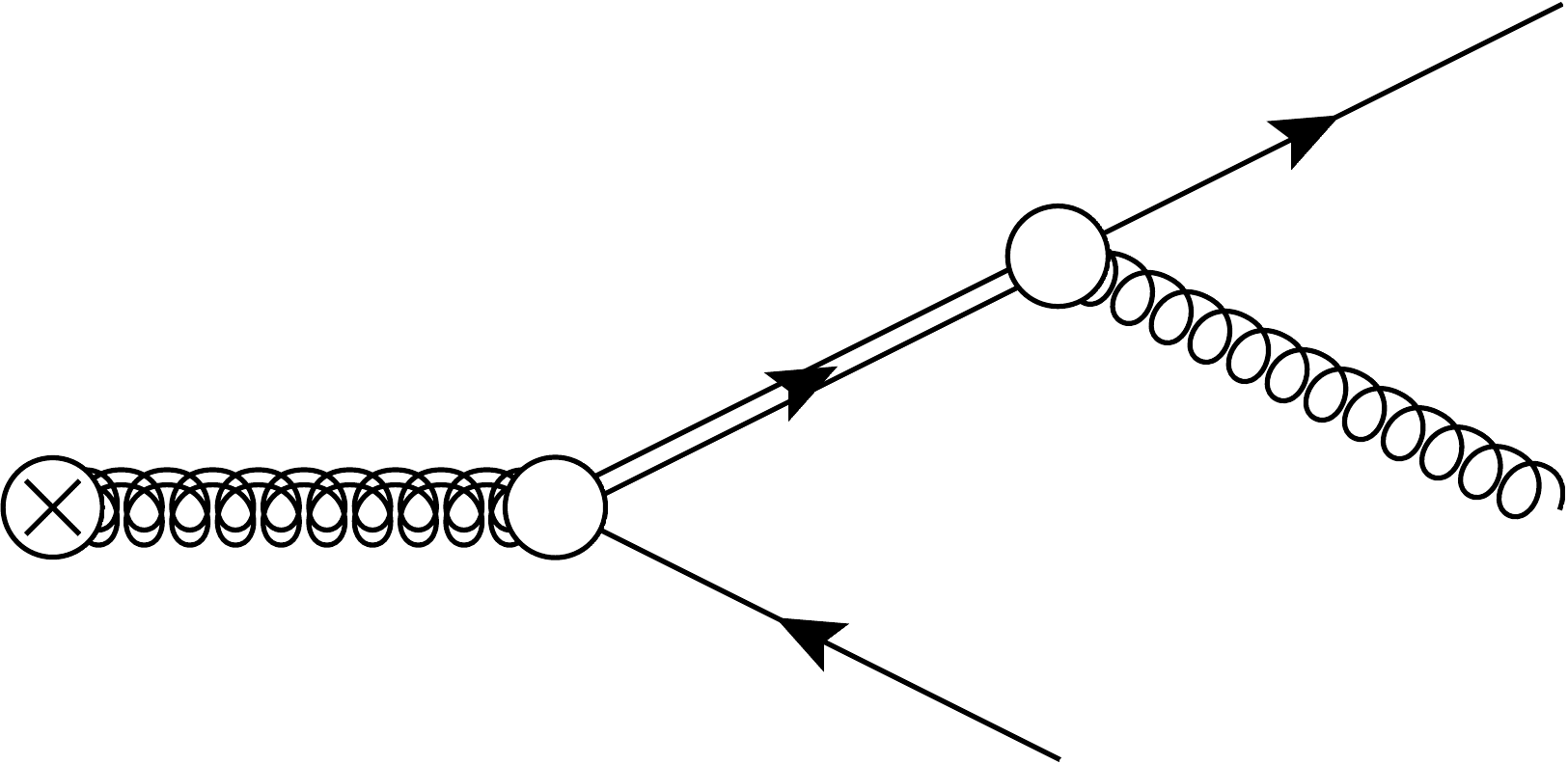}
		\caption*{(d)}
	\end{minipage}%
	\begin{minipage}{.33\textwidth}
		\centering
		\includegraphics[width=.8\linewidth]{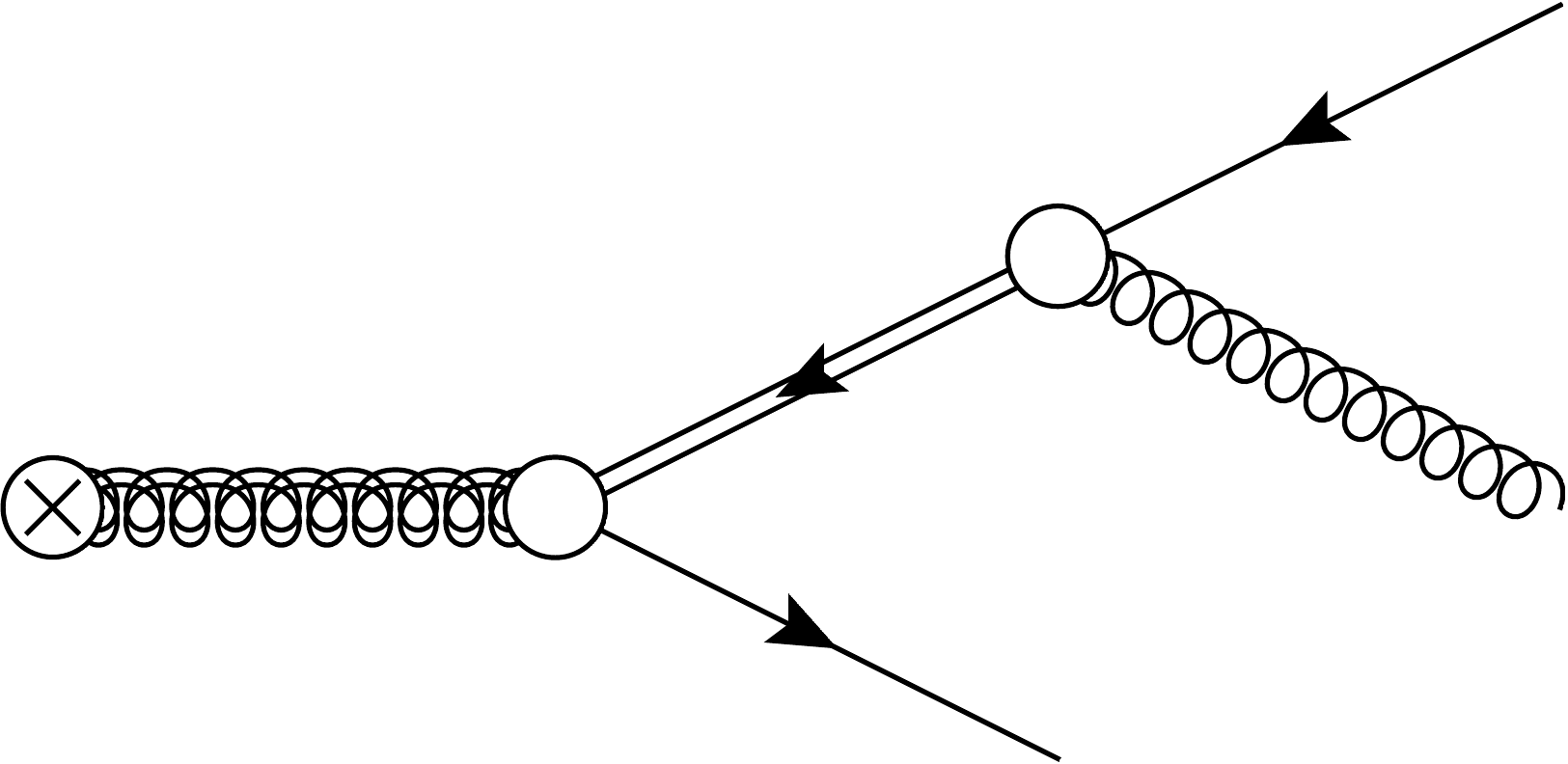}
		\caption*{(e)}
	\end{minipage}%
	\begin{minipage}{.33\textwidth}
		\centering
		\includegraphics[width=.8\linewidth]{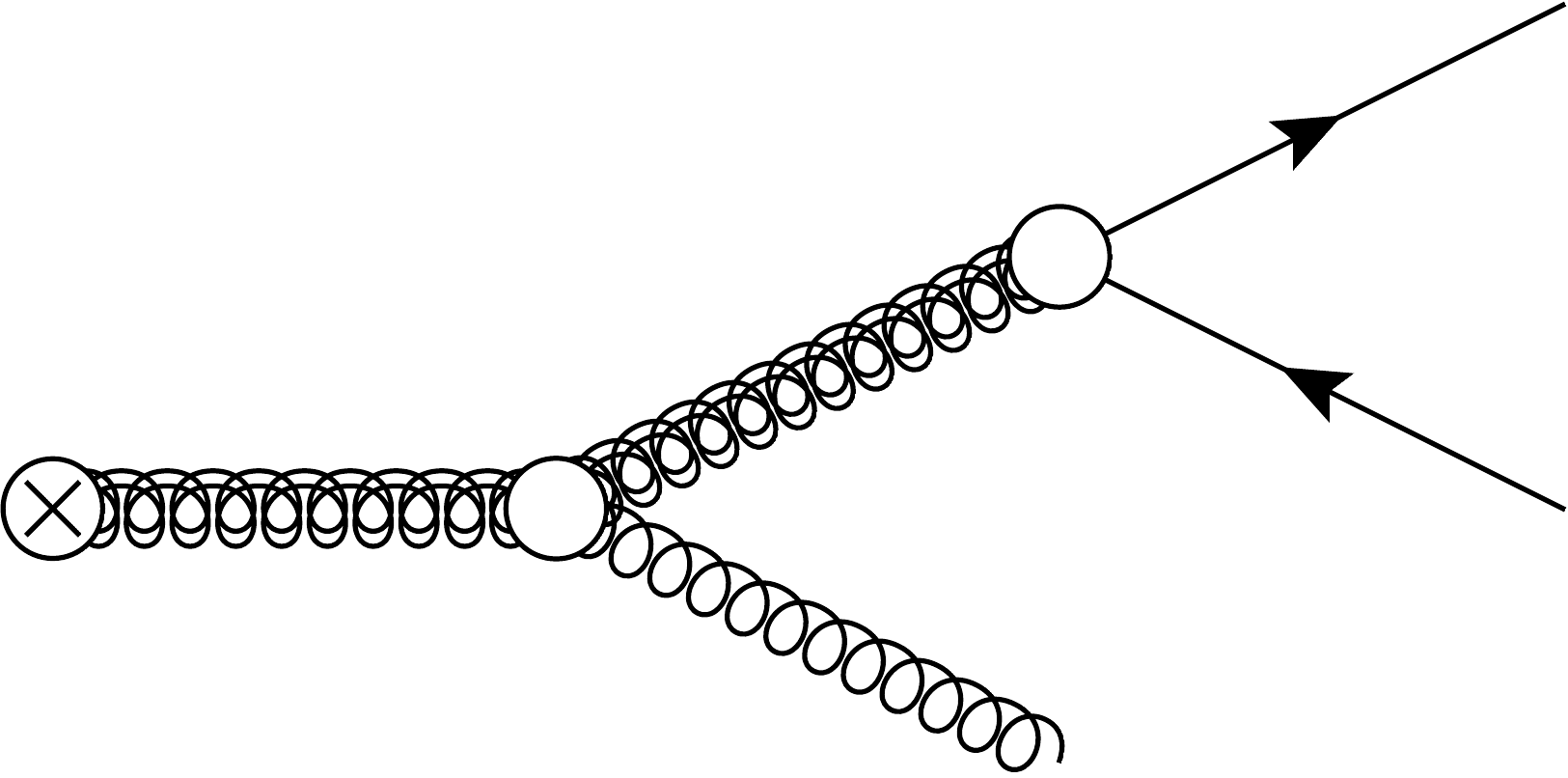}
		\caption*{(f)}
	\end{minipage}
	\caption{Diagrammatic representation of the expansion of the fields in \eqref{eq:perturbativeseriesforfields}. The top row is $\order{g_0}$: (a) represents $\chi_1$ from \eqref{eq:chi1NM}, (b) $\chibar_1$ from \eqref{eq:chi1barNM}, and (c) is a genuine new contribution due to the NMGF, namely $\mcF_{\mu,1}$ from \eqref{eq:F1NM}. The bottom row is $\order{g_0^2}$, and are all terms of $\mcF_{\mu,2}$ from \eqref{eq:F2}: (d) and (e) represent the first line while (f) collectively represents the second line in \eqref{eq:F2}.}
	\label{fig:NMexpansionfirstorders}
\end{figure}

We can modify the flow equation \eqref{eq:NMGF2} by the same Zwanziger term as in \eqref{eq:YMGFmodified} by utilizing a particular $t$-dependent gauge transformation, see appendix \ref{app:gaugetrans}. This gives rise to the modified NMGF equation
\begin{equation}\label{eq:NMGFModified}
\del_t B_\mu(t,x)=D_\nu G_{\nu\mu}(t,x)-g_0j_\mu^a(t,x)T^a+\alpha_0D_\mu\del_\nu B_\nu(t,x),\hspace{1cm}B_\mu(0,x)=A_\mu(x)
\end{equation}
For $\alpha_0=1$ the solution to the modified nonminimal gradient flow equation reads:
\begin{equation}\label{eq:NMGFModifiedSolution}
\begin{split}
B_\mu(t,x)&=e^{t\lapx} A_\mu(x)+\int_0^tds\,e^{(t-s)\lapx} \{ R_\mu(s,x) -g_0 j_\mu^a(s,x)T^a\}\\
&\equiv B_\mu^{YM}(t,x)+\mathcal{F}_\mu(t,x)
\end{split}
\end{equation}
where $R_\mu$ is still the same as in \eqref{eq:Rmu}, though now the $B_\mu$'s are the solution to \eqref{eq:NMGFModified} instead of \eqref{eq:YMGFmodified}, and we have explicitly put the superscript `YM' on the solution \eqref{eq:YMGFmodifiedSolution} of the Yang-Mills gradient flow equation \eqref{eq:YMGFmodified}. Again, we write the fields as a series in powers of the bare coupling:
\begin{equation}\label{eq:perturbativeseriesforfields}
\begin{split}
\chi(t,x)&=\sum_{n=0}^\infty g_0^n \chi_n(t,x),\hspace{1cm}\chibar(t,x)=\sum_{n=0}^\infty g_0^n \chibar_n(t,x),\\
B_\mu(t,x)&=\sum_{n=0}^\infty g_0^n B_{\mu,n}(t,x),\hspace{1cm} B_{\mu,n}(t,x)=B^{YM}_{\mu,n}(t,x)+\mcF_{\mu,n}(t,x)
\end{split}
\end{equation}
and thus we have for the first three orders of $B_\mu$
\begin{align}
B_{\mu,0}^a(t,x)&=B^{YM,a}_{\mu,0}(t,x)=e^{t\lapx}A_\mu^a(x)\\
B_{\mu,1}^a(t,x)&=\int_0^tds\,e^{(t-s)\lapx}\{f^{abc}B_{\nu,0}^b\left(2\del_\nu B_{\mu,0}^c-\del_\mu B_{\nu,0}^c\right)-\chibar_0\gmu T^a \chi_0\}\nonumber\\
&=B_{\mu,1}^{YM,a}(t,x)+\mcF_{\mu,1}^a(t,x)\\
B_{\mu,2}^a(t,x)&=B^{YM,a}_{\mu,2}(t,x)+\mcF_{\mu,2}^a(t,x)
\end{align}
where
\begin{align}
\mcF_{\mu,1}^a(t,x)&=-\int_0^tds\,e^{(t-s)\lapx}\left\{\chibar_0(s,x)\gmu T^a \chi_0(s,x)\right\} \label{eq:F1NM}\\
\mcF_{\mu,2}^a(t,x)&=-\int_0^tds\,e^{(t-s)\lapx}j_{\mu,1}^a(s,x)\nonumber\\
&\hspace{.1cm}+f^{abc}\int_0^tds\,e^{(t-s)\lapx}\{(\del_\mu \mcF_{\nu,1}^b-2\del_\nu\mcF_{\mu,1}^b)B_{\nu,0}^c+(\del_\mu B_{\nu,0}^b-2\del_\nu B_{\mu,0}^b)\mcF_{\nu,1}^c\}\label{eq:F2}
\end{align}
with
\begin{equation}
j_{\mu,1}^a(s,x)=\chibar_1(s,x)\gmu T^a \chi_0(s,x)+\chibar_0(s,x)\gmu T^a \chi_1(s,x)
\end{equation}
and
\begin{align}
\chibar_1(s,x)&=-2\int_0^sdu\,e^{(s-u)\lapx}\{\del_\rho\chibar_0(u,x)B_{\rho,0}(u,x)\} \label{eq:chi1barNM}\\
\chi_1(s,x)&=2\int_0^sdu\,e^{(s-u)\lapx}\{B_{\rho,0}(u,x)\del_\rho\chi_0(u,x)\} \label{eq:chi1NM}
\end{align}
These quantities are diagrammatically represented in figure \ref{fig:NMexpansionfirstorders}.

\subsection{Perturbative calculation of $\braket{E}$}\label{sec:pertCalc<E>NMGF}

\begin{figure}[t]
	\centering
	\begin{minipage}{.24\textwidth}
		\centering
		\includegraphics[width=.8\linewidth]{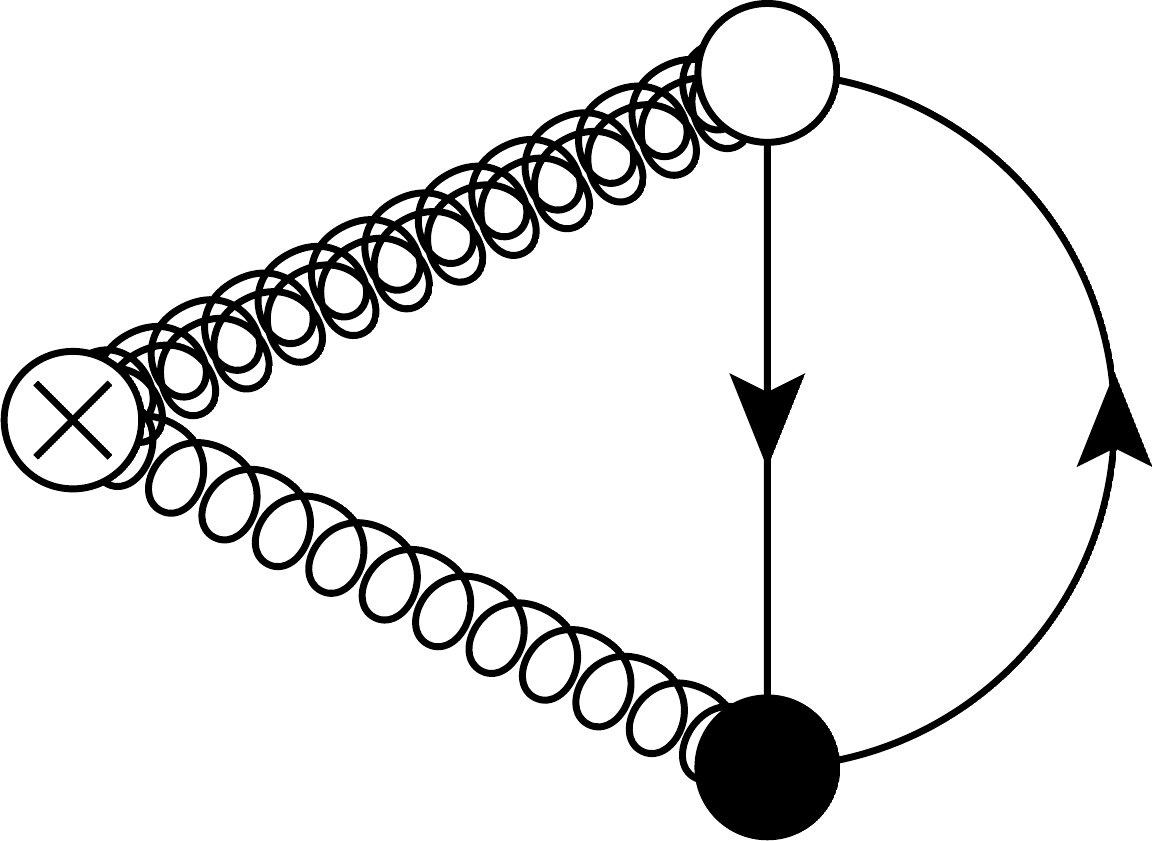}
		\caption*{(a)}
	\end{minipage}%
	\begin{minipage}{.24\textwidth}
		\centering
		\includegraphics[width=.8\linewidth]{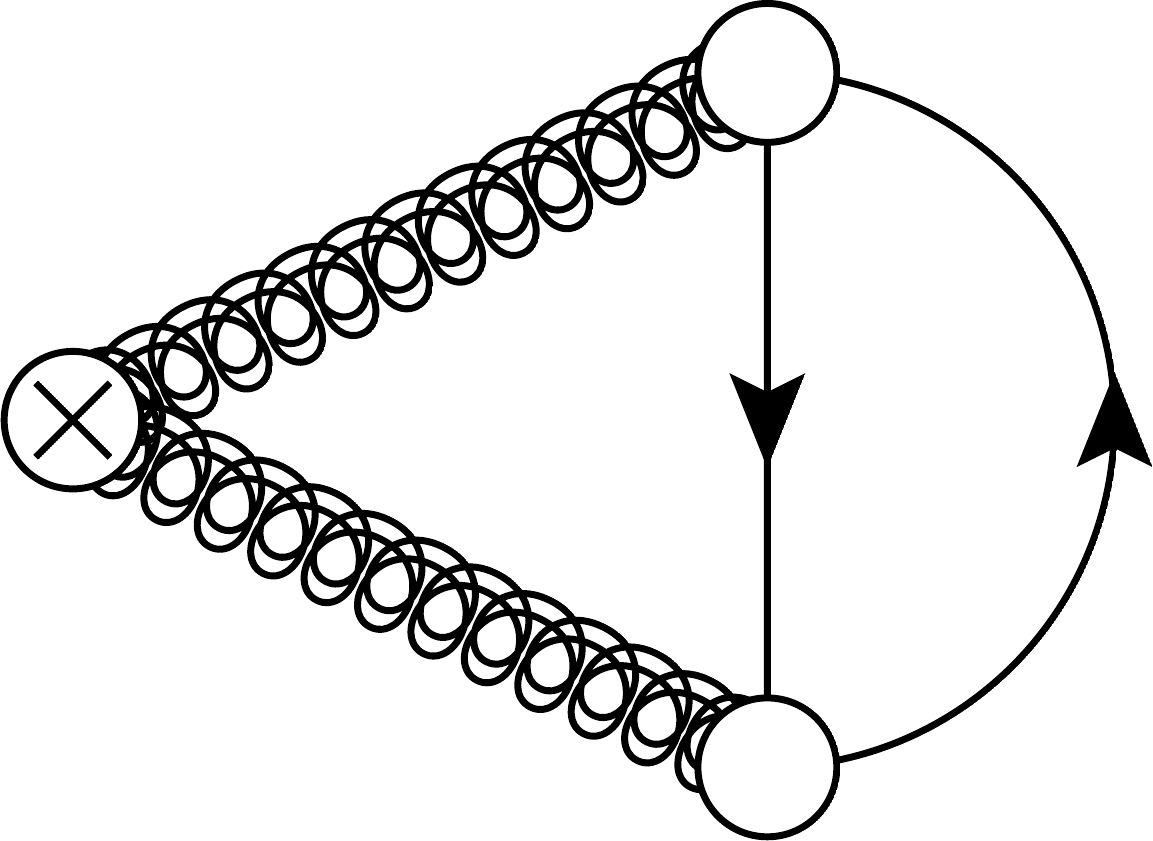}
		\caption*{(b)}
	\end{minipage}%
	\begin{minipage}{.24\textwidth}
		\centering
		\includegraphics[width=.8\linewidth]{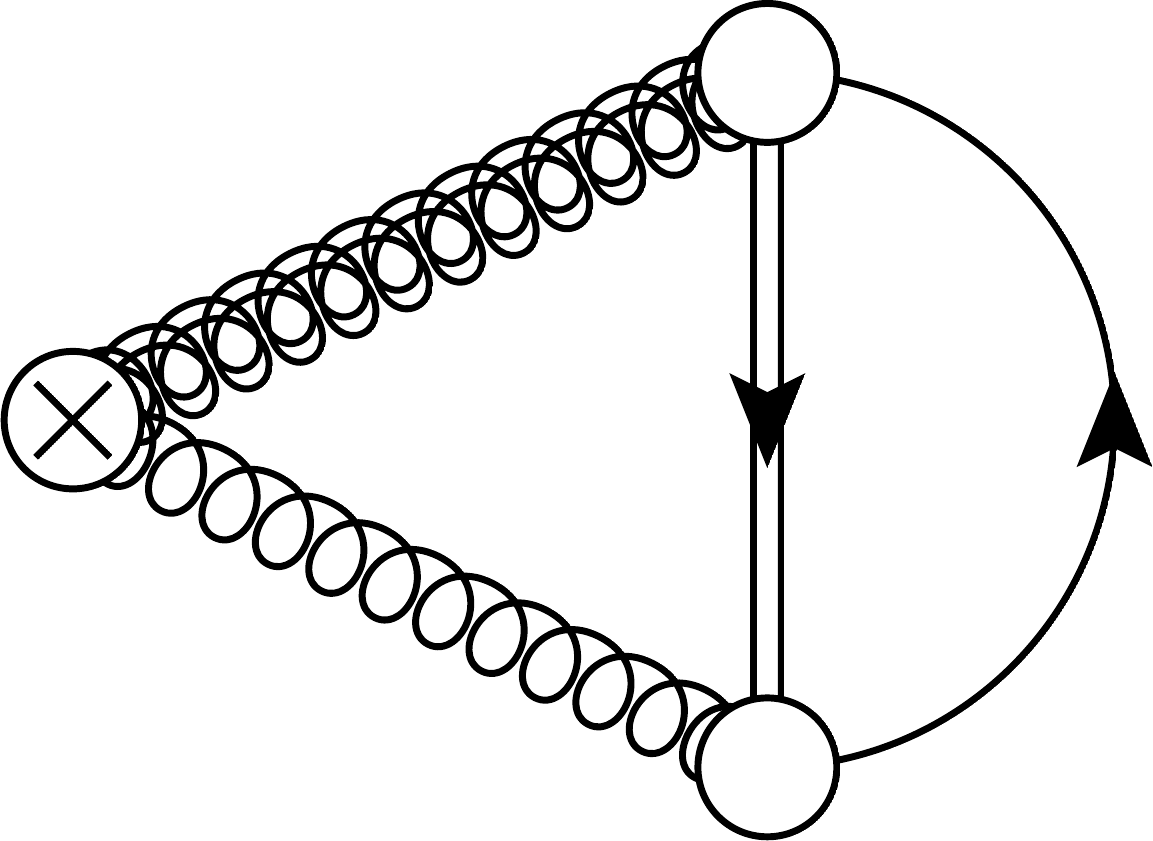}
		\caption*{(c)}
	\end{minipage}
	\begin{minipage}{.24\textwidth}
		\centering
		\includegraphics[width=.8\linewidth]{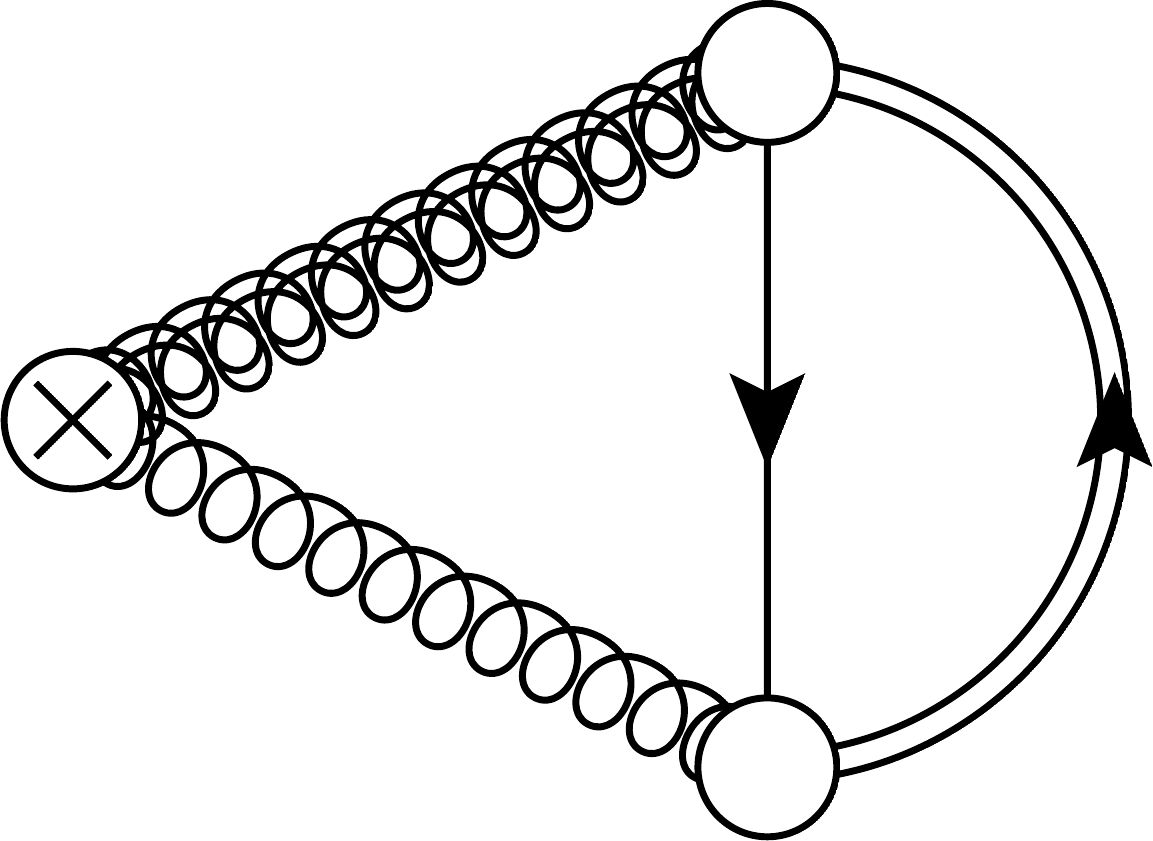}
		\caption*{(d)}
	\end{minipage}%
	\caption{Diagrammatic representation of the contributions to $\braket{E}$ due to the NMGF, whose total value is given in \eqref{eq:totalFermContr}. (a) represents $\mathcal{E}_{\chi,1}$ from \eqref{eq:Echi1}, (b) represents $\mathcal{E}_{\chi,2}$ from \eqref{eq:Echi2}, and the combination of (c) and (d) represents $\mathcal{E}_{\chi,3}$ from \eqref{eq:Echi3}.}
	\label{fig:ENM}
\end{figure}

We present the calculation of the new contributions --- collectively called $\mathcal{E}_\chi$ --- to $\braket{E}$ stemming from the fermionic term $\mcF_\mu$ in $B_\mu$, see \eqref{eq:NMGFModifiedSolution}. The full observable has the same form as in \eqref{eq:EexpandedinB}, though now the $B_\mu$'s represent solutions to the NMGF, so we have
\begin{equation}\label{eq:Eevolved}
\braket{E}=\frac{g_0^2}{4}\braket{G^a_\mn G^a_\mn}=\braket{E}\big|_{B_\mu=B_\mu^{YM}}+\mathcal{E}_\chi
\end{equation}
The new contributions start at $\order{g_0^4}$, and are given by
\begin{align}
\mathcal{E}_{\chi,1}&=g_0^3\braket{\del_\mu\mcF_{\nu,1}^a\left(\del_\mu B_{\nu,0}^a-\del_\nu B_{\mu,0}^a\right)}\nonumber\\
&=4g_0^4 N_f\,\text{tr}(T^aT^a)\Big\{(d-3)I_1+2\left(I_2+2I_3+I_4\right)\Big\} \label{eq:Echi1}\\
{}\nonumber\\
\mathcal{E}_{\chi,2}&=\half g_0^4\braket{\del_\mu\mcF_{\nu,1}^a\left(\del_\mu\mcF_{\nu,1}^a-\del_\nu\mcF_{\mu,1}^a\right)}\nonumber\\
&=4g_0^4N_f\,\text{tr}(T^aT^a)\Big\{I_5+(d-1)I_6+(d-2)I_7\Big\} \label{eq:Echi2}\\
{} \nonumber\\
\mathcal{E}_{\chi,3}&=g_0^4\braket{\del_\mu\mcF_{\nu,2}^a\left(\del_\mu B_{\nu,0}^a-\del_\nu B_{\mu,0}^a\right)}\nonumber\\
&=16g_0^4N_f\,\text{tr}(T^aT^a)\Big\{I_8-I_9\Big\} \label{eq:Echi3}
\end{align}
and the integrals $I_1$--$I_9$ are listed in appendix \ref{app:integralsFermContr}. These contributions are diagrammatically represented in figure \ref{fig:ENM}.
The total fermionic contribution is subsequently given by
\begin{equation}\label{eq:totolFermContrIntegralsSum}
\mathcal{E}_\chi=\sum_{i=1}^3\mathcal{E}_{\chi,i}=4g_0^4N_f\,\text{tr}(T^aT^a)\left((d-2)I_1+I_5+(d-1)I_6+(d-2)I_7+4I_8\right)
\end{equation}
The integrals $I_1$, $I_5$--$I_8$, are solved using Schwinger parametrization:
\begin{equation}
\frac{1}{p^2}=\int_0^\infty du\, e^{-up^2}
\end{equation}
and Gaussian integration
\begin{equation}
\int_{p,q}e^{-\vec{p}^TA\vec{p}}=(4\pi)^{-d}\det(A)^{-d/2}         
\end{equation}
The expressions obtained in this way contain gamma functions, incomplete beta functions and hypergeometric functions, which depend on $d$ only; the $t$ dependence factors out. The $\eps$-expansions with $d=4-2\eps$ of $I_1$, $I_5$--$I_8$ are provided in appendix \ref{app:integralsFermContr}. Adding all contributions, we find for \eqref{eq:totolFermContrIntegralsSum}:
\begin{equation}\label{eq:totalFermContr}
\mathcal{E}_{\chi}=-\half \frac{(N^2-1)}{(8\pi t)^{d/2}}(d-1) N_f\frac{g_0^4}{(4\pi)^2}\left(2+\frac{4}{3}\log 2 -\log 3+\order{\eps}\right)
\end{equation}
which is manifestly finite, as is to be expected anticipating the results of section \ref{sec:renormNMGF}. We now have for the full observable \eqref{eq:Eevolved}:
\begin{equation}
\braket{E}=\half g_0^2\frac{(N^2-1)}{(8\pi t)^{d/2}}(d-1)\left\{1+c'_1g_0^2+\order{g_0^4}\right\}
\end{equation}
\begin{equation}
c'_1=\frac{1}{(4\pi)^2}(4\pi)^{\eps}(8t)^{\eps}\left\{N\left(\frac{11}{3\eps}+\frac{52}{9}-3\log 3\right)-N_f\left(\frac{2}{3\eps}+\frac{22}{9}-\log 3\right)+\order{\eps}\right\}
\end{equation}
Following the same steps as in section \ref{sec:perturbCalcEnonevolved}, i.e.~renormalizing the coupling as in \eqref{eq:couplingRenormalizationMSbar}, expressing the renormalized coupling $\alpha(\mu)$ in terms of the RG invariant coupling $\alpha(q)$ by inverting \eqref{eq:alpha(q)intermsofalpha(mu)} and setting $q=(8t)^{-1/2}$ and $N=3$, we now have:
\begin{equation}\label{eq:<E>finalNMGF}
\braket{E}=\frac{3}{4\pi t^2}\alpha(q)\left\{1+k'_1\alpha(q)+\order{\alpha^2}\right\},\hspace{1cm}k'_1=1.0978-0.1377\times N_f
\end{equation}
Note that the $N_f$ contribution in $k'_1$ is negative, in contrast to the result from \eqref{eq:EN=3Luscher} that is obtained using the Yang-Mills gradient flow. See appendix \ref{app:generalNresults} for the renormalized result for general $N$. We further comment on our result \eqref{eq:<E>finalNMGF} in section \ref{sec:Nfdependence}, where we will make a more elaborate comparison with the YMGF result from \eqref{eq:EN=3Luscher}.

\subsection{Renormalization of evolved fields}\label{sec:renormNMGF}

The result for $\braket{E}$ from the previous section seems to indicate that the nonminimally evolved gauge field remains devoid of any renormalization. However, one might suspect that at the next order in $g_0$, i.e.~$\order{g_0^6}$, divergences arise from the fermionic contribution; at that order the self-energy contribution to the evolved fermionic propagator might have to be canceled with $Z_\chi$ factors. Our aim here is to show that this is not the case, and any correlator consisting only of evolved gauge fields $B_\mu$ --- $g_0 B_\mu$ when canonically normalized --- is still finite, without any renormalization beyond that of the nonevolved $d$ dimensional theory that is being used in the path integral averaging.

The demonstration is a straightforward generalization of the one from \cite{Luscher:2011bx,Luscher:2013cpa}, see also \cite{Hieda:2016xpq}, that we have reviewed in section \ref{sec:renormEvolvedFieldsYM}. We will now show where the changes are with respect to the YMGF case, and why the same reasoning still holds.

First of all, the gauge bulk action $S_{G,fl}$ from \eqref{eq:gaugeBulkActionYM} changes in order to comply with the NMGF equation \eqref{eq:NMGFModified}:
\begin{equation}\label{eq:gaugeBulkActionNMGF}
S_{G,fl}=-2\int_0^\infty dt\int_x\text{tr}\left\{L_\mu(\del_t B_\mu-D_\nu G_{\nu\mu}+g_0j_\mu^aT^a-\alpha_0D_\mu\del_\nu B_\nu)\right\}
\end{equation}
Using the BRST variations provided in appendix \ref{app:BRST}, we establish that the complete bulk action is again invariant:
\begin{equation}
\delta S_{bulk}=0
\end{equation}
Also, the path integral measure has not been changed and is still BRST invariant, therefore $\braket{\delta O}=0$ still holds, where $O$ is any combination of evolved and nonevolved elementary fields. Therefore, also \eqref{eq:BRSTimportantresult} still holds.

The argument for the absence of bulk counterterms translates one-to-one from the YMGF to the NMGF case, with one addition. There now is an additional interaction term proportional to $\chibar L_\mu \chi$, which upon the substitution of \eqref{eq:homogenousbcsubstitutions} yields the bulk interaction $\Psibar L_\mu\Psi$. However, recalling that the only nonzero propagators are $\braket{bL}$, $\braket{\Psi\lambdabar}$, $\braket{\lambda\Psibar}$ and $\braket{d\dbar}$, in order to form a loop in the bulk the presence of the interaction term $\lambdabar B_\mu \lambda$ is essential. It is clear from the fermionic bulk action \eqref{eq:fermBulkAction} that this term is absent, and thus no loops can be formed. Therefore, all bulk correlators still solely consist of trees and no divergences can arise, i.e.~there will be no bulk counterterms.

The allowed boundary counterterms remain to be of the form given in \eqref{eq:boundaryCTs}. Also, since we still have $\braket{\delta O}=0$, and thus in particular \eqref{eq:BRSTimportantresult} still holds, we have $z_1=z_2=0$, i.e.~there is again no field renormalization factor for $L_\mu$, and consequently also no field renormalization factor for $B_\mu$.

\begin{figure}[t]
	\centering
	\begin{minipage}{.33\textwidth}
		\centering
		\includegraphics[width=.8\linewidth]{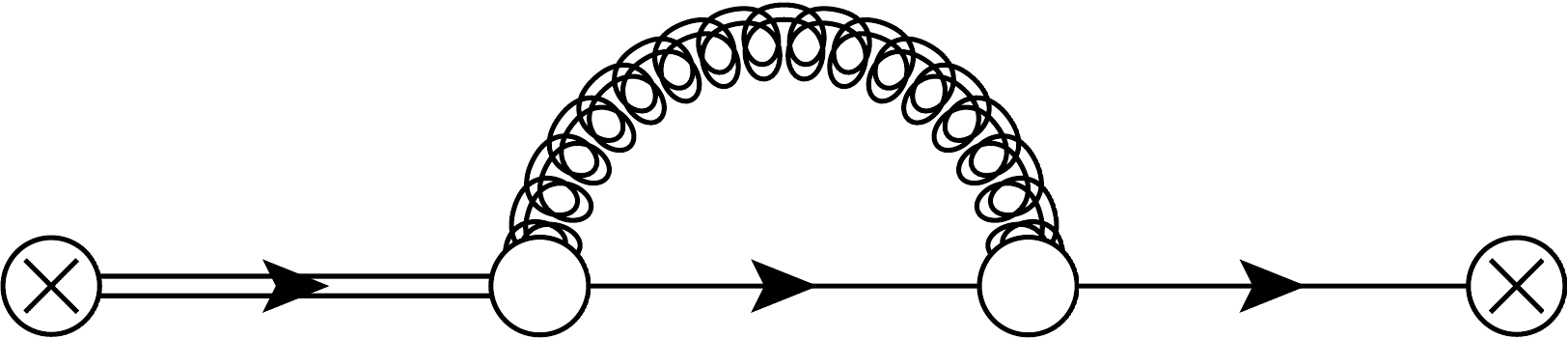}
		\caption*{NM1}
	\end{minipage}%
	\begin{minipage}{.33\textwidth}
		\centering
		\includegraphics[width=.8\linewidth]{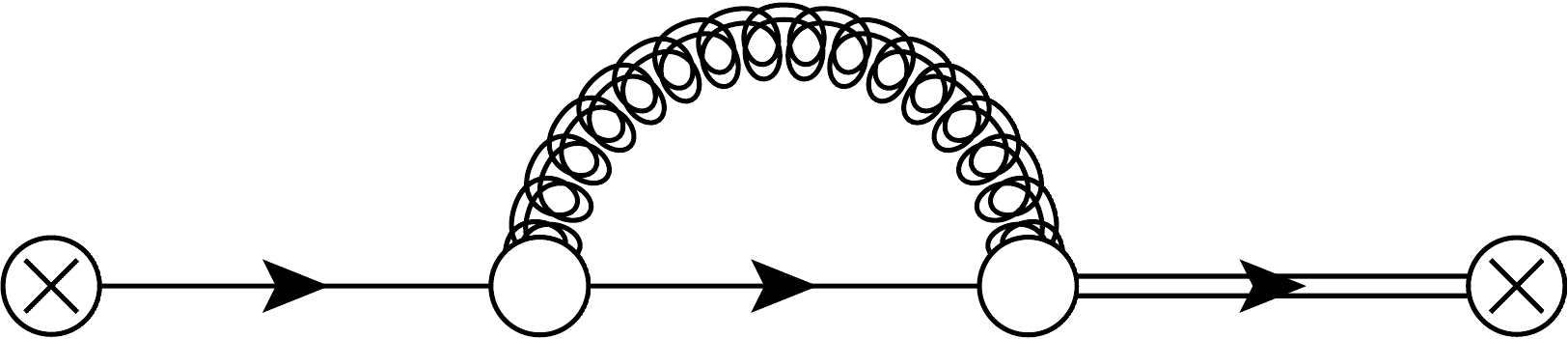}
		\caption*{NM2}
	\end{minipage}%
	\caption{New next-to-leading order diagrams for the NMGF evolved fermion 2-point function. Combined with the diagrams from figure \ref{fig:selfenergyYM} they represent the full NMGF evolved fermion 2-point function, which is rendered finite by multiplying with the renormalization factor $Z_\chi$ from \eqref{eq:ZchiNM}.}
	\label{fig:selfenergyNM}
\end{figure}

Now, returning to the issue raised at the beginning of this section: might it be that the $\chi$ and $\chibar$ field in $j_\mu^a$ in the solution \eqref{eq:NMGFModifiedSolution} of the NMGF equation \eqref{eq:NMGFModified} have to be renormalized in order to render correlators of $B_\mu$ finite? It is now clear that the answer is no. A $Z_\chi^{-1}$ factor in the flow equation \eqref{eq:NMGFModified} would imply a $Z_\chi^{-1}$ factor multiplying $j_\mu^a$ in the gauge bulk action \eqref{eq:gaugeBulkActionNMGF}. This in turn would imply a bulk counterterm, unless $L_\mu=Z_\chi(L_R)_\mu$ i.e. that $L_\mu$ renormalizes. The bulk counterterm is excluded due to the absence of loops in the bulk, and the renormalization of $L_\mu$ is excluded due to BRST symmetry.
\\
The discussion presented above does not make any statement about the value of $Z_\chi$, and there is no reason why it should remain the same as when using the YMGF, \eqref{eq:ZchiYM}. The new vertex at $\order{g_0}$ due to the NMGF --- see \eqref{eq:F1NM} and corresponding figure \ref{fig:NMexpansionfirstorders}.c --- gives rise to a new contribution to $Z_\chi$. In figure \ref{fig:selfenergyNM} we show the diagrams that contribute to the evolved fermion 2-point function, and their values are presented in table \ref{tab:ZchiNMGF} in units of $C_2(R)\frac{g_0^2}{(4\pi)^2}S(\xbar_t-\ybar_t)$, where $S(\xbar_t-\ybar_t)$ is the evolved fermion propagator:
\begin{equation}\label{eq:evolved fermion propagater}
S(\xbar_t-\ybar_t)\equiv e^{t(\Delta_x+\Delta_y)}S(x-y)=\int_pe^{ip(x-y)}e^{-2tp^2}\frac{-i\pslash}{p^2}
\end{equation}
Combining these new contributions with the self-energy diagrams presented in appendix \ref{app:ZchiYMGF} we find
\begin{equation}\label{eq:ZchiNM}
\text{NMGF}:\hspace{1cm}Z_{\chi}(g(\mu),\eps)=1+C_2(R)\frac{2}{\eps}\frac{g^2(\mu)}{(4\pi)^2}+\order{g^4}
\end{equation}

\begin{table}[t]
	\caption{UV divergence of the diagrams from figure \ref{fig:selfenergyNM} in units of $C_2(R)\frac{g_0^2}{(4\pi)^2}S(\xbar_t-\ybar_t)$.}
	\label{tab:ZchiNMGF}
	\begin{center}
		\begin{tabular}{ ll }
			\hline
			diagram & value \\
			\hline
			{} & {} \\
			NM1 + NM2 \hspace{1cm} & $\frac{1}{\eps}+\order{\eps^0}$ \\
			{} & {} \\
			\hline
		\end{tabular}
	\end{center}
\end{table}

\section{Using $\slashed{D}^2$ in the fermion flow equations}\label{sec:Dslash^2}

Next we present the fermion flow equation using $\slashed{D}^2$ instead of $D^2$, and calculate its contribution to $\braket{E}$. The well-known relation between these operators is:
\begin{equation}\label{eq:Dslash^2}
\slashed{D}^2= D^2+\frac{g_0}{2}\sigma_\mn G_\mn
\end{equation}
with $\sigma_\mn=\half[\gmu,\gnu]$. The modified fermion flow equations from \eqref{eq:gfeqnfermions} now read
\begin{equation}\label{eq:DslashGFeqnfermions}
\begin{split}
\del_t\chi(t,x)&=\{\slashed{\Delta}-\alpha_0 g_0\del_\mu B_\mu(t,x)\}\,\chi(t,x),\hspace{1cm}\chi(0,x)=\psi(x)\\
\del_t\chibar(t,x)&=\chibar(t,x)\{\overleftarrow{\slashed{\Delta}}+\alpha_0g_0\del_\mu B_\mu(t,x)\},\hspace{1cm}\chibar(0,x)=\psibar(x)
\end{split}
\end{equation}
with $\slashed{\Delta}=\slashed{D}^2$, $\overleftarrow{\slashed{\Delta}}=\overleftarrow{\slashed{D}}^2$. The solutions to these flow equations are
\begin{equation}\label{eq:DslashGFeqnfermionssolution}
\begin{split}
\chi(t,x)&=e^{t\lapx}\psi(x)+\int_0^tds\,e^{(t-s)\lapx}\{\slashed{\Delta}' \chi(s,x)\}\\
\chibar(t,x)&=e^{t\lapx}\psibar(x)+\int_0^tds\,e^{(t-s)\lapx}\{\chibar(s,x)\overleftarrow{\slashed{\Delta}}'\}
\end{split}
\end{equation}
with
\begin{equation}
\slashed{\Delta}'=\Delta'+\frac{g_0}{2}\sigma_\mn G_\mn,\hspace{1cm} \overleftarrow{\slashed{\Delta}}'=\overleftarrow{\Delta}'+\frac{g_0}{2}\sigma_\mn G_\mn
\end{equation}
where $\Delta'$ and $\overleftarrow{\Delta}'$ are given in \eqref{eq:Deltaprime}. The combination of these fermion flow equations and the nonminimal gauge field flow from section \ref{sec:NMGF} will be referred to as the `slashed' nonminimal gradient flow (sNMGF).

It is convenient to define the extra contribution in $\chi(t,x)$ stemming from the $\sigma_\mn$ term in \eqref{eq:Dslash^2}:
\begin{equation}
\begin{split}
\chi_{\slashed{\Delta}}(t,x)&=\chi_{\Delta}(t,x)+\chi^*(t,x)\\
\chibar_{\slashed{\Delta}}(t,x)&=\chibar_{\Delta}(t,x)+\chibar^*(t,x)
\end{split}
\end{equation}
where $\chi_{\slashed{\Delta}}$ is representing the solution from \eqref{eq:DslashGFeqnfermionssolution}, and $\chi_{\Delta}$ the solution from \eqref{eq:gffermionsmodifiedsolution}. The extra contributions $\chi^*,\chibar^*$ are given by
\begin{equation}
\begin{split}
\chi^*(t,x)&=\int_0^tds\,e^{(t-s)\lapx}\{\Delta'\chi^*(s,x)+\frac{g_0}{2}\sigma_\mn G_\mn(s,x)\chi_{\slashed{\Delta}}(s,x)\}\\
\chibar^*(t,x)&=\int_0^tds\,e^{(t-s)\lapx}\{\chibar^*(s,x)\overleftarrow{\Delta}'+\frac{g_0}{2}\chibar_{\slashed{\Delta}}(s,x)\sigma_\mn G_\mn(s,x)\}
\end{split}
\end{equation}
\begin{figure}[t]
	\centering
	\begin{minipage}{.33\textwidth}
		\centering
		\includegraphics[width=.8\linewidth]{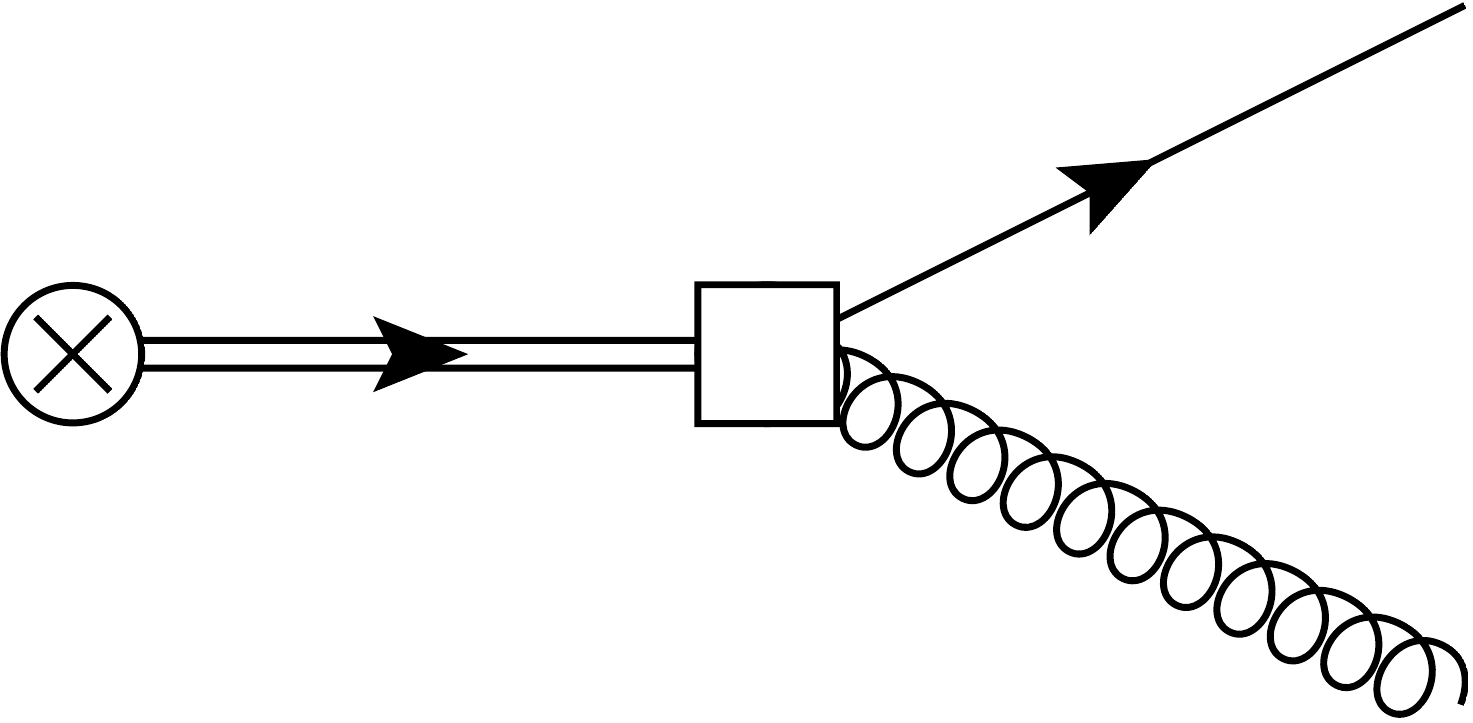}
		\caption*{(a)}
	\end{minipage}%
	\begin{minipage}{.33\textwidth}
		\centering
		\includegraphics[width=.8\linewidth]{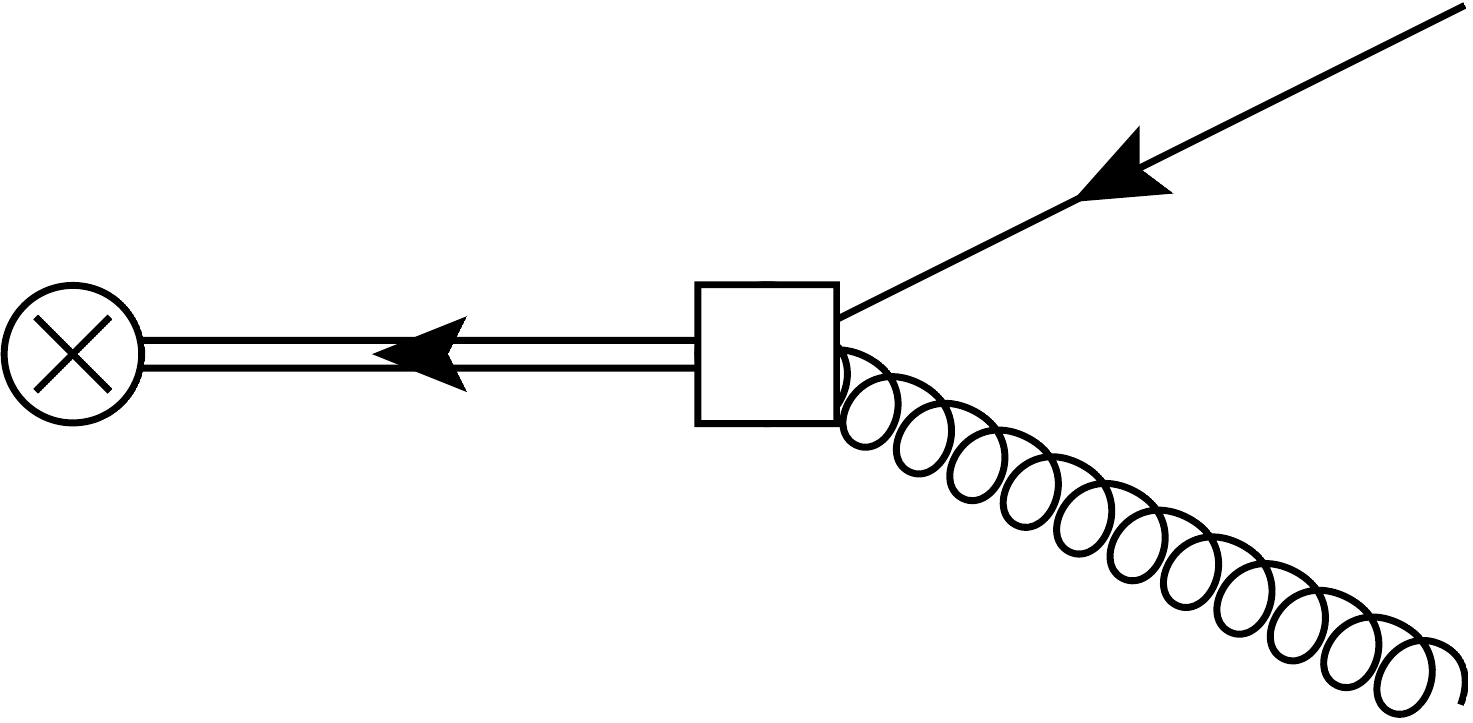}
		\caption*{(b)}
	\end{minipage}%
	\begin{minipage}{.33\textwidth}
		\centering
		\includegraphics[width=.8\linewidth]{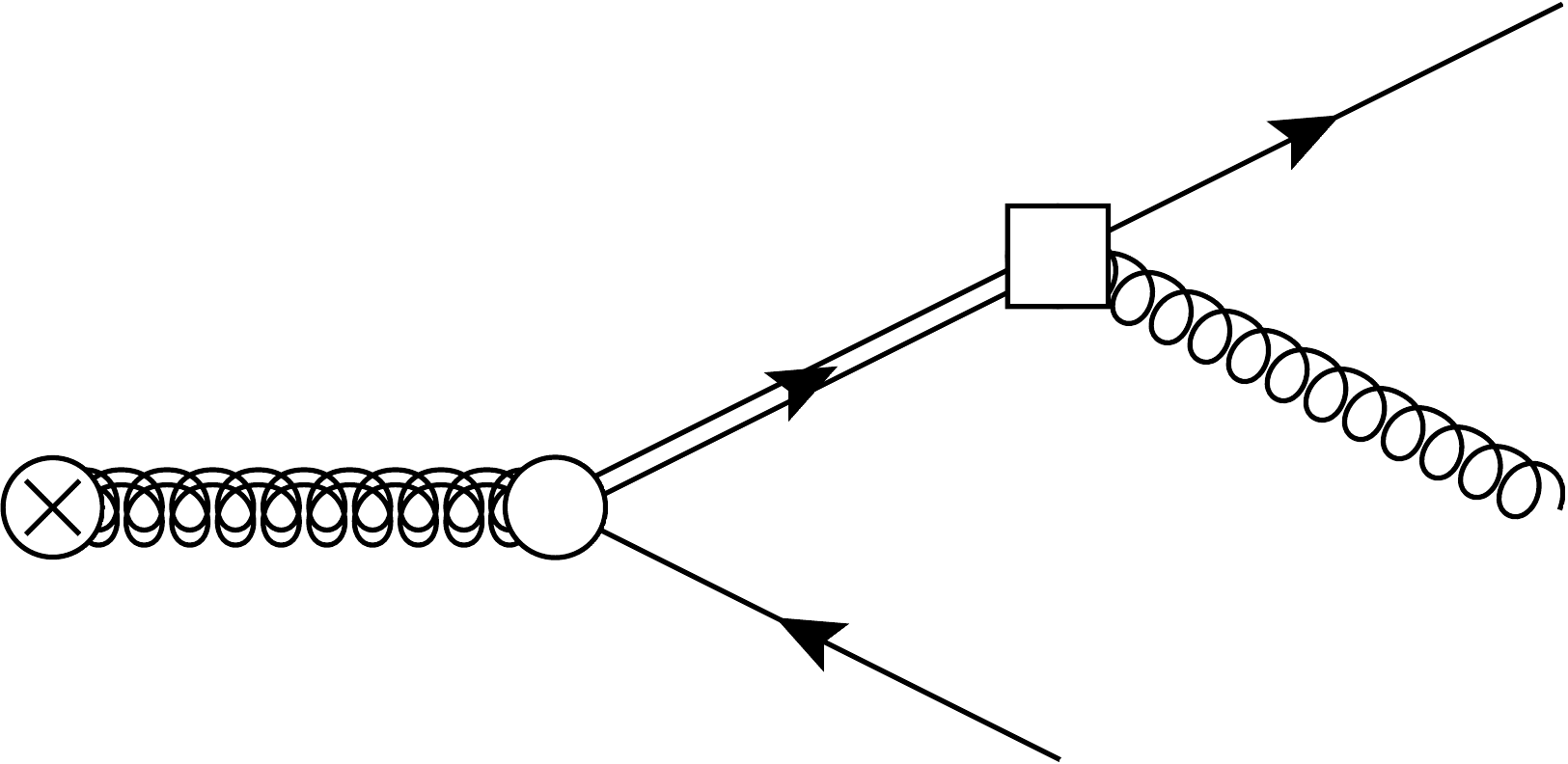}
		\caption*{(c)}
	\end{minipage}
	\begin{minipage}{.33\textwidth}
		\centering
		\includegraphics[width=.8\linewidth]{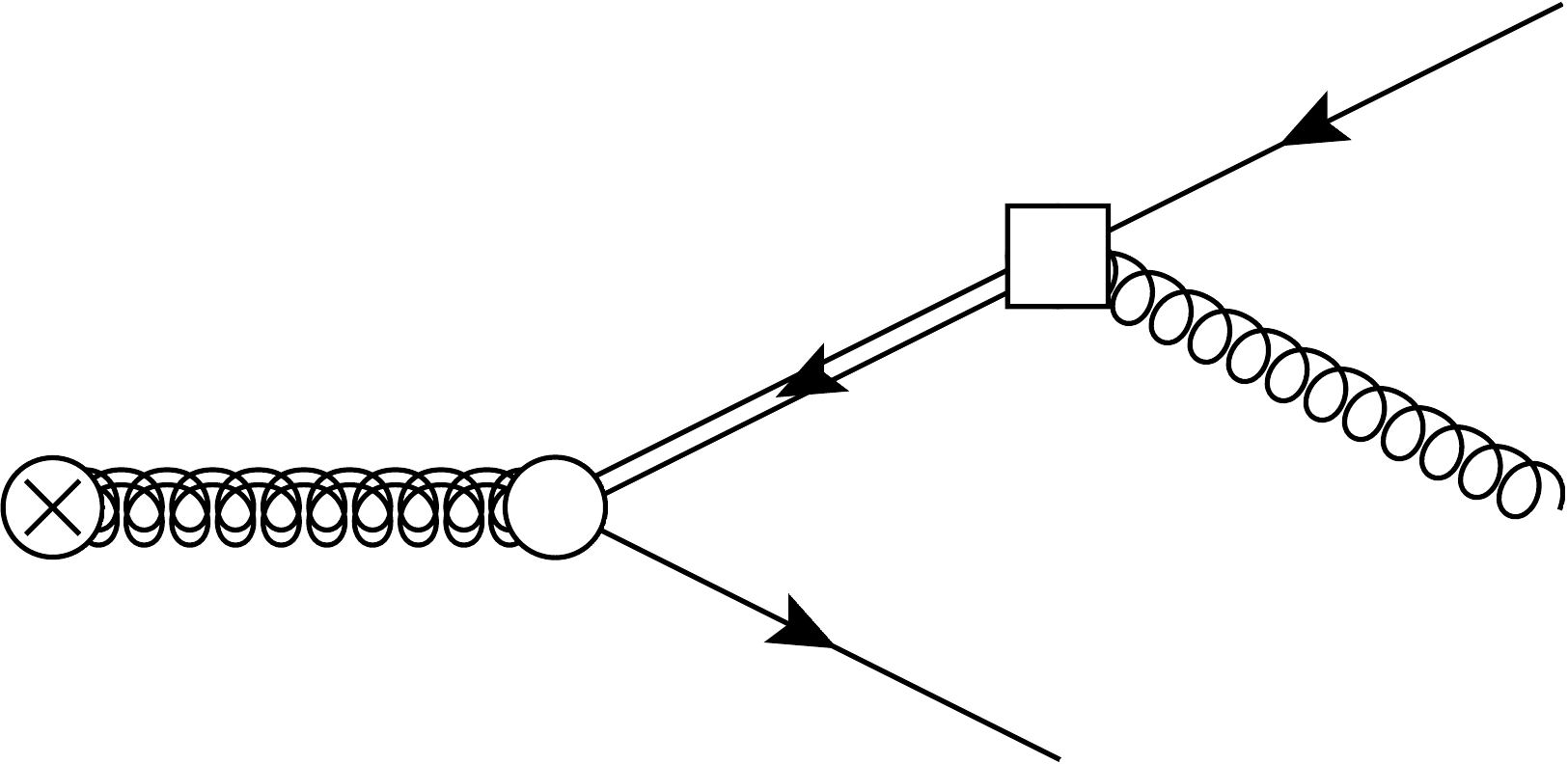}
		\caption*{(d)}
	\end{minipage}%
	\caption{Diagrammatic representation of the new terms due to the sNMGF in the expansion of the fields in \eqref{eq:seriessNMGFchis}. The new interaction vertices due to the sNMGF are represented by white blocks instead of blobs. (a) and (b) are $\order{g_0}$ and represent $\chi_1^*$ and $\chibar_1^*$ from \eqref{eq:sNMchipertexp}, while (c) and (d) are $\order{g_0^2}$ and collectively represent $\mcF_{\mu,2}^*$ from \eqref{eq:F2sNM}.}
	\label{fig:sNMexpansions}
\end{figure}
Writing these contributions as a series in powers of $g_0$
\begin{equation}\label{eq:seriessNMGFchis}
\chi^*(t,x)=\sum_{n=0}^\infty g_0^n\chi^*_n(t,x),\hspace{1cm}\chibar^*(t,x)=\sum_{n=0}^\infty g_0^n\chibar^*_n(t,x)
\end{equation}
we have for the first two orders:
\begin{equation}\label{eq:sNMchipertexp}
\begin{split}
\chi^*_0(t,x)&=\chibar^*_0(t,x)=0\\
\chi^*_1(t,x)&=\frac{1}{2}\sigma_\mn\int_0^tds\,e^{(t-s)\lapx}\{(\del_\mu B_{\nu,0}(s,x)-\del_\nu B_{\mu,0}(s,x))\chi_0(s,x)\}\\
\chibar^*_1(t,x)&=\frac{1}{2}\int_0^tds\,e^{(t-s)\lapx}\{\chibar_0(s,x)(\del_\mu B_{\nu,0}(s,x)-\del_\nu B_{\mu,0}(s,x))\}\sigma_\mn
\end{split}
\end{equation}
The first --- and at this order in $g_0$ only --- nonzero new contribution to $B_\mu$ from \eqref{eq:NMGFModifiedSolution} is then given by
\begin{equation}\label{eq:F2sNM}
\mathcal{F}^{*a}_{\mu,2}(t,x)=-\int_0^tds\,e^{(t-s)\lapx}j_{\mu,1}^{*a}(s,x)
\end{equation}
where
\begin{equation}
j^{*a}_{\mu,1}(t,x)=\chibar_1^*(t,x)\gmu T^a\chi_0(t,x)+\chibar_0(t,x)\gmu T^a \chi^*_1(t,x)
\end{equation}
These new contributions are diagrammatically represented in figure \ref{fig:sNMexpansions}. Note that the new interaction vertices due to the sNMGF are represented by white blocks instead of blobs.

\subsection{Perturbative calculation of $\braket{E}$}
\begin{figure}[t]
	\centering
	\begin{minipage}{.24\textwidth}
		\centering
		\includegraphics[width=.8\linewidth]{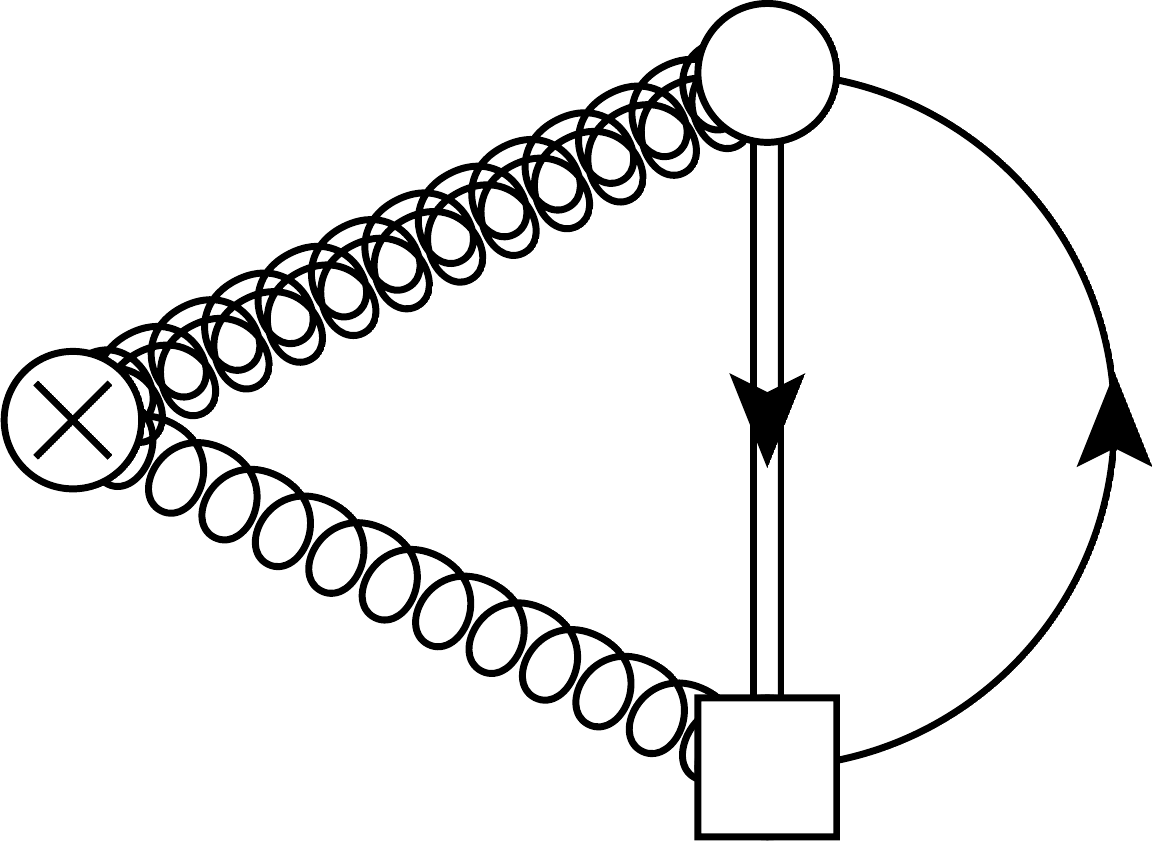}
		\caption*{(a)}
	\end{minipage}%
	\begin{minipage}{.24\textwidth}
		\centering
		\includegraphics[width=.8\linewidth]{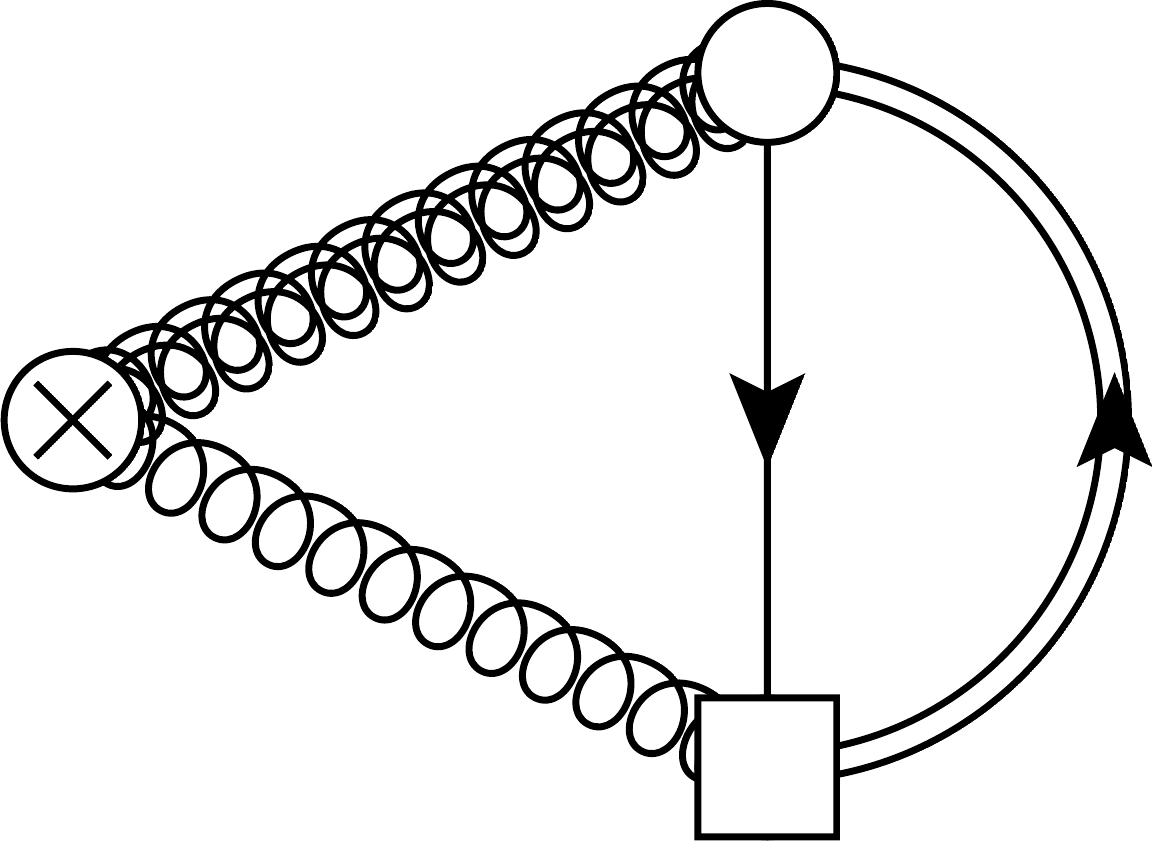}
		\caption*{(b)}
	\end{minipage}%
	\caption{Diagrammatic representation of the contribution to $\braket{E}$ due to the sNMGF, whose total value is given in \eqref{eq:sNMcontr}. (a) and (b) collectively represent $\mathcal{E}_\chi^*$ from \eqref{eq:sNMEstar}.}
	\label{fig:sNME}
\end{figure}

The calculation of $\braket{E}$ now consists of calculating one additional contribution with respect to the calculation presented in section \ref{sec:pertCalc<E>NMGF}. Comparing to \eqref{eq:Eevolved}, we now have:
\begin{equation}\label{eq:EevolvedDslash}
\braket{E}=\braket{E}\big|_{B_\mu=B_\mu^{YM}}+\mathcal{E}_\chi+\mathcal{E}_\chi^*
\end{equation}
with $\mathcal{E}_\chi$ as presented in section \ref{sec:pertCalc<E>NMGF}, and
\begin{equation}\label{eq:sNMEstar}
\mathcal{E}^*_\chi=g_0^4\braket{\del_\mu\mathcal{F}_{\nu,2}^{*a}\left(\del_\mu B_{\nu,0}^a-\del_\nu B_{\mu,0}^a\right)}
\end{equation}
which is diagrammatically represented in figure \ref{fig:sNME}.
Employing the expressions presented in the previous section, we evaluate this term as
\begin{equation}\label{eq:fermioncontrDslashintegrals}
\begin{split}
\mathcal{E}^*_\chi&=-8(d-1)g_0^4N_f\,\text{tr}(T^aT^a)\int_0^tds\int_0^sdu\int_{p,q}e^{-2sp^2-2tq^2-2(s-u)p\cdot q}\frac{p\cdot q}{p^2}\\
&\equiv -4(d-1)g_0^4N_f\,\text{tr}(T^aT^a)\{I_{10}-I_{11}\}
\end{split}
\end{equation}
The integrals $I_{10},I_{11}$ and their $\eps$-expansions are listed in appendix \ref{app:integralsFermContr}. The new contribution is then given by
\begin{equation}\label{eq:sNMcontr}
\mathcal{E}^*_\chi=-\half \frac{(N^2-1)}{(8\pi t)^{d/2}}(d-1)N_f\frac{g_0^4}{(4\pi)^2}\left(4\log 2-2\log 3+\order{\eps}\right)
\end{equation}
and, similarly to $\mathcal{E}_\chi$ from \eqref{eq:totalFermContr}, this is manifestly finite. Again this is to be expected, anticipating the arguments provided in section \ref{sec:renormsNMGF}.

We obtain for the total fermionic contribution:
\begin{equation}\label{eq:fermioncontrDslashepsexp}
\mathcal{E}_\chi+\mathcal{E}_{\chi}^*=-\half \frac{(N^2-1)}{(8\pi t)^{d/2}}(d-1)N_f\frac{g_0^4}{(4\pi)^2}\left(2+\frac{16}{3}\log 2-3\log 3+\order{\eps}\right)
\end{equation}
We therefore now have for the full observable from \eqref{eq:EevolvedDslash}:
\begin{equation}
\braket{E}=\half g_0^2\frac{(N^2-1)}{(8\pi t)^{d/2}}(d-1)\left\{1+\tilde{c}_1g_0^2+\order{g_0^4}\right\}
\end{equation}
\begin{equation}
\tilde{c}_1=\frac{1}{(4\pi)^2}(4\pi)^{\eps}(8t)^{\eps}\left\{N\left(\frac{11}{3\eps}+\frac{52}{9}-3\log 3\right)-N_f\left(\frac{2}{3\eps}+\frac{22}{9}+4\log 2-3\log 3\right)+\order{\eps}\right\}
\end{equation}
Repeating the procedure from sections \ref{sec:perturbCalcEnonevolved} and \ref{sec:pertCalc<E>NMGF}, i.e.~renormalizing the coupling as in \eqref{eq:couplingRenormalizationMSbar}, expressing the renormalized coupling $\alpha(\mu)$ in terms of the RG invariant coupling $\alpha(q)$ by inverting \eqref{eq:alpha(q)intermsofalpha(mu)} and setting $q=(8t)^{-1/2}$ and $N=3$, we now obtain:
\begin{equation}\label{eq:<E>finalDslash}
\braket{E}=\frac{3}{4\pi t^2}\alpha(q)\left\{1+\tilde{k}_1\alpha(q)+\order{\alpha^2}\right\},\hspace{1cm}\tilde{k}_1=1.0978-0.1835\times N_f
\end{equation}
Again, for the renormalized result for general $N$ we refer the reader to appendix \ref{app:generalNresults}. We will compare the sNMGF result \eqref{eq:<E>finalDslash} with the NMGF and YMGF results \eqref{eq:<E>finalNMGF} and \eqref{eq:EN=3Luscher}, respectively, in section \ref{sec:Nfdependence}.

\subsection{Renormalization of evolved fields}\label{sec:renormsNMGF}

\begin{figure}[t]
	\centering
	\begin{minipage}{.33\textwidth}
		\centering
		\includegraphics[width=.8\linewidth]{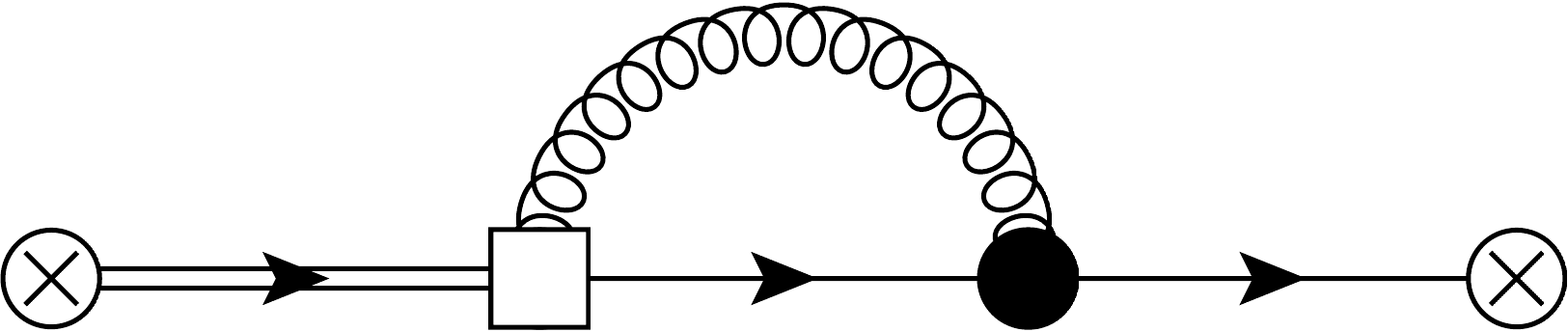}
		\caption*{sNM1}
	\end{minipage}%
	\begin{minipage}{.33\textwidth}
		\centering
		\includegraphics[width=.8\linewidth]{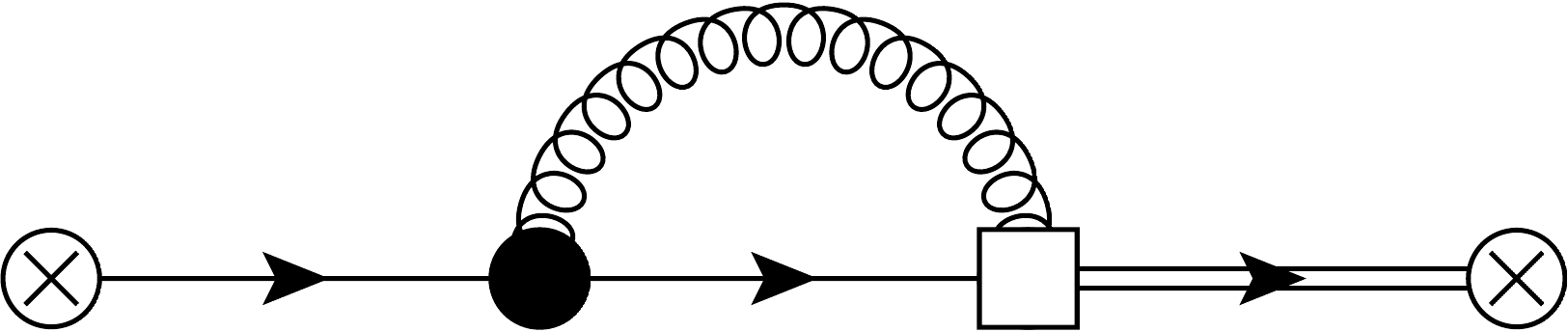}
		\caption*{sNM2}
	\end{minipage}%
	\begin{minipage}{.33\textwidth}
		\centering
		\includegraphics[width=.8\linewidth]{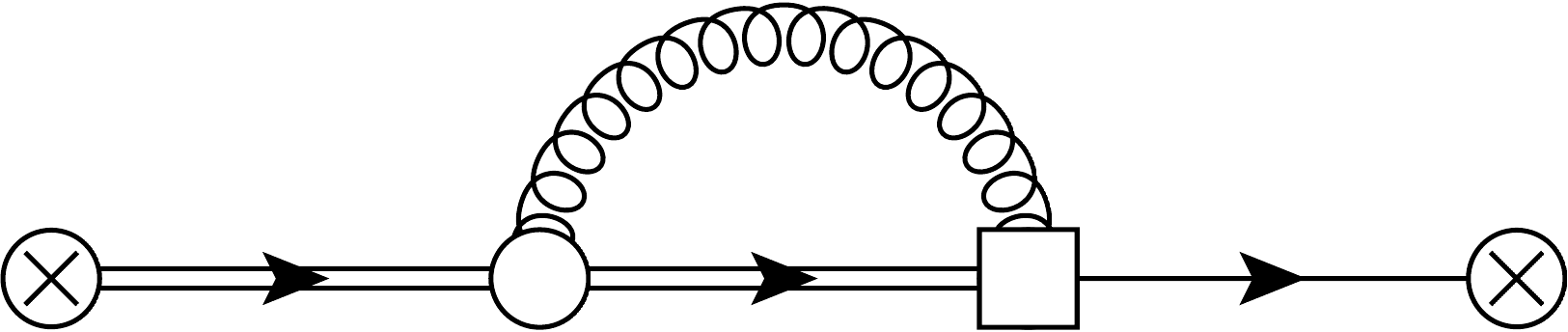}
		\caption*{sNM3}
	\end{minipage}
	\begin{minipage}{.33\textwidth}
		\centering
		\includegraphics[width=.8\linewidth]{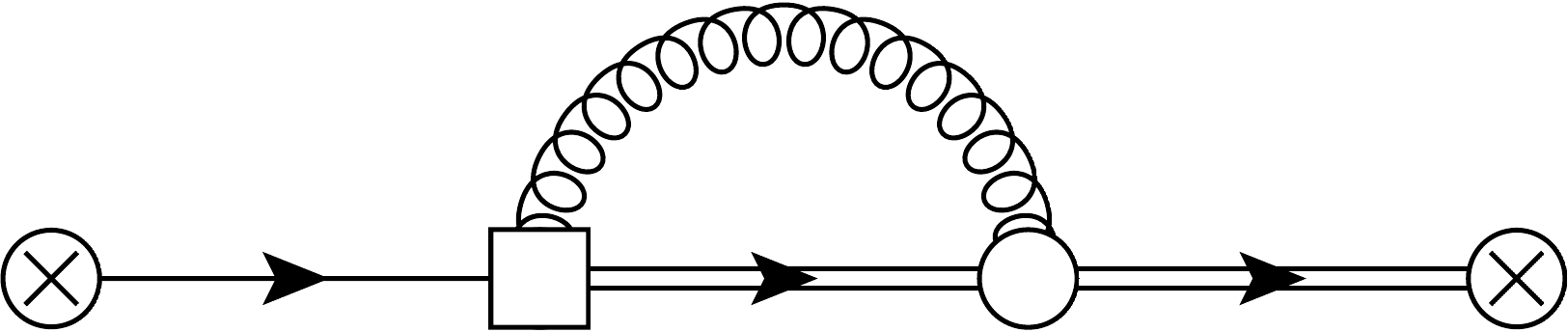}
		\caption*{sNM4}
	\end{minipage}%
	\begin{minipage}{.33\textwidth}
		\centering
		\includegraphics[width=.8\linewidth]{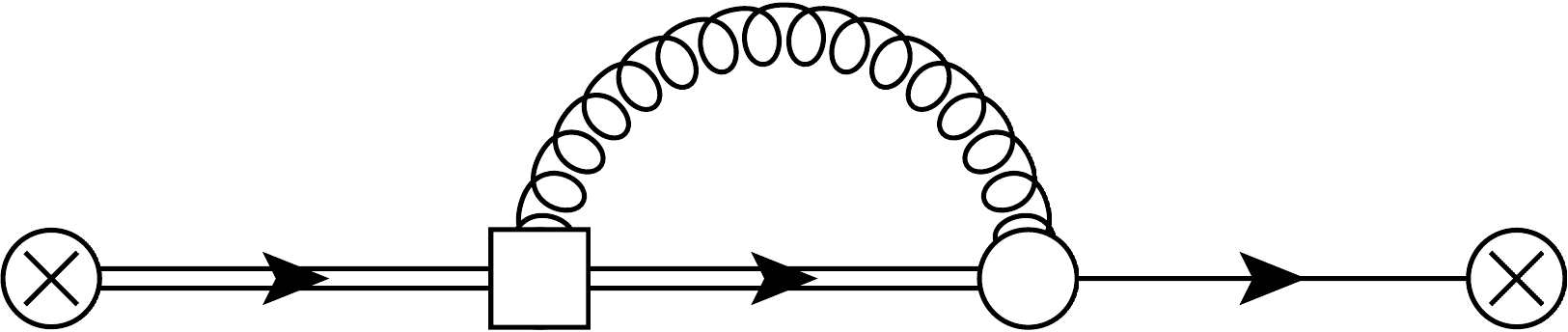}
		\caption*{sNM5}
	\end{minipage}%
	\begin{minipage}{.33\textwidth}
		\centering
		\includegraphics[width=.8\linewidth]{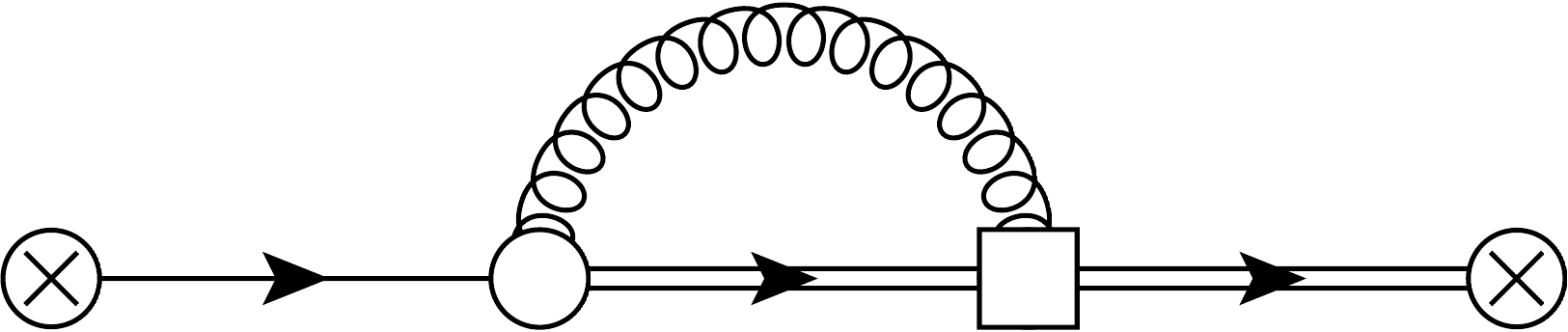}
		\caption*{sNM6}
	\end{minipage}
	\begin{minipage}{.33\textwidth}
		\centering
		\includegraphics[width=.8\linewidth]{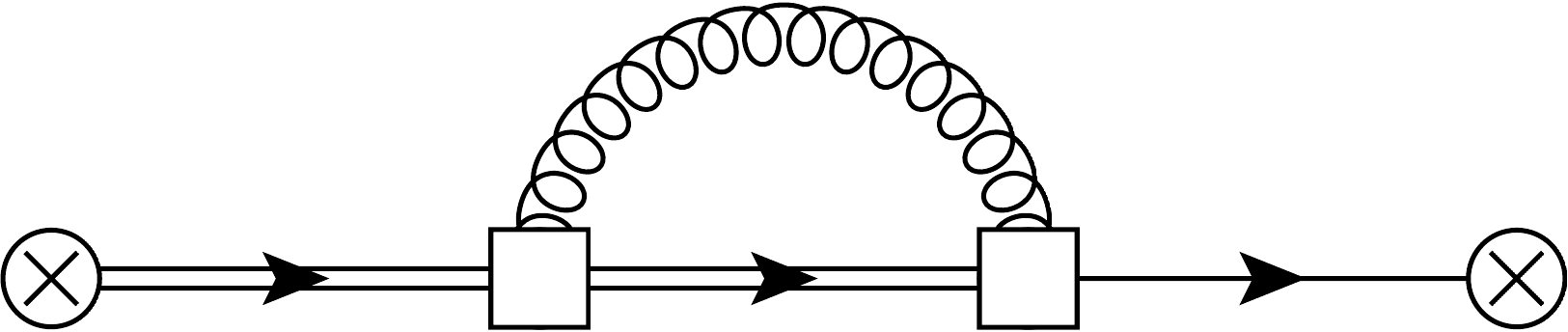}
		\caption*{sNM7}
	\end{minipage}%
	\begin{minipage}{.33\textwidth}
		\centering
		\includegraphics[width=.8\linewidth]{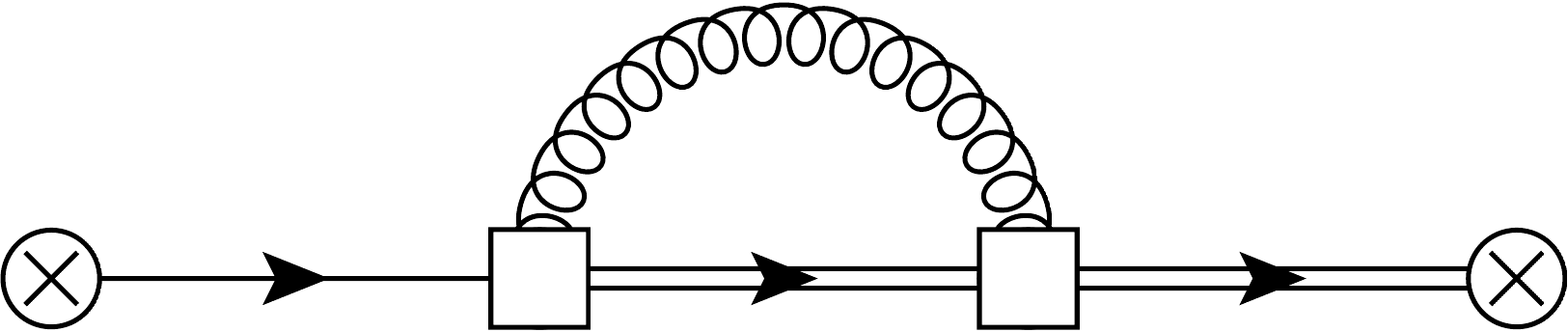}
		\caption*{sNM8}
	\end{minipage}%
	\begin{minipage}{.33\textwidth}
		\centering
		\includegraphics[width=.8\linewidth]{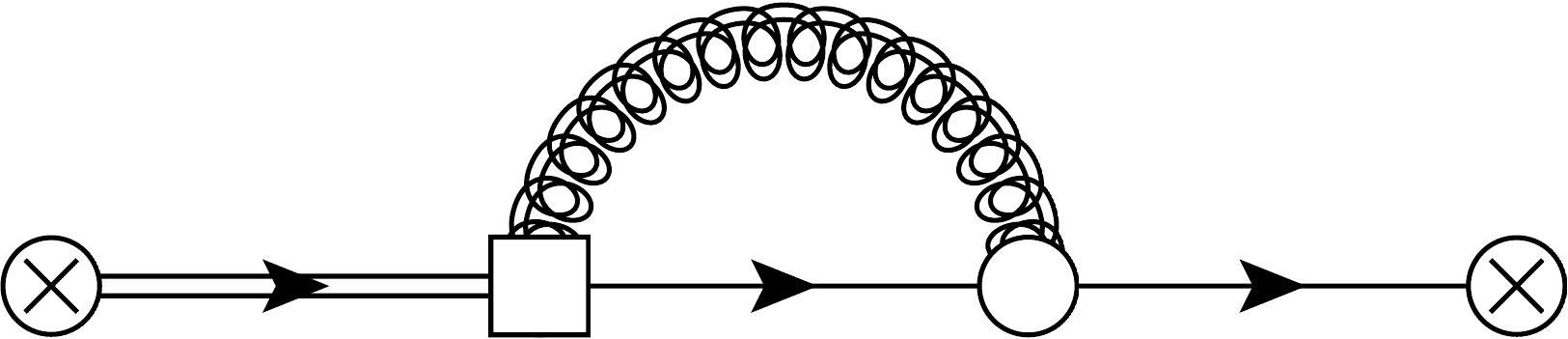}
		\caption*{sNM9}
	\end{minipage}%
	
	\begin{minipage}{.33\textwidth}
		\centering
		\includegraphics[width=.8\linewidth]{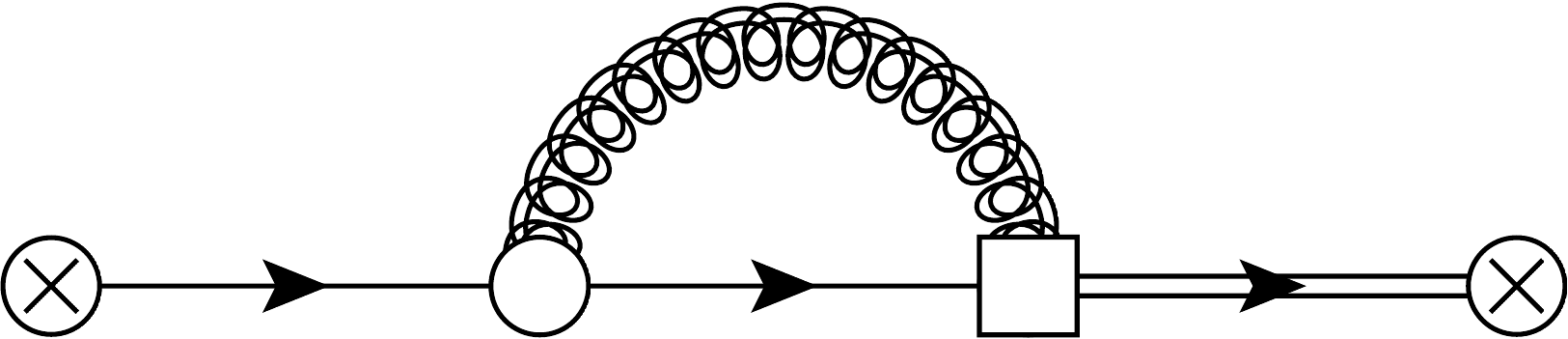}
		\caption*{sNM10}
	\end{minipage}%
	\caption{New next-to-leading order diagrams for the sNMGF evolved fermion 2-point function. Combined with the diagrams from figure \ref{fig:selfenergyNM} and figure \ref{fig:selfenergyYM} they represent the full sNMGF evolved fermion 2-point function, which is rendered finite by multiplying with the renormalization factor $Z_\chi$ from \eqref{eq:Z_chi sNMGF}.}
	\label{fig:selfenergy sNM}
\end{figure}

The arguments presented in section \ref{sec:renormNMGF} are fully applicable to the case presented here. The added gamma matrix structure in \eqref{eq:DslashGFeqnfermions} does not change any of the BRST transformation properties presented in appendix \ref{app:BRST}, and we still have $\delta S_{bulk}=0$. The appearance of the new terms in the fermionic bulk action \eqref{eq:fermBulkAction} --- now with $\slashed{\Delta}$ instead of $\Delta$ --- proportional to $\lambdabar\sigma_\mn G_\mn\chi$ and $\chibar\sigma_\mn G_\mn\lambda$ also does not change any of the reasoning presented in sections \ref{sec:renormEvolvedFieldsYM} and \ref{sec:renormNMGF}. Thus, all conclusions of those sections carry over.
\\ \\
With regard to the renormalization factor $Z_\chi$, there is no reason for it not to change, and indeed it does. The new $\order{g_0}$ sNMGF vertex represented by the white block in figure \ref{fig:sNMexpansions}.a and figure \ref{fig:sNMexpansions}.b, combined with the QCD vertex and the $\order{g_0}$ vertices from figure \ref{fig:NMexpansionfirstorders}, yields 10 new contributions to the evolved fermion 2-point function. These are represented in figure \ref{fig:selfenergy sNM}, and their values are given in table \ref{tab:ZchisNMGF}.
\begin{table}[H]
	\caption{UV divergences of the diagrams from figure \ref{fig:selfenergy sNM} in units of $C_2(R)\frac{g_0^2}{(4\pi)^2}S(\xbar_t-\ybar_t)$.}
	\label{tab:ZchisNMGF}
	\begin{center}
		\begin{tabular}{ ll }
			\hline
			diagram & value \\
			\hline
			{} & {} \\
			sNM1 + sNM2\hspace{1cm} & $\frac{3}{\eps}+\order{\eps^0}$ \\
			{} & {} \\
			sNM3 + sNM4 & $\order{\eps^0}$ \\
			{} & {} \\
			sNM5 + sNM6 & $\order{\eps^0}$ \\
			{} & {} \\
			sNM7 + sNM8 & $\frac{3}{2\eps}+\order{\eps^0}$ \\
			{} & {} \\
			sNM9 + sNM10 & $\frac{3}{2\eps}+\order{\eps^0}$ \\
			{} & {} \\
			\hline
		\end{tabular}
	\end{center}
\end{table}
\noindent
The full sNMGF evolved fermion 2-point function is given by the combined contributions from figures \ref{fig:selfenergyYM}, \ref{fig:selfenergyNM} and \ref{fig:selfenergy sNM}, and it is rendered finite by multiplying with the sNMGF evolved fermion renormalization factor
\begin{equation}\label{eq:Z_chi sNMGF}
\text{sNMGF}:\hspace{1cm} Z_\chi(g(\mu),\eps)=1-C_2(R)\frac{4}{\eps}\frac{g^2(\mu)}{(4\pi)^2}+\order{g^4}
\end{equation}

\section{$N_f$ dependence of $\braket{E}$ for the different flow equations}
\label{sec:Nfdependence}

\begin{figure}[t!]
	\centering
	\begin{minipage}{.42\linewidth}
		\begin{figure}[H]
			\includegraphics[width=\linewidth]{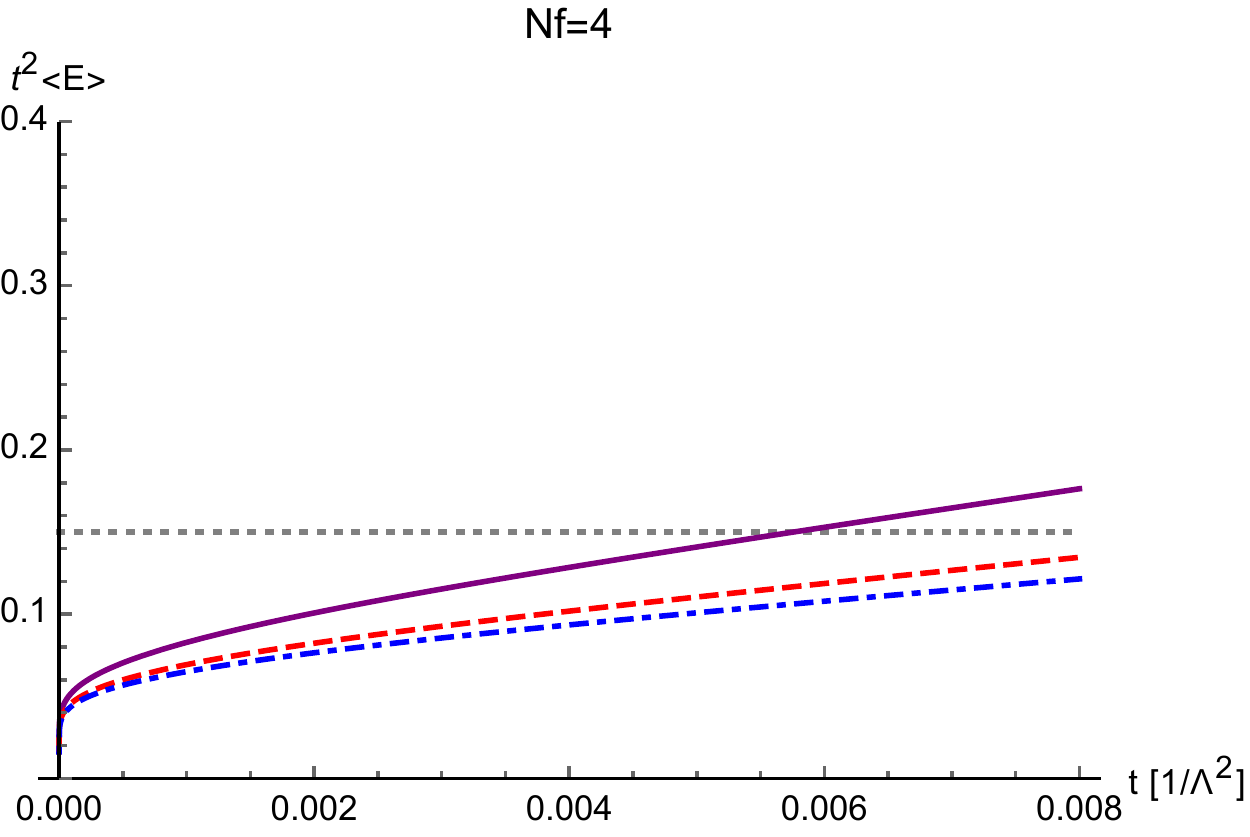}
		\end{figure}
	\end{minipage}\hspace{.6cm}
	\begin{minipage}{.52\linewidth}
		\begin{figure}[H]
			\includegraphics[width=\linewidth]{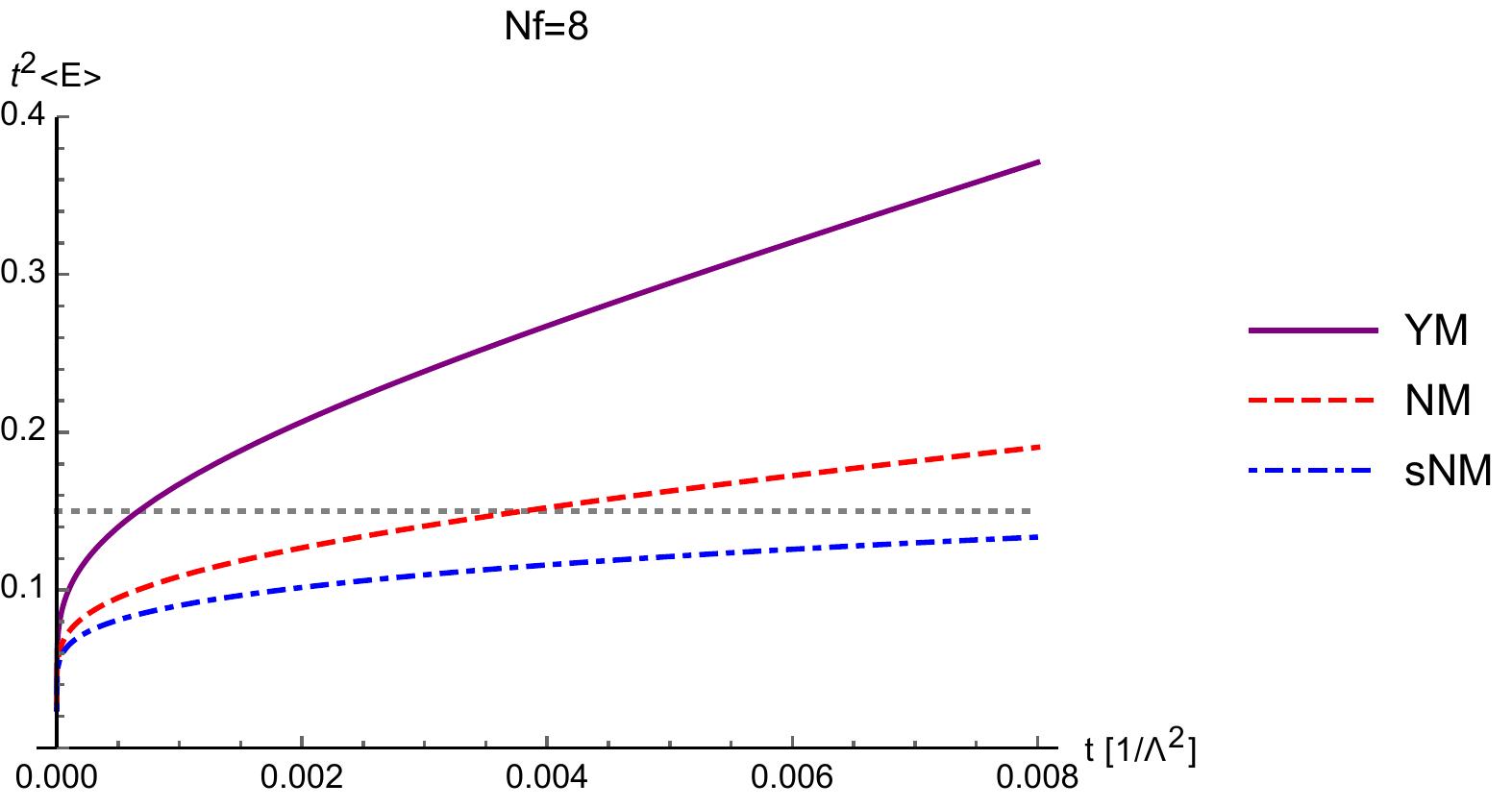}
		\end{figure}
	\end{minipage}
	\caption{Behaviour of the different $t^2\braket{E}_i$'s for $N_f=4,8$, as a function of $t$ in units of $\LQCD^{-2}$, see \eqref{eq:<E>_iintermsofLambbdaN=3}. From top to bottom: the YM (solid purple), the NM (dashed red), and the sNM (blue dot-dashed) gradient flow cases. The horizontal gray dotted line is there to better visualize the relative changes.}
	\label{fig:plots}
\end{figure}

Next we compare the $N_f$ dependence of the different expressions for $\braket{E}$ with $N=3$, namely \eqref{eq:EN=3Luscher}, \eqref{eq:<E>finalNMGF} and \eqref{eq:<E>finalDslash}, which are respectively the Yang-Mills (YM), the nonminimal (NM), and the `slashed' nonminimal (sNM) gradient flow cases. For general $N$ see appendix \ref{app:generalNresults}. Collecting the results, again setting $q=(8t)^{-1/2}$, we have
\begin{equation}\label{eq:<E>_i}
\braket{E}_i=\frac{3}{4\pi t^2}\alpha(q)\left\{1+k_{i}\alpha(q)+\order{\alpha^2}\right\}
\end{equation}
with $\alpha(q)$ the renormalization group invariant $\MSbar$ coupling, $i=\{YM,NM,sNM\}$, and
\begin{subequations}\label{eq:k_is}
\begin{align}
k_{YM}&=1.0978+0.0075\times N_f\\
k_{NM}&=1.0978-0.1377\times N_f\\
k_{sNM}&=1.0978-0.1835\times N_f
\end{align}
\end{subequations}
It is convenient to express $\alpha(q)$ in terms of the fundamental scale $\LQCD$\footnote{In general the renormalization group invariant scale $\LQCD$ depends on $N$, $N_f$, the value of the coupling at some given scale, and the renormalization scheme used. We will solely use it as a measure for the flow time $t$ and will not worry about its value or how it changes with $N_f$; when comparing theories with different $N_f$ we will keep the dimensionless product $t\LQCD^2$ fixed.} using the universal 2-loop renormalization group improved UV asymptotic expression \cite{Caswell:1974gg}, see \eqref{eq:couplingInTermsOfLambda}. We expand the coupling in powers of $1/l$, with $l=-\log\left(8t\LQCD^2\right)$, and we obtain for \eqref{eq:<E>_i}:
\begin{equation}\label{eq:<E>_iintermsofLambbdaN=3}
\braket{E}_i=\frac{3}{t^2\beta_0 l}\left\{1+\frac{1}{\beta_0 l}\left(4\pi k_i-\frac{\beta_1}{\beta_0}\log(l)\right)\right\}+\mathcal{O}\left(l^{-3}\right)
\end{equation}
The values for $t^2\braket{E(t)}_i$ are plotted for $N_f=4, 8$ in figure \ref{fig:plots}.

In \cite{Luscher:2010iy} it was found, using lattice simulations, that for $N=3$, $N_f=0$, beyond the perturbative regime, $t^2\braket{E(t)}$ grows roughly linearly with $t$. There, as a possible explanation for this behaviour the author alluded to the fact that the gradient flow drives the gauge fields towards stationary points of the YM action. For these configurations the right hand side of the flow equation \eqref{eq:YMGF} is small, implying that $E$ will change relatively little with $t$.

When considering cases with $N_f\neq 0$, this same reasoning should result in a more pronounced slowing down of $t^2\braket{E(t)}$ in the (s)NM cases compared to the YMGF, since the latter now no longer drives the gauge field towards the stationary points of the full action of the theory.

The graphs in figure \ref{fig:plots} seem to confirm this reasoning. Moreover, as $N_f$ is increased, the difference between the use of the different flow equations becomes more clear. It would be interesting to see this behaviour being reproduced in lattice simulations when the different flows are implemented.

\subsection{Sensitivity to change in $N_f$}

\begin{figure}[t!]
	\centering
	\includegraphics[width=.52\linewidth]{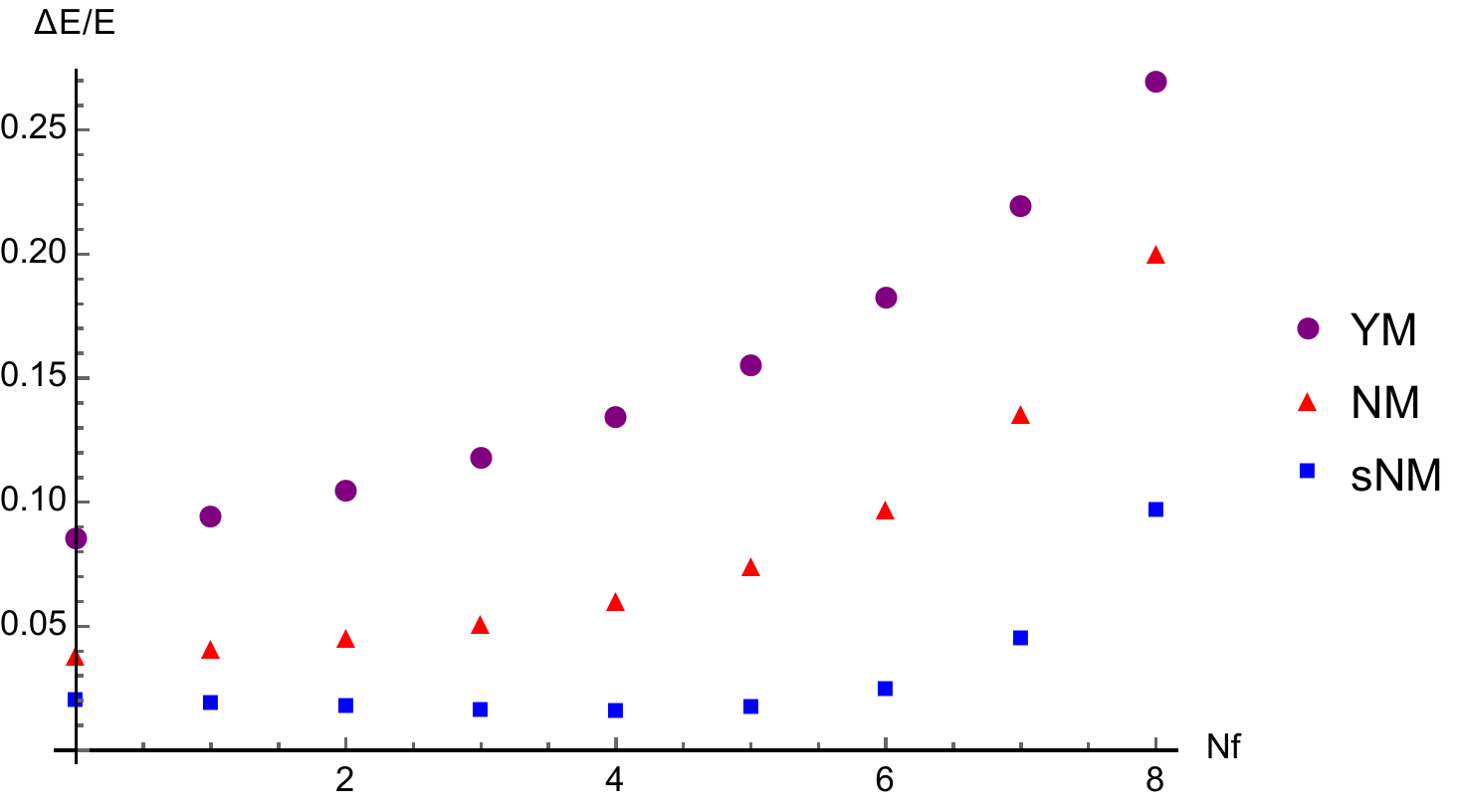}
	\caption{Relative change for the different $\braket{E}_i$'s from \eqref{eq:<E>_i}, \eqref{eq:k_is} at the scale $t=0.008/\LQCD^2$ when $N_f$ is increased by 1, see \eqref{eq:relative change}. Differences decrease for smaller values of $t\LQCD^2$, but the qualitative structure remains the same.}
	\label{fig:relchange}
\end{figure}

We can study the sensitivity to changes in $N_f$ of the different $\braket{E}_i$'s from \eqref{eq:<E>_i}, \eqref{eq:k_is} by considering the relative change:
\begin{equation}\label{eq:relative change}
\frac{\Delta \braket{E}_{N_f}}{\braket{E}_{N_f}}=\frac{\braket{E}_{N_f+1}-\braket{E}_{N_f}}{\braket{E}_{N_f}}
\end{equation}
The results are plotted in figure \ref{fig:relchange}. It shows that the sNMGF provides a much more stable value for $\braket{E}$ under changes of $N_f$ than the YMGF. For example, for $N_f=4 \rightarrow5$, at $t=0.008/\LQCD^2$, the sNMGF result changes by $2\%$, while for the YMGF we have a change of around $13\%$.

We would again like to stress that although $\LQCD$ will also change with $N_f$, the dimensionless product $t\LQCD^2$ of the flow time, or `smearing radius' $\sqrt{t}$, and the fundamental physical scale $\LQCD$ is kept fixed.

\section{Generalized flows}\label{sec:generalized flows}

The extensions of the Yang-Mills gradient flow presented in sections \ref{sec:NMGF} and \ref{sec:Dslash^2} are motivated by the similarities between the Langevin equations used in stochastic quantization and the gradient flow, and by the observation that the YMGF does not drive the gauge field towards the stationary points of the full action of the QCD-like theory when fermionic matter is present.

However, we also noted that in principle any gradient flow equation respecting all the symmetries of the nonevolved theory is equally valid from a purely mathematical point of view. Therefore we also present the most general flow equations:
\begin{subequations}\label{eq:generalized flow equations}
\begin{align}
\del_t B_\mu(t,x) &= D_\nu G_{\nu\mu}(t,x) - a_1g_0 j_\mu^a(t,x) T^a + \alpha_0 D_\mu\del_\nu B_\nu(t,x),\hspace{.6cm}B_\mu(0,x)=A_\mu(x)\\
\del_t\chi(t,x)&=\left\{\Delta+a_2\frac{g_0}{2}\sigma_\mn G_\mn(t,x)-\alpha_0 g_0\del_\mu B_\mu(t,x)\right\}\chi(t,x),\hspace{.5cm}\chi(0,x)=\psi(x)\\
\del_t\chibar(t,x)&=\chibar(t,x)\left\{\overleftarrow{\Delta}+a_2\frac{g_0}{2}\sigma_\mn G_\mn(t,x)+\alpha_0g_0\del_\mu B_\mu(t,x)\right\},\hspace{.4cm}\chibar(0,x)=\psibar(x)
\end{align}
\end{subequations}
where we include the modification terms proportional to $\alpha_0$, and $a_1,a_2$ are real constants parametrizing the amount of `nonminimality'. The YMGF is the minimal case: $(a_1,a_2)=(0,0)$, the NMGF presented in section \ref{sec:NMGF} is obtained setting $(a_1,a_2)=(1,0)$, and the sNMGF from section \ref{sec:Dslash^2} corresponds to $(a_1,a_2)=(1,1)$. It should be clear from the discussions in sections \ref{sec:renormEvolvedFieldsYM}, \ref{sec:renormNMGF} and \ref{sec:renormsNMGF} that the qualitative renormalization properties remain the same in this generalized case.

The solutions to the generalized flow equations from \eqref{eq:generalized flow equations} are straightforward generalizations of those given in sections \ref{sec:NMGF} and \ref{sec:Dslash^2}, and read (setting $\alpha_0=1$)
\begin{subequations}\label{eq:GGF solutions}
\begin{align}
B_\mu(t,x)&=e^{t\lapx} A_\mu(x)+\int_0^tds\,e^{(t-s)\lapx} \{ R_\mu(s,x) -a_1g_0 j_\mu^a(s,x)T^a\}\\
\chi(t,x)&=e^{t\lapx}\psi(x)+\int_0^tds\,e^{(t-s)\lapx}\left\{\left(\Delta'+a_2\frac{g_0}{2}\sigma_\mn G_\mn\right)\chi(s,x)\right\}\\
\chibar(t,x)&=e^{t\lapx}\psibar(x)+\int_0^tds\,e^{(t-s)\lapx}\left\{\chibar(s,x)\left(\overleftarrow{\Delta}'+a_2\frac{g_0}{2}\sigma_\mn G_\mn\right)\right\}
\end{align}
\end{subequations}
Using these solutions we calculate the expectation value of the observable $E(t,x)$ from \eqref{eq:E}. For general $N$ we obtain
\begin{equation}\label{eq:GGF E}
\braket{E(a_1,a_2;t)}=\frac{3(N^2-1)}{32\pi t^2}\alpha(\mu)\left\{1+k(a_1,a_2;\mu^2t)\alpha(\mu)+\order{\alpha^2}\right\}
\end{equation}
where
\begin{equation}\label{eq:GGF k}
\begin{split}
k(a_1,a_2;\mu^2t)&=k_{YM}(\mu^2t)-\frac{N_f\,a_1}{12\pi}\bigg\{4+40\log 2-24\log 3\\
&\hspace{1cm}+a_1\left(2-36\log 2+21\log 3\right)+6a_2\left(2\log 2-\log 3\right)\bigg\}
\end{split}
\end{equation}
and $k_{YM}(\mu^2t)$ is given in \eqref{eq:k_isgeneralN}. The $k_i$'s from \eqref{eq:k_is} are special cases of $k(a_1,a_2)$: $k_{YM}=k(0,0)=k(0,1)$, $k_{NM}=k(1,0)$ and $k_{sNM}=k(1,1)$.

The evolved fermion renormalization factor is also calculated using the generalized flow equations, and reads
\begin{equation}\label{eq:GGF Z_chi}
Z_\chi(a_1,a_2;g(\mu),\eps)=1+\frac{3-a_1(1+6a_2)}{\eps}C_2(R)\frac{g^2(\mu)}{(4\pi)^2}+\order{g^4}
\end{equation}
Thus the anomalous dimension associated with the evolved fermion fields is given by
\begin{equation}
\gamma_\chi(a_1,a_2;g)\equiv-\frac{1}{2}\frac{d\log Z_\chi}{d\log\mu}=\left(3-a_1\left(1+6a_2\right)\right)C_2(R)\frac{g^2(\mu)}{(4\pi)^2}+\order{g^4}
\end{equation}
Note that when $a_1(1+6a_2)=3$ the first coefficient of the anomalous dimension vanishes. This implies that there exists a renormalization scheme such that $\gamma_\chi$ vanishes to all orders in the gauge coupling, see e.g.~\cite{Collins:1984xc}.

\section{Conclusion and outlook}\label{sec:conclusion}

In this paper we have investigated some of the properties and consequences of nonminimal gradient flows in non-Abelian gauge theories containing fermions, i.e.~flow equations which are defined using the full action, instead of only the Yang-Mills action. The most important new feature is that the flow equation for the gauge field $B_\mu$ acquires a purely fermionic term proportional to the gauge covariant vector current. In addition to the conventional fermion flow equations based on the operator $D^2$, we also studied those based on $\slashed{D}^2$.

We found that the same qualitative renormalization properties hold for the nonminimally evolved fields as in the Yang-Mills gradient flow case. That is, any correlator --- possibly containing evolved composite operators --- consisting of $n$ (s)NMGF evolved fermion fields $\chi$ and $n$ (s)NMGF evolved fields $\chibar$ and any number of (s)NMGF evolved gauge fields $B_\mu$ with Wilsonian normalization is rendered finite by multiplicative renormalization with $Z_\chi^{n}$, supplemented with the usual gauge coupling renormalization, where the different $Z_\chi$ factors for the different flows are given by
\begin{equation}
Z_\chi(g(\mu),\eps)=1+\left\{\begin{array}{lr}
\text{YM:} & +3\\
\text{NM:} & +2\\
\text{sNM:} & -4 
\end{array}
\right\}\times \frac{1}{\eps} C_2(R)\frac{g^2(\mu)}{(4\pi)^2}+\order{g^4(\mu)}
\end{equation}

Furthermore, we have extended the nonminimal flows to the most general form compliant with the symmetries of the nonevolved theory. We have shown that in general $Z_\chi$ depends on two constants $a_1,a_2$, and that there exists a particular combination of these constants for which the $\order{g^2}$ term vanishes. This implies that there exists a one-parameter family of flow systems for which there is a renormalization scheme in which the anomalous dimension of the evolved fermion field vanishes to all orders in the gauge coupling.

Additionally, we calculated $\braket{E(t)}$ up to next-to-leading order in the coupling using the generalized nonminimal gradient flow, of which the YMGF, NMGF and sNMGF are special cases. We found that out of these three different cases $\braket{E(t)}$ calculated with the sNMGF has the smallest dependence on $N_f$. Also, for fixed $N_f\neq 0$, $t^2\braket{E(t)}_{sNM}$ grows slower with $t$ than $t^2\braket{E(t)}_{NM}$, which in turn grows slower with $t$ than the conventionally employed $t^2\braket{E(t)}_{YM}$. This behaviour might be attributed to the fact that the gauge fields are now driven towards the stationary points of the full action, instead of to the stationary points of the Yang-Mills action only.

\subsection*{Possible lattice applications}
It may be interesting to see if the above mentioned properties of $\braket{E(t)}$ will persist beyond the perturbative regime. If so, the nonminimal gradient flows might have useful applications in lattice studies involving fermions.

First of all, the gradient flow derived scales $t_0$ and $w_0$ (see e.g.~\cite{Sommer:2014mea}) might have a smaller $N_f$ dependence when derived using the nonminimal flows. A much used definition for $t_0$ is
\begin{equation}\label{eq:GFscale}
t_0^2\braket{E(t_0)}=0.3
\end{equation}
and figure \ref{fig:plots} seems to imply that when using the YMGF at $N_f=8$ this definition leads to a relatively small value for the dimensionless quantity $t_0\LQCD^2$. The nonminimal flows will lead to a significantly larger value for $t_0\LQCD^2$ when using the same definition \eqref{eq:GFscale} --- more comparable to the pure Yang-Mills case \cite{Luscher:2010iy} --- possibly making them practically more useful in studies including fermions.

Secondly, the application of the generalized nonminimal gradient flow to the lattice will give the scales $t_0,w_0$ a continuous dependence on the parameters $a_1,a_2$ which are defining the flow. This might provide additional tests on lattice implementations of the gradient flow.

Finally, it may be interesting to investigate the effect of the nonminimal flows on the GF evolved topological charge\footnote{We thank the reviewer for pointing this out.}. In the chiral limit at finite volume these charges should be absent, yet a recent study \cite{Hasenfratz:2020vta} shows that on the lattice the gradient flow sometimes promotes gauge field vacuum fluctuations to instanton-like objects, resulting in a nonzero net topological charge. 
The nonminimal flows might be helpful in investigating this effect and eventually minimizing it.

\section*{Acknowledgments}
The author would like to thank Elisabetta Pallante for useful discussions and suggestions. This work has been supported by the foundation of Dutch research institutes (NWO-I).

\appendix

\section{Notation and conventions}\label{app:notationconventions}
We work with anti-Hermitian generators $T^a$, $a=1,...,N^2-1$,
\begin{equation}
[T^a,T^b]=f^{abc}T^c
\end{equation}
with real structure constants $f^{abc}$. For the normalization of the trace we use
\begin{equation}
\text{tr}_R(T^aT^b)=-T(R)\delta^{ab},\hspace{1cm}\text{tr}_R(1)=d(R)
\end{equation}
where $T(R)$ and $d(R)$ are the index resp. dimension of representation $R$. We usually omit the subscript $R$, since we have (unless stated otherwise):
\begin{equation}
\text{tr}(T^aT^b)=\text{tr}_F(T^aT^b)=-\half\delta^{ab},\hspace{1cm} \text{tr}(1)=N
\end{equation}
with $F$ denoting the (anti)fundamental representation. The quadratic Casimir of representation $R$ is related to the generators via
\begin{equation}
T^aT^a=-C_2(R)1
\end{equation}
and thus
\begin{equation}
C_2(R)=-\frac{\text{tr}_R(T^aT^a)}{d(R)}=\frac{T(R)d(G)}{d(R)}
\end{equation}
where $d(G)=d(adj)=N^2-1$, $d(F)=N$, $T(F)=\half$.
\\ \\
We work with Euclidean metric $\delta_\mn$, $\delta_{\mu\mu}=d$. The gamma matrices satisfy
\begin{equation}
\{\gmu,\gnu\}=2\delta_\mn,\hspace{1cm}\gmu^\dagger=\gmu
\end{equation}
\\
For the Euclidean space- and momentum integrals we use the abbreviations:
\begin{equation}
\int_x\equiv\int d^dx,\hspace{1cm}\int_p\equiv\int \frac{d^dp}{(2\pi)^d}
\end{equation}

\section{Euclidean QCD-like action}\label{app:euclideanQCDaction}

The Euclidean gauge-fixed $d$-dimensional action of QCD-like theories with massless fermions and canonically normalized gauge fields is given by
\begin{equation}
S_{QCD}=S_{YM}+S_{F}+S_{gf}+S_{c\cbar}
\end{equation}
where
\begin{align}
S_{YM}&=-\half \int_x\text{tr}\left(F_\mn F_\mn\right)\\
S_{F}&=\half\int_x \left( \psibar \slashed{D} \psi - \psibar \overleftarrow{\slashed{D}}\psi\right)\\
S_{gf}&=-\lambda_0\int_x \text{tr}\left\{(\del_\mu A_\mu)^2\right\}\\
S_{c\bar{c}}&=-2\int_x\text{tr}\left(\del_\mu\bar{c}D_\mu c\right)
\end{align}
with
\begin{equation}
F_\mn=\del_\mu A_\nu -\del_\nu A_\mu + g_0 [A_\mu,A_\nu]
\end{equation}
and
\begin{equation}
D_\mu\psi=(\del_\mu+g_0A_\mu)\psi,\hspace{1cm}\psibar \overleftarrow{D}_\mu=\del_\mu\psibar- g_0 \psibar A_\mu,\hspace{1cm} D_\mu c = \del_\mu c +g_0[A_\mu,c]
\end{equation}

\section{Gauge transformations}\label{app:gaugetrans}

We parametrize a gauge transformation by $\Lambda\in SU(N)$:
\begin{equation}
\Lambda(t,x)=e^{\omega^a(t,x)T^a}
\end{equation}
\subsection*{Nonevolved case: $t=0$}\label{app:gaugetransnonevolved}
This is the well-known case, repeated here for convenience with our conventions:
\begin{equation}
\begin{split}
\psi'&=\Lambda \psi\\
\psibar'&=\psibar\Lambda^{-1}\\
A_\mu'&=\Lambda A_\mu \Lambda^{-1}-\frac{1}{g_0}(\del_\mu\Lambda)\Lambda^{-1}=\Lambda A_\mu \Lambda^{-1}+\frac{1}{g_0}\Lambda\del_\mu\Lambda^{-1}
\end{split}
\end{equation}
and the matter current $j_\mu^aT^a=\psibar\gmu T^a\psi T^a$ transforms as:
\begin{equation}
{j_\mu^a}'T^a=\Lambda j_\mu^a T^a \Lambda^{-1}
\end{equation}

\subsection*{Evolved case: $t>0$}\label{app:gaugetransevolved}

The modified flow equations are given in \eqref{eq:YMGFmodified} and \eqref{eq:NMGFModified}. For the fermions the two different modified flow equations are given in \eqref{eq:gfeqnfermionsmodified} and \eqref{eq:DslashGFeqnfermions}.
These flow equations are invariant under the $t$-dependent gauge transformations $\Lambda(t,x)$, under which the solutions of these flow equations transform as:
\begin{equation}\label{eq:gaugeTransNonzeroT}
\begin{split}
B_\mu'&=\Lambda B_\mu\Lambda^{-1} +\frac{1}{g_0} \Lambda\del_\mu\Lambda^{-1}\\
\chi'&=\Lambda\chi\\
\chibar'&=\chibar\Lambda^{-1}
\end{split}
\end{equation}
and $\Lambda(t,x)$ must satisfy
\begin{equation}\label{eq:gaugeTransNonzeroTFlow}
\del_t\Lambda=\alpha_0 D_\nu\del_\nu\Lambda, \hspace{1cm}\del_t\Lambda^{-1}=\alpha_0D_\nu\del_\nu\Lambda^{-1}
\end{equation}
Note that the boundary condition at $t=0$ is unconstrained, and thus these transformations generalize the gauge symmetry of the $SU(N)$ theory to all flow times.

\subsection*{Modified flow equations}\label{app:gaugetransmodifiedfeq}

The modified flow equations \eqref{eq:YMGFmodified}, \eqref{eq:NMGFModified}, \eqref{eq:gfeqnfermionsmodified} and \eqref{eq:DslashGFeqnfermions} are related to the flow equations without the modification term by a gauge transformation with a different `flow equation' than \eqref{eq:gaugeTransNonzeroTFlow}. Let the primed fields be the solutions to the modified flow equations, and the nonprimed fields the solutions to the nonmodified flow equations. The latter are obtained from the former by performing the transformations \eqref{eq:gaugeTransNonzeroT} with
\begin{equation}
\begin{split}
\del_t\Lambda&=-\alpha_0g_0(\del_\nu B_\nu') \Lambda\\
\del_t\Lambda^{-1}&=\alpha_0g_0 \Lambda^{-1} (\del_\nu B_\nu')
\end{split}
\end{equation}
and it is then straightforward to verify that 
\begin{equation}
\begin{split}
\del_t B_\mu'=D_\nu' G_{\nu\mu}'-g_0{j_\mu^a}' T^a+\alpha_0 D_\mu'\del_\nu B_\nu'\,\,\,\,\,\,&\Longleftrightarrow\,\,\,\,\,\, \del_t B_\mu=D_\nu G_{\nu\mu}-g_0j_\mu^aT^a\\
\del_t{\chi}'=\Delta'\chi'-\alpha_0g_0\del_\nu B_\nu'\chi'\,\,\,\,\,\,&\Longleftrightarrow\,\,\,\,\,\,\del_t{\chi}=\Delta\chi\\
\del_t{\bar{\chi}}'=\chibar'\overleftarrow{\Delta}'+\alpha_0g_0\chibar'\del_\nu B_\nu'\,\,\,\,\,\,&\Longleftrightarrow\,\,\,\,\,\,\del_t{\bar{\chi}}=\chibar\overleftarrow{\Delta}
\end{split}
\end{equation}
The same holds when $\Delta$ and $\overleftarrow{\Delta}$ are replaced by $\slashed{\Delta}$ and $\overleftarrow{\slashed{\Delta}}$, as in \eqref{eq:DslashGFeqnfermions}.

\section{BRST transformations}\label{app:BRST}
The BRST variations of the $d$-dimensional unrenormalized fields are given by
\begin{align}
\delta A_\mu &= D_\mu c\\
\delta c&=-g_0c^2\\
\delta \cbar &= \lambda_0 \del_\mu A_\mu\\
\delta \psi&=-g_0 c \psi\\
\delta \psibar&=-g_0 \psibar c
\end{align}
where $\delta$ can be viewed as a nilpotent operator, which anticommutes with grassmann variables, i.e. the (anti-)ghosts and (anti-)quarks.

The BRST variations of the bulk fields are given by
\begin{align}
\delta B_\mu &= D_\mu d\\
\delta \chi &= -g_0 d \chi\\
\delta \chibar &= -g_0 \chibar d\\
\delta L_\mu &= g_0 [L_\mu,d]\\
\delta \lambda &=-g_0 d \lambda\\
\delta \lambdabar &= -g_0 \lambdabar d\\
\delta d &= -g_0d^2\\
\delta \dbar &= D_\mu L_\mu - g_0\{d,\dbar\}+ g_0\lambdabar T^a \chi T^a-g_0\chibar T^a \lambda T^a 
\end{align}
The nontrivial variation of $\dbar$ is chosen such that $\delta S_{bulk}=0$, and $\delta^2\dbar=0$.

Useful identities in checking $\delta S_{bulk}=0$ are obtained by defining
\begin{align}
E_\mu&=\del_t B_\mu-D_\nu G_{\nu\mu} +g_0 j_\mu^a T^a -\alpha_0 D_\mu\del_\nu B_\nu \label{eq:Emudef}\\
e&=\del_t d -\alpha_0D_\mu\del_\mu d\\
f&=(\del_t-\Delta+\alpha_0 g_0\del_\nu B_\nu)\chi\\
\fbar&=\chibar(\overleftarrow{\del}_t-\overleftarrow{\Delta}-\alpha_0 g_0\del_\nu B_\nu)
\end{align}
and noting
\begin{align}
\delta E_\mu &= g_0[E_\mu,d]+D_\mu e \label{eq:Emuvar}\\
\delta e &= - g_0 \{e,d\}\\
\delta f &= -g_0df -g_0 e\chi\\
\delta \fbar &= -g_0 \fbar d-g_0\chibar e
\end{align}
The same identities hold when $\Delta$ and $\overleftarrow{\Delta}$ are replaced by $\slashed{\Delta}$ and $\overleftarrow{\slashed{\Delta}}$, as in section \ref{sec:Dslash^2}.

In the Yang-Mills gradient flow case the $g_0j_\mu^aT^a$ term is absent in \eqref{eq:Emudef}, but \eqref{eq:Emuvar} will still hold.

\section{$Z_\chi$ using the Yang-Mills gradient flow}\label{app:ZchiYMGF}

\begin{figure}[t]
	\centering
	\begin{minipage}{.33\textwidth}
		\centering
		\includegraphics[width=.8\linewidth]{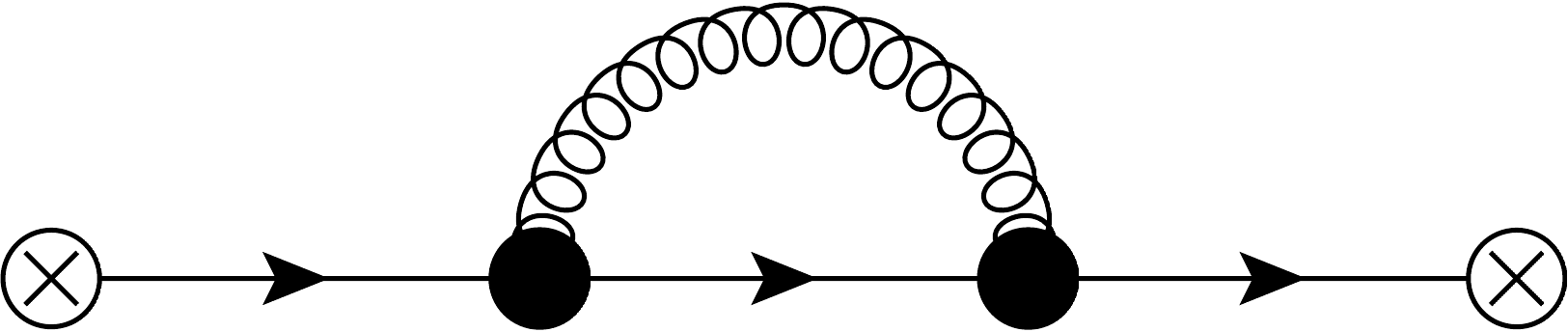}
		\caption*{YM1}
	\end{minipage}%
	\begin{minipage}{.33\textwidth}
		\centering
		\includegraphics[width=.8\linewidth]{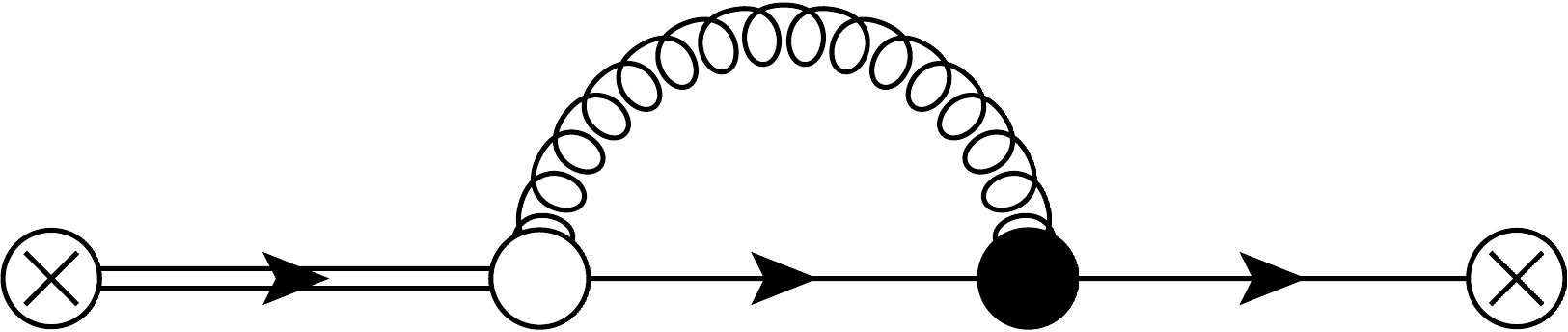}
		\caption*{YM2}
	\end{minipage}%
	\begin{minipage}{.33\textwidth}
		\centering
		\includegraphics[width=.8\linewidth]{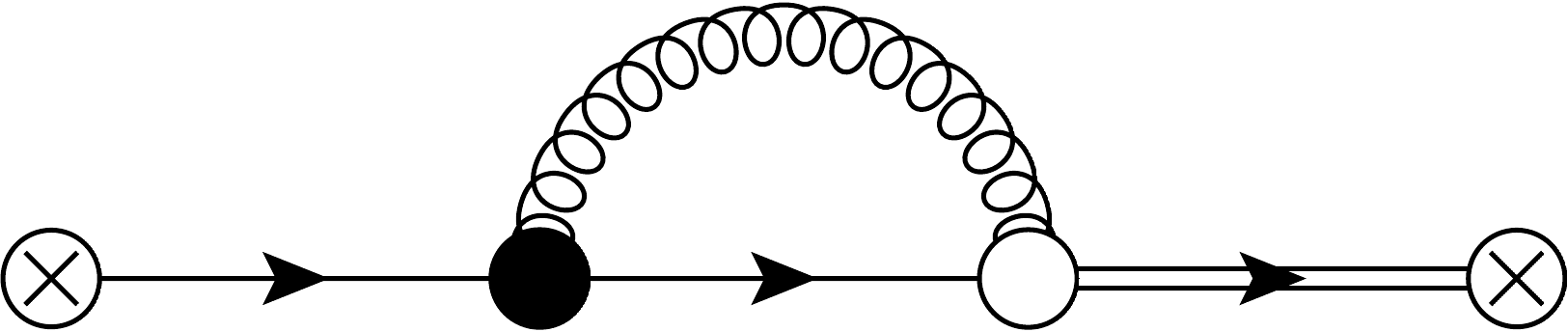}
		\caption*{YM3}
	\end{minipage}
	\begin{minipage}{.33\textwidth}
		\centering
		\includegraphics[width=.8\linewidth]{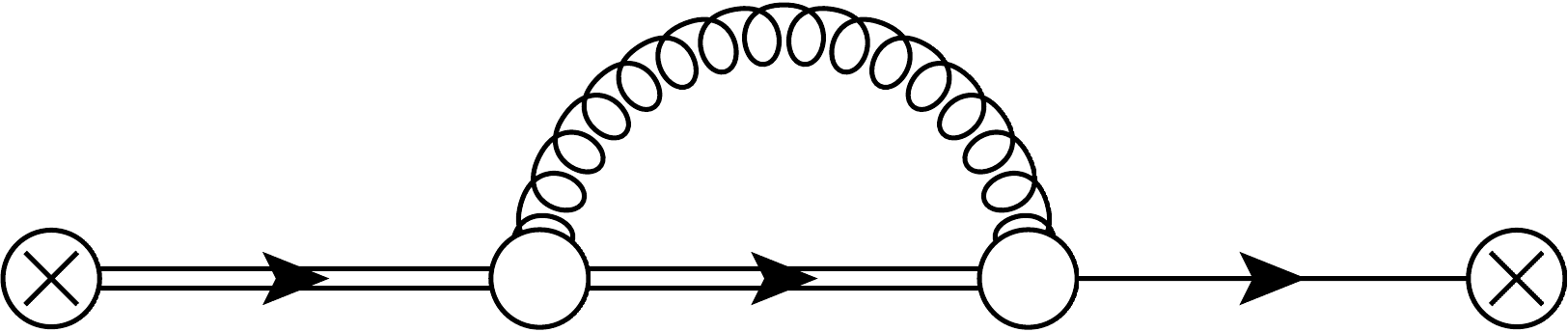}
		\caption*{YM4}
	\end{minipage}%
	\begin{minipage}{.33\textwidth}
		\centering
		\includegraphics[width=.8\linewidth]{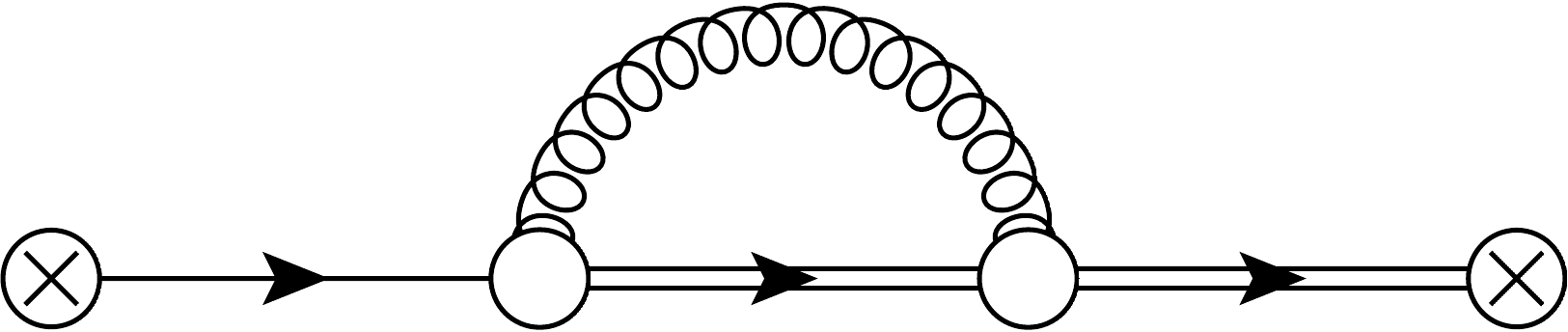}
		\caption*{YM5}
	\end{minipage}%
	\begin{minipage}{.33\textwidth}
		\centering
		\includegraphics[width=.8\linewidth]{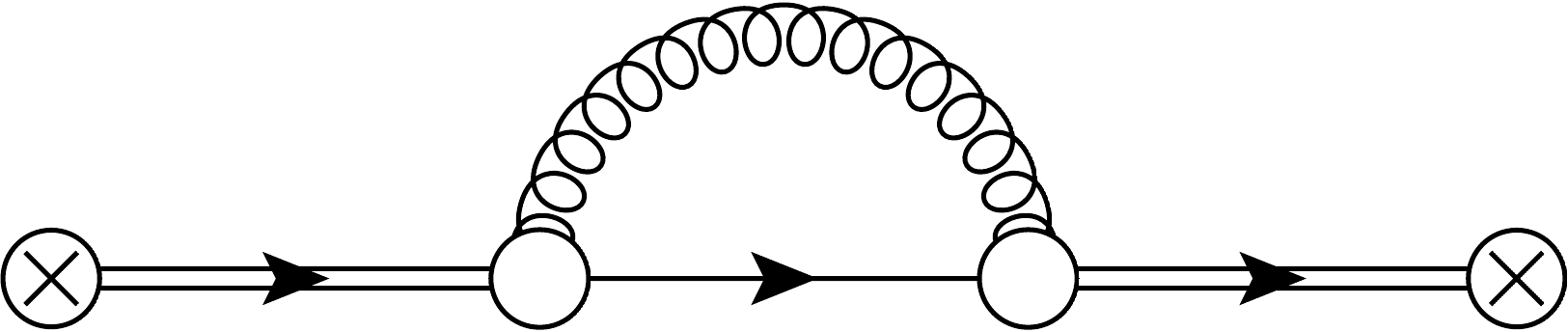}
		\caption*{YM8}
	\end{minipage}
	\begin{minipage}{.33\textwidth}
		\centering
		\includegraphics[width=.8\linewidth]{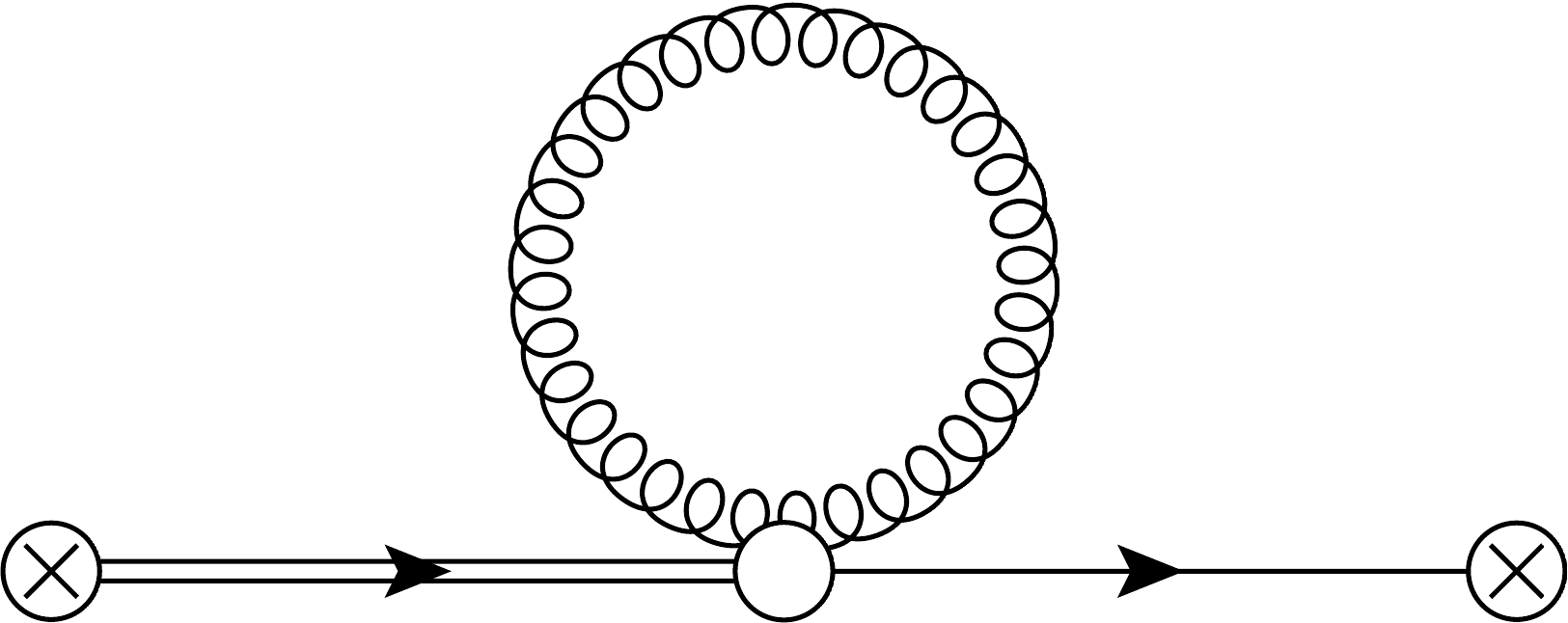}
		\caption*{YM6}
	\end{minipage}%
	\begin{minipage}{.33\textwidth}
		\centering
		\includegraphics[width=.8\linewidth]{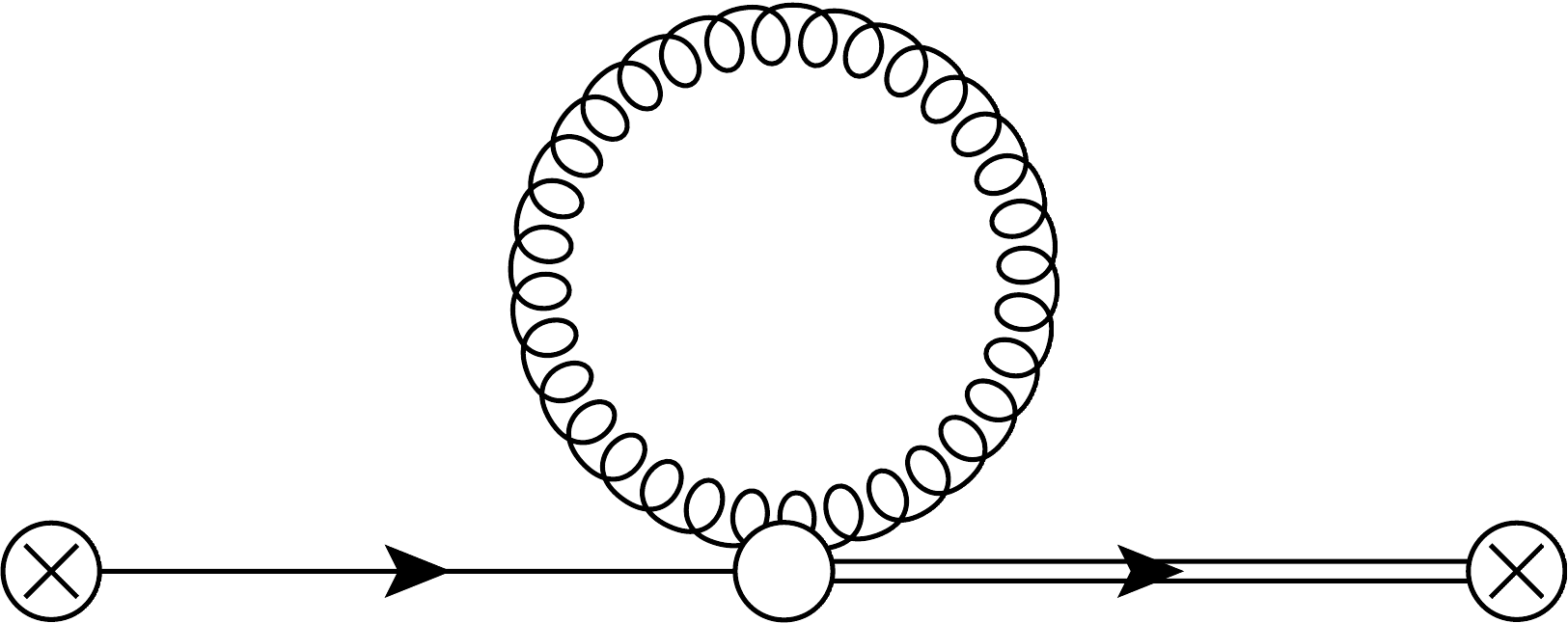}
		\caption*{YM7}
	\end{minipage}%
	\caption{Next-to-leading order diagrams for the YMGF evolved fermion 2-point function. White blobs represent GF vertices, black blobs the usual QCD vertices, and double lines indicate a kernel including integration over the associated flow time.}
	\label{fig:selfenergyYM}
\end{figure}

We present the results of the calculation of the evolved fermion renormalization factor $Z_\chi$ using the evolved fermion 2-point function $\braket{\chi(t,x)\chibar(t,y)}$. We find $Z_\chi$ by imposing:
\begin{equation}
\braket{\chi_R(t,x)\chibar_R(t,y)}=Z_\chi\braket{\chi(t,x)\chibar(t,y)}=\order{\eps^0}
\end{equation}
The diagrams contributing at $\order{g_0^2}$ are shown in figure \ref{fig:selfenergyYM}.
We collect the results in table~\ref{tab:ZchiYMGF}, where $S(\xbar_t-\bar{y}_t)$ is the evolved fermion propagator given in \eqref{eq:evolved fermion propagater}.
\begin{table}[t]
	\caption{UV divergences of the diagrams from figure \ref{fig:selfenergyYM} in units of $C_2(R)\frac{g_0^2}{(4\pi)^2}S(\xbar_t-\ybar_t)$.}
	\label{tab:ZchiYMGF}
	\begin{center}
		\begin{tabular}{ ll }
			\hline
			diagram & value \\
			\hline
			{} & {} \\
			YM1 & $-\frac{1}{\eps}+\order{\eps^0}$ \\
			{} & {} \\
			YM2 + YM3 & $-\frac{2}{\eps}+\order{\eps^0}$ \\
			{} & {} \\
			YM4 + YM5 & $\frac{1}{\eps}+\order{\eps^0}$ \\
			{} & {} \\
			YM6 + YM7\hspace{1cm} & $-\frac{1}{\eps}+\order{\eps^0}$ \\
			{} & {} \\
			YM8 & $\order{\eps^0}$ \\
			{} & {} \\
			\hline
		\end{tabular}
	\end{center}
\end{table}
\noindent
Adding the results from table \ref{tab:ZchiYMGF}, we have at $\order{g_0^2}$ in $d=4-2\eps$:
\begin{equation}
\braket{\chi(t,x)\chibar(t,y)}\big|_{\order{g_0^2}}=-\frac{3}{\eps}C_2(R)\frac{g_0^2}{(4\pi)^2}S(\xbar_t-\ybar_t)+\order{\eps^0}
\end{equation}
and therefore the renormalization factor for fermions evolved with the flow equation \eqref{eq:gfeqnfermions}, while the gauge fields are evolved with the Yang-Mills gradient flow \eqref{eq:YMGF}, is given by
\begin{equation}
Z_\chi(g(\mu),\eps)=1+C_2(R)\frac{3}{\eps}\frac{g^2(\mu)}{(4\pi)^2}+\order{g^4}
\end{equation}
as is given in \eqref{eq:ZchiYM}. This calculation is generalized to the case where the gauge fields are evolved with the nonminimal gradient flow from \eqref{eq:NMGF2} in section \ref{sec:renormNMGF}, and the case where the fermions are evolved by the `slashed' flow equation \eqref{eq:DslashGFeqnfermions} together with the gauge fields evolved by the nonminimal gradient flow in section \ref{sec:renormsNMGF}. The generalized case is presented in \eqref{eq:GGF Z_chi}.

\section{Integrals from fermionic contribution to $\braket{E}$}\label{app:integralsFermContr}
The integrals are given by
\begin{align}
I_1&=\int_0^tds\int_{p,q}e^{-2t(p^2+q^2)-(2t-s)2p\cdot q}\frac{p\cdot q}{p^2q^2}\\
I_2&=\int_0^tds\int_{p,q}e^{-2t(p^2+q^2)-(2t-s)2p\cdot q}\frac{1}{(p+q)^2}\\
I_3&=\int_0^tds\int_{p,q}e^{-2t(p^2+q^2)-(2t-s)2p\cdot q}\frac{p\cdot q}{p^2(p+q)^2}\\
I_4&=\int_0^tds\int_{p,q}e^{-2t(p^2+q^2)-(2t-s)2p\cdot q}\frac{(p\cdot q)^2}{p^2q^2(p+q)^2}=\half I_1-I_3\\
I_5&=\int_0^tds_1\int_0^tds_2\int_{p,q}e^{-2t(p^2+q^2)-(2t-s_1-s_2)2p\cdot q}\\
I_6&=\int_0^tds_1\int_0^tds_2\int_{p,q}e^{-2t(p^2+q^2)-(2t-s_1-s_2)2p\cdot q}\frac{p\cdot q}{p^2}\\
I_7&=\int_0^tds_1\int_0^tds_2\int_{p,q}e^{-2t(p^2+q^2)-(2t-s_1-s_2)2p\cdot q}\frac{(p\cdot q)^2}{p^2q^2}\\
I_8&=\int_0^tds\int_0^sdu\int_{p,q}e^{-2tp^2-2sq^2-(s-u)2p\cdot q}\\
I_9&=\int_0^tds\int_0^sdu\int_{p,q}e^{-2tp^2-2sq^2-(s-u)2p\cdot q}\frac{(p\cdot q)^2}{p^2q^2}=\half I_2+\half I_3\\
I_{10}&=\int_0^tds\int_{p,q}e^{-2sp^2-2tq^2}\frac{1}{p^2}\\
I_{11}&=\int_0^tds\int_{p,q}e^{-2sp^2-2tq^2-s2p\cdot q}\frac{1}{p^2}
\end{align}
where we related different integrals by using
\begin{equation}
p\cdot q=\half((p+q)^2-p^2-q^2),\hspace{1cm}p\cdot q \, e^{s2p\cdot q}=\half\frac{d}{ds}\,e^{s2p\cdot q}
\end{equation}
Performing the integrals we obtain
\begin{align}
I_1&=-\frac{(4\pi)^{-d}t^{2-d}}{2(d-2)(4-d)}\left\{\frac{(2^{5-d}-2^3)}{(d-2)}\Gamma\left(\frac{d}{2}\right)\Gamma\left(3-\frac{d}{2}\right)+3^{2-\frac{d}{2}}{}_2F_1\left(1,1,3-\frac{d}{2},\frac{3}{4}\right)\right\}\\
&{}\nonumber\\
I_5&=-2^{2-2d}(4\pi)^{-d}t^{2-d}\bigg\{\beta\left(\frac{1}{4},1-\frac{d}{2},1-\frac{d}{2}\right)-4\beta\left(\frac{1}{4},2-\frac{d}{2},1-\frac{d}{2}\right)\nonumber\\
&\hspace{1cm}-2\beta\left(\half,1-\frac{d}{2},1-\frac{d}{2}\right)+4\beta\left(\half,2-\frac{d}{2},1-\frac{d}{2}\right)\\
&\hspace{1cm}-\frac{2^d}{(d-2)(4-d)}{}_2F_1\left(1-\frac{d}{2},\frac{d}{2},3-\frac{d}{2},\frac{1}{4}\right)\bigg\}\nonumber\\
&{}\nonumber\\
I_6&=(4\pi)^{-d}t^{2-d}\bigg\{\frac{3^{-1-\frac{d}{2}}}{2(d-2)}\left[3F_1\left(2,\frac{d}{2},\frac{d}{2},3,\frac{1}{3},-1\right)-F_1\left(3,\frac{d}{2},\frac{d}{2},4,\frac{1}{3},-1\right)\right]\nonumber\\
&\hspace{1cm}-\frac{2^{4-2d}}{(d-3)}\beta\left(\frac{1}{4},2-\frac{d}{2},1-\frac{d}{2}\right)+\frac{3^{1-\frac{d}{2}}}{2(d-2)(d-3)}\\
&\hspace{1cm}-\frac{2^{1-2d}3}{(d-2)}\left[\beta\left(\frac{1}{4},1-\frac{d}{2},1-\frac{d}{2}\right)-\beta\left(\half,1-\frac{d}{2},1-\frac{d}{2}\right)\right]
\bigg\}\nonumber\\
&{}\nonumber\\
I_7&=(4\pi)^{-d}t^{2-d}\bigg\{\frac{2^{2-d}}{(d-2)^2}-\frac{(2^2-2^{3-d})}{(d-2)^2(4-d)}\Gamma\left(\frac{d}{2}\right)\Gamma\left(3-\frac{d}{2}\right)\nonumber\\
&\hspace{1cm}+\frac{3^{2-\frac{d}{2}}}{2(d-2)(4-d)}{}_2F_1\left(1,1,3-\frac{d}{2},\frac{3}{4}\right)\bigg\}\\
&{}\nonumber\\
I_8&=(4\pi)^{-d}t^{2-d}\bigg\{\frac{2^{4-2d}}{(d-3)}\beta\left(\frac{1}{4},2-\frac{d}{2},1-\frac{d}{2}\right)-\frac{2^{1-d}}{(d-3)}{}_2F_1\left(\half,\frac{d}{2},\frac{3}{2},\frac{1}{4}\right)\bigg\}\\
&{}\nonumber\\
I_{10}&=(4\pi)^{-d}t^{2-d}\frac{2^{3-d}}{(4-d)(d-2)}\\
&{}\nonumber\\
I_{11}&=(4\pi)^{-d}t^{2-d}\bigg\{\frac{3^{1-\frac{d}{2}}}{(4-d)(d-3)}{}_2F_1\left(1,3-d,3-\frac{d}{2},\frac{1}{4}\right)-\frac{3^{1-\frac{d}{2}}}{(d-3)(d-2)}\bigg\}
\end{align}
where we use the incomplete beta function, the hypergeometric function and the Appell hypergeometric function, whose integral representations are given by
\begin{equation}
\beta(x,a,b)=\int_0^xdt\,t^{a-1}(1-t)^{b-1}=\frac{x^a}{a}{}_2F_1\left(a,1-b,a+1,x\right)
\end{equation}
\begin{equation}
{}_2F_1\left(a,b,c,z\right)=\frac{\Gamma(c)}{\Gamma(b)\Gamma(c-b)}\int_0^1dt\,\frac{t^{b-1}(1-t)^{c-b-1}}{(1-zt)^a}
\end{equation}
\begin{equation}
F_1(a,b_1,b_2,c,z_1,z_2)=\frac{\Gamma(c)}{\Gamma(a)\Gamma(c-a)}\int_0^1dt\,\frac{t^{a-1}(1-t)^{c-a-1}}{(1-z_1t)^{b_1}(1-z_2t)^{b_2}}
\end{equation}
Next we set $d=4-2\eps$ and expand in $\eps$. The following identities have been derived using results from \cite{Ancarani:2009zz}
\begin{align}
\frac{d}{dc}\,{}_2F_1\left(1,1,c,\frac{3}{4}\right)\bigg|_{c=1}&=-8\log 2\label{eq:derivativeofhypergeom}\\
\frac{d}{db}{}_2F_1\left(1,b,1,\frac{1}{4}\right)\bigg|_{b=-1}&=\frac{3}{2}\log 2 -\frac{3}{4}\log 3\\
F_1\left(2,2,2,3,\frac{1}{3},-1\right)&=\frac{9}{16}\log 3\\
\beta\left(\frac{1}{2};1-\frac{d}{2},1-\frac{d}{2}\right)&=\frac{2}{\eps}-2+\order{\eps}\\
\beta\left(\frac{1}{4};1-\frac{d}{2},1-\frac{d}{2}\right)&=\frac{2}{\eps}-\frac{14}{3}-2\log 3+\order{\eps}\\
\beta\left(\frac{1}{2};2-\frac{d}{2},1-\frac{d}{2}\right)&=\frac{1}{\eps}+1+\order{\eps}\\
\beta\left(\frac{1}{4};2-\frac{d}{2},1-\frac{d}{2}\right)&=\frac{1}{\eps}+\frac{1}{3}-\log 3+\order{\eps}
\end{align}
Employing these expressions, together with the usual gamma function expansions, we obtain
\begin{align}
I_1
&=-\frac{1}{8}t^{2-d}(4\pi)^{-d}\left\{\frac{1}{\eps}+1+4\log 3-6\log 2 +\order{\eps}\right\}\\
I_5&=\frac{1}{16}t^{2-d}(4\pi)^{-d}\left\{\frac{1}{\eps}+1+4\log 2-\log 3+\order{\eps}\right\}\\
I_6
&=-\frac{1}{16}t^{2-d}(4\pi)^{-d}\left\{\frac{1}{\eps}+1-2\log 3 +4\log 2+\order{\eps}\right\}\\
I_7
&=\frac{1}{16}t^{2-d}(4\pi)^{-d}\left\{\frac{1}{\eps}+2-14\log 2 +8\log 3+\order{\eps}\right\}\\
I_8&=\frac{1}{16}t^{2-d}(4\pi)^{-d}\left\{\frac{1}{\eps}+1+4\log 2-2\log 3 +\order{\eps}\right\}\\
I_{10}&=\frac{1}{8}t^{2-d}(4\pi)^{-d}\left\{\frac{1}{\eps}+1+2\log 2+\order{\eps}\right\}\\
I_{11}&=\frac{1}{8}t^{2-d}(4\pi)^{-d}\left\{\frac{1}{\eps}+1+4\log 2-\log 3+\order{\eps}\right\}
\end{align}
Plugging these expansions into \eqref{eq:totolFermContrIntegralsSum} --- optionally supplemented with \eqref{eq:fermioncontrDslashintegrals} --- we arrive at the total fermionic contribution \eqref{eq:totalFermContr} resp.~\eqref{eq:fermioncontrDslashepsexp}.

\section{$\braket{E}$ results for general $N$}\label{app:generalNresults}

Here we collect the results for $\braket{E}$ for the different gradient flow cases for general $N$ and $N_f$:
\begin{equation}\label{eq:<E>_igeneralN}
\braket{E}_i=\frac{3(N^2-1)}{32\pi t^2}\alpha(\mu)\left\{1+k_i(\mu^2t)\alpha(\mu)+\order{\alpha^2}\right\}
\end{equation}
with $\alpha(\mu)=\frac{g^2(\mu)}{4\pi}$ the renormalized $\MSbar$ coupling, see \eqref{eq:couplingRenormalizationMSbar}, and
\begin{subequations}\label{eq:k_isgeneralN}
\begin{align}
k_{YM}(\mu^2t)&=\frac{1}{4\pi}\left\{N\left(\frac{11}{3}L+\frac{52}{9}-3\log 3\right)-N_f\left(\frac{2}{3}L+\frac{4}{9}-\frac{4}{3}\log 2\right)\right\}\\
k_{NM}(\mu^2t)&=\frac{1}{4\pi}\left\{N\left(\frac{11}{3}L+\frac{52}{9}-3\log 3\right)-N_f\left(\frac{2}{3}L+\frac{22}{9}-\log 3\right)\right\}\\
k_{sNM}(\mu^2t)&=\frac{1}{4\pi}\left\{N\left(\frac{11}{3}L+\frac{52}{9}-3\log 3\right)-N_f\left(\frac{2}{3}L+\frac{22}{9}+4\log 2-3\log 3\right)\right\}
\end{align}
\end{subequations}
where $L=\log(8\mu^2t)+\gamma_E$.

Rewriting \eqref{eq:couplingRenormalizationMSbar} in terms of $\alpha$ we have
\begin{equation}
\alpha_0=\mu^{2\eps}\left(4\pi e^{-\gamma_E}\right)^{-\eps}\alpha(\mu)Z_\alpha(\alpha,\eps)
\end{equation}
with $Z_\alpha(\alpha,\eps)=Z_g^2(g,\eps)$ from \eqref{eq:b0etc}.
The renormalization group (RG) invariant coupling $\alpha(q)$ is implicitly defined by integrating the $d=4-2\eps$ beta function:
\begin{equation}
\int_{\alpha(\mu)}^{\alpha(q)}\frac{d\alpha}{\alpha\beta(\alpha,\eps)}=\log\left(\frac{q^2}{\mu^2}\right)        
\end{equation}
where
\begin{equation}
\beta(\alpha,\eps)=\frac{d\log\alpha}{d\log\mu^2}=-\eps+\beta(\alpha),\hspace{1cm} \beta(\alpha)=-\frac{d\log Z_\alpha}{d\log\mu^2}=-\frac{\beta_0}{4\pi}\alpha(\mu)+\dots
\end{equation}
with $\beta_0$ given below in \eqref{eq:beta0beta1}. The 1-loop perturbative expression for the RG invariant coupling $\alpha(q)$ in $d=4$ then reads
\begin{equation}\label{eq:alpha(q)intermsofalpha(mu)}
\alpha(q)=\alpha(\mu)-\frac{\beta_0}{4\pi}\alpha^2(\mu)\log\left(\frac{q^2}{\mu^2}\right)+\order{\alpha^3}
\end{equation}
We write \eqref{eq:<E>_igeneralN} in terms of the RG invariant coupling $\alpha(q)$ by inverting \eqref{eq:alpha(q)intermsofalpha(mu)}, and subsequently setting $q=(8t)^{-1/2}$ and $N=3$ we obtain \eqref{eq:<E>_i} and \eqref{eq:k_is}.

Equation \eqref{eq:<E>_igeneralN} can be expressed in terms of the fundamental scale of the nonevolved theory $\LQCD$ by inserting the 2-loop universal RG improved UV asymptotic expression for the coupling \cite{Caswell:1974gg}
\begin{equation}\label{eq:couplingInTermsOfLambda}
\alpha(q)=\frac{4\pi}{\beta_0\log\left(\frac{q^2}{\LQCD^2}\right)}\left\{1-\frac{\beta_1}{\beta_0^2}\frac{\log \log\left(\frac{q^2}{\LQCD^2}\right)}{\log\left(\frac{q^2}{\LQCD^2}\right)}\right\}+\mathcal{O}\left(\log\left(\frac{q^2}{\LQCD^2}\right)^{-3}\right)
\end{equation}
with
\begin{equation}\label{eq:beta0beta1}
\beta_0=\frac{11}{3}N-\frac{2}{3}N_f,\hspace{1cm}\beta_1=\frac{34}{3}N^2-\frac{10}{3}NN_f-\frac{N^2-1}{N}N_f
\end{equation}
thus yielding
\begin{equation}
\begin{split}
\braket{E}_i&=\frac{3(N^2-1)}{8t^2}\frac{1}{\beta_0\log\left(\frac{q^2}{\LQCD^2}\right)}\left\{1+\frac{1}{\beta_0\log\left(\frac{q^2}{\LQCD^2}\right)}\left(4\pi k_i(q^2t)-\frac{\beta_1}{\beta_0}\log\log\left(\frac{q^2}{\LQCD^2}\right)\right)\right\}\\
&\hspace{.5cm}+\mathcal{O}\left(\log\left(\frac{q^2}{\LQCD^2}\right)^{-3}\right)
\end{split}
\end{equation}
Again setting $N=3$ and $q=(8t)^{-1/2}$ we obtain \eqref{eq:<E>_iintermsofLambbdaN=3}.

\bibliography{mybib}{}

\providecommand{\href}[2]{#2}\begingroup\raggedright\begin{thebibliography}{10}

\bibitem{Atiyah:1982fa}
M.~Atiyah and R.~Bott, \emph{{The Yang-Mills equations over Riemann surfaces}},
  {\emph{Phil. Trans. Roy. Soc. Lond. A} {\bfseries 308} (1982) 523}.

\bibitem{Donaldson:1985zz}
S.~Donaldson, \emph{{Anti Self-Dual Yang-Mills Connections Over Complex
  Algebraic Surfaces and Stable Vector Bundles}},
  \href{https://doi.org/10.1112/plms/s3-50.1.1}{\emph{Proc. Lond. Math. Soc.}
  {\bfseries 50} (1985) 1}.

\bibitem{Narayanan:2006rf}
R.~Narayanan and H.~Neuberger, \emph{{Infinite N phase transitions in continuum
  Wilson loop operators}},
  \href{https://doi.org/10.1088/1126-6708/2006/03/064}{\emph{JHEP} {\bfseries
  03} (2006) 064} [\href{https://arxiv.org/abs/hep-th/0601210}{{\ttfamily
  hep-th/0601210}}].

\bibitem{Luscher:2009eq}
M.~Luscher, \emph{{Trivializing maps, the Wilson flow and the HMC algorithm}},
  \href{https://doi.org/10.1007/s00220-009-0953-7}{\emph{Commun. Math. Phys.}
  {\bfseries 293} (2010) 899}
  [\href{https://arxiv.org/abs/0907.5491}{{\ttfamily 0907.5491}}].

\bibitem{Luscher:2010iy}
M.~Lüscher, \emph{{Properties and uses of the Wilson flow in lattice QCD}},
  \href{https://doi.org/10.1007/JHEP08(2010)071}{\emph{JHEP} {\bfseries 08}
  (2010) 071} [\href{https://arxiv.org/abs/1006.4518}{{\ttfamily 1006.4518}}].

\bibitem{Lohmayer:2011si}
R.~Lohmayer and H.~Neuberger, \emph{{Continuous smearing of Wilson Loops}},
  \href{https://doi.org/10.22323/1.139.0249}{\emph{PoS} {\bfseries LATTICE2011}
  (2011) 249} [\href{https://arxiv.org/abs/1110.3522}{{\ttfamily 1110.3522}}].

\bibitem{Luscher:2011bx}
M.~Luscher and P.~Weisz, \emph{{Perturbative analysis of the gradient flow in
  non-abelian gauge theories}},
  \href{https://doi.org/10.1007/JHEP02(2011)051}{\emph{JHEP} {\bfseries 02}
  (2011) 051} [\href{https://arxiv.org/abs/1101.0963}{{\ttfamily 1101.0963}}].

\bibitem{Hieda:2016xpq}
K.~Hieda, H.~Makino and H.~Suzuki, \emph{{Proof of the renormalizability of the
  gradient flow}},
  \href{https://doi.org/10.1016/j.nuclphysb.2017.02.017}{\emph{Nucl. Phys. B}
  {\bfseries 918} (2017) 23}
  [\href{https://arxiv.org/abs/1604.06200}{{\ttfamily 1604.06200}}].

\bibitem{Luscher:2013cpa}
M.~Luscher, \emph{{Chiral symmetry and the Yang--Mills gradient flow}},
  \href{https://doi.org/10.1007/JHEP04(2013)123}{\emph{JHEP} {\bfseries 04}
  (2013) 123} [\href{https://arxiv.org/abs/1302.5246}{{\ttfamily 1302.5246}}].

\bibitem{Sommer:2014mea}
R.~Sommer, \emph{{Scale setting in lattice QCD}},
  \href{https://doi.org/10.22323/1.187.0015}{\emph{PoS} {\bfseries LATTICE2013}
  (2014) 015} [\href{https://arxiv.org/abs/1401.3270}{{\ttfamily 1401.3270}}].

\bibitem{Fodor:2017gtj}
Z.~Fodor, K.~Holland, J.~Kuti, D.~Nogradi and C.H.~Wong, \emph{{Extended
  investigation of the twelve-flavor $\beta$-function}},
  \href{https://doi.org/10.1016/j.physletb.2018.02.008}{\emph{Phys. Lett. B}
  {\bfseries 779} (2018) 230}
  [\href{https://arxiv.org/abs/1710.09262}{{\ttfamily 1710.09262}}].

\bibitem{Hasenfratz:2017mdh}
A.~Hasenfratz, C.~Rebbi and O.~Witzel, \emph{{Testing Fermion Universality at a
  Conformal Fixed Point}},
  \href{https://doi.org/10.1051/epjconf/201817503006}{\emph{EPJ Web Conf.}
  {\bfseries 175} (2018) 03006}
  [\href{https://arxiv.org/abs/1708.03385}{{\ttfamily 1708.03385}}].

\bibitem{Ramos:2015dla}
A.~Ramos, \emph{{The Yang-Mills gradient flow and renormalization}},
  \href{https://doi.org/10.22323/1.214.0017}{\emph{PoS} {\bfseries LATTICE2014}
  (2015) 017} [\href{https://arxiv.org/abs/1506.00118}{{\ttfamily
  1506.00118}}].

\bibitem{Parisi:1980ys}
G.~Parisi and Y.-s.~Wu, \emph{{Perturbation Theory Without Gauge Fixing}},
  {\emph{Sci. Sin.} {\bfseries 24} (1981) 483}.

\bibitem{Damgaard:1987rr}
P.H.~Damgaard and H.~Huffel, \emph{{Stochastic Quantization}},
  \href{https://doi.org/10.1016/0370-1573(87)90144-X}{\emph{Phys. Rept.}
  {\bfseries 152} (1987) 227}.

\bibitem{Tzani:1986ps}
R.~Tzani, \emph{{Evaluation of the Chiral Anomaly by the Stochastic
  Quantization Method}},
  \href{https://doi.org/10.1103/PhysRevD.33.1146}{\emph{Phys. Rev. D}
  {\bfseries 33} (1986) 1146}.

\bibitem{Harlander:2016vzb}
R.V.~Harlander and T.~Neumann, \emph{{The perturbative QCD gradient flow to
  three loops}}, \href{https://doi.org/10.1007/JHEP06(2016)161}{\emph{JHEP}
  {\bfseries 06} (2016) 161}
  [\href{https://arxiv.org/abs/1606.03756}{{\ttfamily 1606.03756}}].

\bibitem{Symanzik:1981wd}
K.~Symanzik, \emph{{Schrodinger Representation and Casimir Effect in
  Renormalizable Quantum Field Theory}},
  \href{https://doi.org/10.1016/0550-3213(81)90482-X}{\emph{Nucl. Phys. B}
  {\bfseries 190} (1981) 1}.

\bibitem{Caswell:1974gg}
W.E.~Caswell, \emph{{Asymptotic Behavior of Nonabelian Gauge Theories to Two
  Loop Order}}, \href{https://doi.org/10.1103/PhysRevLett.33.244}{\emph{Phys.
  Rev. Lett.} {\bfseries 33} (1974) 244}.

\bibitem{Collins:1984xc}
J.C.~Collins, \emph{{Renormalization}: {An Introduction to Renormalization, The
  Renormalization Group, and the Operator Product Expansion}}, vol.~26 of
  \emph{Cambridge Monographs on Mathematical Physics}, Cambridge University
  Press, Cambridge (1986),
  \href{https://doi.org/10.1017/CBO9780511622656}{10.1017/CBO9780511622656}.

\bibitem{Hasenfratz:2020vta}
A.~Hasenfratz and O.~Witzel, \emph{{Dislocations under gradient flow and their
  effect on the renormalized coupling}},
  \href{https://arxiv.org/abs/2004.00758}{{\ttfamily 2004.00758}}.

\bibitem{Ancarani:2009zz}
L.~Ancarani and G.~Gasaneo, \emph{{Derivatives of any order of the Gaussian
  hypergeometric function (2)F1(a, b, c z) with respect to the parameters a, b
  and c}}, \href{https://doi.org/10.1088/1751-8113/42/39/395208}{\emph{J. Phys.
  A} {\bfseries 42} (2009) 395208}.

\end{thebibliography}\endgroup
\bibliographystyle{JHEP}

\end{document}